\pdfoutput=1
\documentclass[final]{elsarticle}
\usepackage[T1]{fontenc}
\usepackage[utf8]{inputenc}
\usepackage{courier}
\usepackage{amsmath,amssymb}  

\usepackage[text={6.5in,9.0in},centering]{geometry}

\usepackage{textcomp}
\usepackage{wasysym}
\usepackage{tabularx}
\usepackage{multicol}
\usepackage{longtable,booktabs,colortbl}
\usepackage[vcentermath]{youngtab}
\usepackage{cancel}

\usepackage[pdftex,table,svgnames,hyperref]{xcolor}

\definecolor{darkred}{rgb}{0.5 0 0}
\definecolor{darkgreen}{rgb}{0.5 .5 0}
\definecolor{darkblue}{rgb}{0 0 .5}
\usepackage[hyphens]{url}
\usepackage[pdftex,bookmarksopen]{hyperref}
 \hypersetup{%
    pdftitle={LieART 2.0 - A Mathematica Application for Lie Algebras and Representation Theory}, %%
    pdfauthor={Robert Feger},%
    pdfsubject={},%
    pdfcreator={pdfTeX},%
    pdfproducer={Robert Feger},%
    pdfkeywords={},%
    colorlinks=true%
  }
\biboptions{sort&compress}
  
\colorlet{tableoverheadcolor}{gray!37.5}
\colorlet{tableheadcolor}{gray!40}
\colorlet{tablerowcolor}{gray!20}

\usepackage{lieart}
\usepackage{mathematica}
\usepackage{lieartpaper}
\usepackage{import}

\numberwithin{table}{section}

\journal{Computer Physics Communications}

\begin{document}

\begin{frontmatter}

\title{LieART 2.0 -- A Mathematica Application for Lie Algebras and Representation Theory}

\author[DWD]{Robert Feger\corref{author}}
\ead{robert.feger@gmail.com}
\cortext[author] {Corresponding author}
\author[VU]{Thomas W. Kephart}
\ead{tom.kephart@gmail.com}
\author[VU]{Robert J. Saskowski
\footnote{{present address: Leinweber Center for Theoretical Physics,  U. of Michigan, Ann Arbor, MI 48109}}}
\ead{robert.j.saskowski@vanderbilt.edu }
\address[DWD]{Deutscher Wetterdienst, Frankfurter Str. 135, 63067 Offenbach am Main, Germany}
\address[VU]{Department of Physics and Astronomy, Vanderbilt University, Nashville, TN 37235}

\begin{abstract}
We present LieART 2.0 which contains substantial extensions to the Mathematica application LieART (\underline{Lie}
\underline{A}lgebras and \underline{R}epresentation \underline{T}heory) for
computations frequently encountered in Lie algebras and representation theory,
such as tensor product decomposition and subalgebra branching of irreducible
representations. 
The basic procedure is unchanged---it computes root systems of Lie algebras, weight systems and several other properties of irreducible representations, but new features and procedures have been included to
allow the extensions to be seamless. The new version of LieART continues to be user friendly.
New extended tables of properties, tensor products and branching rules of
irreducible representations are included in the appendix for use without Mathematica software. LieART 2.0 now includes the branching rules to special subalgebras for all classical and exceptional Lie algebras up to and including rank 15.
\end{abstract}


\end{frontmatter}

\newpage
\tableofcontents
\newpage

%% This list environment is used for the references in the
%% Program Summary
%%
\newcounter{bla}
\newenvironment{refnummer}{%
\list{[\arabic{bla}]}%
{\usecounter{bla}%
 \setlength{\itemindent}{0pt}%
 \setlength{\topsep}{0pt}%
 \setlength{\itemsep}{0pt}%
 \setlength{\labelsep}{2pt}%
 \setlength{\listparindent}{0pt}%
 \settowidth{\labelwidth}{[9]}%
 \setlength{\leftmargin}{\labelwidth}%
 \addtolength{\leftmargin}{\labelsep}%
 \setlength{\rightmargin}{0pt}}}
 {\endlist}

% Computer program descriptions should contain the following
% PROGRAM SUMMARY.

{\bf PROGRAM SUMMARY}
  %Delete as appropriate.

\begin{small}
\noindent
{\em Authors:}
	Robert Feger,  Thomas W. Kephart and Robert J. Saskowski\\
{\em Program Title:}
	LieART 2.0\\
% {\em Journal Reference:}                                      \\
%   %Leave blank, supplied by Elsevier.
% {\em Catalogue identifier:}                                   \\
%   %Leave blank, supplied by Elsevier.
{\em Licensing provisions:}
  GNU Lesser General Public License (LGPL)\\
{\em Programming language:}
	Mathematica\\
{\em Computer:}
	x86, x86\_64, PowerPC\\
{\em Operating system:}
	cross-platform\\
{\em RAM:} 
	$\geq 4\,\text{GB}$ recommended. Memory usage depends strongly on the Lie algebra's 
	rank and type, as well as the dimensionality of the representations in the 
	computation.
{\em Classification:}
	4.2, 11.1\\
{\em External routines/libraries:}
	Wolfram Mathematica 8--12\\
{\em Nature of problem:}\\
	The use of Lie algebras and their representations is widespread in physics, 
	especially in particle physics. The description of nature in terms of gauge 
	theories requires the assignment of fields to representations of compact Lie 
	groups and their Lie algebras. Mass and interaction terms in the Lagrangian give 
	rise to the need for computing tensor products of representations of Lie 
	algebras. The mechanism of spontaneous symmetry breaking leads to the 
	application of subalgebra decomposition. This computer code was designed for the 
	purpose of Grand Unified Theory (GUT) model building, (where compact Lie groups 
	beyond the \U1, \SU2 and \SU3 of the Standard Model of particle physics are 
	needed), but it has found use in a variety of other applications. Tensor product 
	decomposition and subalgebra decomposition have been 
	implemented for all classical Lie groups \SU{N}, \SO{N} and \Sp{2N} and all the 
	exceptional groups \E6, \E7, \E8, \F4 and	\G2. This includes both regular and irregular 
	(special) subgroup decomposition of all Lie groups up through rank 15, and many more.\\
{\em Solution method:}\\
	LieART generates the weight system of an irreducible representation (irrep) of a 
	Lie algebra by exploiting the Weyl reflection groups, which is inherent in all 
	simple Lie algebras. Tensor products are computed by the application of Klimyk's 
	formula, except for \SU{N}'s, where the Young-tableaux algorithm is used. 
	Subalgebra decomposition of \SU{N}'s are performed by projection matrices, which 
	are generated from an algorithm to determine maximal subalgebras as originally 
	developed by Dynkin \cite{Dynkin:1957um,Dynkin:1957dm}. We generate projection matrices by the Dynkin procedure, i.e., removing dots from the Dynkin or extended Dynkin diagram, for regular subalgebras, and we implement explicit projection matrices for special subalgebras. \\
{\em Restrictions:}\\
	Internally irreps are represented by their unique Dynkin label. LieART's default 
	behavior in \texttt{TraditionalForm} is to print the dimensional name, which is 
	the labeling preferred by physicist. Most Lie algebras can have more than
	one irrep of the same dimension and different irreps with the same dimension are usually 
	distinguished by one or more primes~(e.g.~$\irrep{175}$ and $\irrep[1]{175}$ of \A4). 
	To determine the need for one or more primes of an irrep a brute-force loop over 
	other irreps must be performed to search for irreps with the same 
	dimensionality. Since Lie algebras have an infinite number of irreps, this loop 
	must be cut off, which is done by limiting the maximum Dynkin digit in the loop. 
	In rare cases for irreps of high dimensionality in high-rank algebras the used 
	cutoff is too low and the assignment of primes is incorrect. However, this only 
	affects the display of the irrep. All computations involving this irrep are 
	correct, since the internal unique representation of Dynkin labels is used.\\
{\em Running time:}\\
	From less than a second to hours depending on the Lie algebra's rank and type 
	and/or the dimensionality of the representations in the computation.\\
\end{small}
\newpage

\section{Introduction}

The application of group theory to symmetries associated with physical systems is one of the most important
tools of modern science. Examples abound in physics, chemistry and beyond into biology 
and engineering. Continuous groups in the form
of Lie groups are all important in physics, with smaller Lie groups 
entering quantum mechanics, physical chemistry and
condensed matter physics, while the full spectrum of Lie groups, i.e., the classical
groups \SU{N}, \SO{N} and \Sp{2N} and the exceptional groups \E6, \E7, \E8, \F4 and
\G2, have all appeared in particle physics.
Lie groups have had many other application in theoretical physics from
gravity, to string theory, M-theory, F-theory, and beyond. 
In LieART \cite{Feger:2015bs} the focus is on Lie algebras of compact Lie groups that
are most useful for particle physics. However, most of our results could
easily be extended to the non-compact forms. 

LieART is a convenient easy to use Mathematica application to explore Lie groups and their algebras.
It was designed to provide most of the Lie algebra information needed for particle physics model building.
In particular it can find tensor product rules and irreducible representation (irrep) decompositions for Lie algebras 
found in Grand Unified Theories (GUTs) and other common extensions of the standard model. But
LieART has also found application in many other areas of theoretical physics. LieART 2.0 is a response to the need for easy access to a more complete list of decompositions from Lie algebras to their maximal subalgebras. The remaining regular
maximal subalgebra decompositions not contained in LieART 1.0 have been added as well as the full list of irregular (special)
maximal subalgebra decompositions. The complete set of maximal subalgebra decompositions
is now included in the code through rank 15 of the initial algebra, and LieART 2.0 can compute almost all branching rules beyond rank 15 as well.

The gauge group of the Standard Model of particle physics is
$\SU{3}_\text{C}{\times}\SU{2}_\text{L}{\times}\U{1}_\text{Y}$.
Although the Standard Model is in agreement with most data, it is thought to be incomplete
with many unexplained free parameters and hints of new physics beyond the standard model. 
To determine these parameters, many extensions 
of the Standard-Model gauge group have been proposed. Low energy flavor 
models usually extend the symmetry, often with a discrete
 factor group but sometimes with a continuous--either global or local continuous group.

To understand high energy behavior, Grand Unified Theories 
were proposed, where extensions are via higher gauge symmetry, typically \SU5 \cite{Georgi:1974sy}, \SO{10}
\cite{Georgi:1974xy,Fritzsch:1974nn} or \E6 \cite{Gursey:1975ki}, although many other choices have
been explored. Reviews of Lie algebras can be found in \cite{Slansky:1981yr,
McKay:99021}, including tables of irreducible representations (irreps) and their
invariants. There are also numerous textbooks on the topic, e.g., see
\cite{Wybourne,Georgi:1982jb,Ramond:2010zz,cahn1984semi,Humphreys:1980dw} and a variety of software packages are available \cite{Schur,Lie,SimpLie,Nazarov:2011mv}.

While extensive tables 
already exist for building GUT models \cite{Slansky:1981yr,McKay:99021}, (for recent comprehensive results see \cite{Yamatsu:2015npn}) it has sometimes been necessary to go
beyond what is tabulated in the literature \cite{Feger:2015bs}. Our purpose here is to 
provide the software to handle any situation that may arise.
In describing this software we will incorporate a short review of much of the necessary
group-theory background including root and weight systems, the associated
Weyl groups for Lie algebras, orthogonal basis
systems, and group orbits, which we use to calculate
tensor products and irrep decompositions. More details can be found in \cite{Feger:2015bs}.

LieART has been used in many applications. Since it can provide Dynkin indices of irreps it is useful in obtaining renormalization group equations \cite{Lyonnet:2016xiz,Apruzzi:2018xkw,Bednyakov:2018cmx}.
In model building it has been used for extensions of the Standard Model in the way of BSM models \cite{Ferretti:2013kya,Mojaza:2014ppe,Matsumoto:2014ila,Barnard:2014tla,Lim:2014zfa,Hollands:2016kgm,ElHedri:2016onc}, grand unified models \cite{Albright:2012zt,Antipin:2015xia,Fonseca:2015aoa,Feger:2015xqa,Maurer:2015zpa,GonzaloVelasco:2015hki,Albright:2016lpi,Bajc:2016efj,Schwichtenberg:2017xhv,Yamatsu:2017mei,Benli:2017eld,Ernst:2018bib,Chala:2018ari,Yamatsu:2018tnv}, and other applications including SUSY models \cite{Fonseca:2013qka,Chen:2018wep,Agarwal:2018oxb}, supergravity models \cite{Abzalov:2015ega} and general particle physics models \cite{Panico:2015jxa,Kephart:2015oaa}. In UV completions of particle physics models, LieART has found uses in string theory \cite{Danielsson:2015rca,Nibbelink:2016wms}, holography \cite{Ahn:2017noo,Apruzzi:2017nck,Ahn:2018lqx}, (super-)conformal field theory \cite{deMedeiros:2013mca,Agarwal:2013uga,Chacaltana:2014jba,Hwang:2014uwa,Mitev:2014jza,Chacaltana:2016shw,Cordova:2016emh,Karateev:2017yoq,Jefferson:2017ahm}, M-theory \cite{Kim:2014nqa,Danielsson:2015tsa} and F-theory \cite{Martucci:2015oaa,Cvetic:2018xaq,Raghuram:2018hjn}. Beyond these applications LieART has been used in studying quantum groups \cite{Ahmed:2017mqq,Nepomechie:2017hgw,Nepomechie:2018dsn,Nepomechie:2018wzp}, instantons and other defects \cite{Kim:2018gjo}, condensed matter physics \cite{Bazzanella:2014vla}, as well as many other aspects of mathematical physics \cite{Kim:2012gu,Lemos:2012ph,Cremonesi:2013lqa,Fortunato:2013cta,Hanany:2014dia,Kim:2015jxc,Lehman:2015via,Hanany:2015hxa,Hoehn:2015zom,Matassa:2015gla,Linander:2016jyf,Urichuk:2016xau,Allys:2016kbq,Allys:2016hfl,Watanabe:2017bmi,Hayashi:2017jze,Niehoff:2017mbk,Moriyama:2017nbw,Mukhi:2017ugw,Shahlaei:2018lth,Liendo:2018ukf,Cheng:2018wll}. We have received valuable feedback from a number of researchers working in these areas and we have done our best to fix any bugs and incorporate suggested improvements where ever possible.

LieART's code exploits the Weyl reflection group to
make computations fast and economical with respect to memory. We 
focus on the usability of LieART with the user in mind:
Irreps can be entered by their dimensional name, or the more unique Dynkin label. 
LieART can display results in the form of \LaTeX\ commands for easy inclusion in publications (see the supplemental \LaTeX\ style file).

The paper is organized as follows: In Section~\ref{sec:DownloadAndInstallation} 
we give updated instructions for downloading and installing LieART, as well as locating 
its documentation integrated in Mathematica's help system. 
Section~\ref{sec:QuickStart} contains a quick-start tutorial for LieART, 
introducing the most important functions for the most common tasks in an 
example-based fashion. 
Section~\ref{sec:Implementation} explains the extension of the subalgebra branching and the handling of multiple branching rules.
Section~\ref{sec:TheoryAndImplementation} presents a 
self-contained overview of the Lie algebra theory used in LieART and gives notes 
on its implementation. A list of all LieART commands is provided. 
Section~\ref{sec:Results} summarizes and tabulates all regular and special subalgebras up to rank 15, which are implemented in LieART. Section~\ref{sec:Benchmarks} gives benchmarks for a few tensor-product decompositions and a subalgebra decomposition of a large irrep. 
In Section~\ref{LaTeXPackage} we present a \LaTeX\ style file included in LieART 
for displaying weights, roots and irreps properly. In 
Section~\ref{ConclusionsAndOutlook} we conclude and give an outlook on future 
versions. The appendix contains a collection of tables with 
properties of irreps, tensor products and branching rules for all maximal subalgebras of algebras through rank 15. These tables follow 
\cite{Slansky:1981yr} in style, but extend most of the 
results. The tables can be used to conveniently lookup many frequently used results and do not require the direct use of LieART.

\subsection*{Post Publication Note}
There are some minor disagreements between the LieART 2.0.2 code and the published paper
[R. Feger,T. W. Kephart and R. J. Saskowski,  Comput. Phys. Commun. 257, 107490 (2020)]
that describes it. The code has been altered to better align the normalizations and branching rules used by Slansky and Yamatsu. The code should be seen as taking precedent over the paper.

\pagebreak

\section{Download and Installation}
\label{sec:DownloadAndInstallation}

\subsection{Download}
LieART is hosted by Hepforge, IPPP Durham. The LieART project home page is

\href{http://lieart.hepforge.org/}{\texttt{http://lieart.hepforge.org/}}

and the LieART Mathematica application can be downloaded as archive from

\href{http://lieart.hepforge.org/downloads/}{\texttt{http://lieart.hepforge.org/downloads/}}

\subsection{Automatic Installation}

Start Mathematica and in the front end select the menu entry
\newcommand\nextstep{$\mathtt{\;\to\;}$}

   \texttt{File}\nextstep\texttt{Install\ldots}

In the appearing dialog select \texttt{Application} as \texttt{Type of Item to
Install} and the archive in the open file dialog from \texttt{Source}.
(It is not necessary to extract the archive since Mathematica does this automatically.)
Choose whether you want to install LieART for an individual user or system wide. For
a system-wide installation you might be asked for the superuser password.

\subsection{Manual Installation}

If problems with the automatic installation occur, proceed with a manual one as follows:
Extract the archive to the subdirectory \texttt{Applications} of the
directory to which \texttt{\$UserBaseDirectory} is set for a user-only
installation. For a system-wide installation place it in the subdirectory 
\texttt{AddOns/Applications} of \texttt{\$InstallationDirectory}. Restart Mathematica to allow
it to integrate LieART's documentation in its help system.

\subsection{Documentation}

The documentation of LieART is integrated in Mathematica's help system. After
restarting Mathematica the following path should lead to LieART's documentation:

\texttt{Help}\newline
\nextstep\texttt{Wolfram Documentation}\newline
\hphantom{\nextstep}\nextstep\texttt{Add-ons\:\&\:Packages} (at the bottom)\newline
\hphantom{\nextstep\nextstep}\nextstep\texttt{LieART}, Button labeled "\texttt{Documentation}"

(Alternatively, a search for ``LieART'' (with the correct case) in the Documentation Center leads to the same
page.) The displayed page serves as the documentation home of LieART and includes links
to the descriptions of its most important functions.

The documentation of LieART includes a \texttt{Quick Start Tutorial} for the impatient,
which can be found on the top right of LieART's documentation home under the
drop down \texttt{Tutorials}.

Tables of representation properties, tensor products and branching rules
generated by LieART can be found in the section \texttt{Tables} at the bottom of
LieART's documentation home.

\subsection{\LaTeX\ Package}

LieART comes with a \LaTeX\ package that defines commands to display irreps, roots and weights properly.
The style file \texttt{lieart.sty} can be found in the subdirectory \texttt{latex/} of the LieART project tree. Please copy it to a location where your \LaTeX\ installation can find it, which may be the directory
of your \LaTeX\ source file using it.

\section{Quick Start}
\label{sec:QuickStart}
\newcommand{\mmastring}[1]{\textcolor{gray}{#1}}

This section provides a tutorial introducing the most important and frequently used functions of LieART for Lie-algebra and representation-theory related calculations.
The functions are introduced based on simple examples that can easily be modified and extended to the user's desired application.
Most examples use irreducible representations (irreps) of \SU5, which most textbooks use in examples since is less trivial than \SU3, but small enough to return results
almost instantly on any recent computer. Also, \SU5 frequently appears in unified model building since the Standard-Model gauge group is one of its maximal subgroups.
This tutorial can also be found in the LieART documentation integrated into the Wolfram Documentation as \texttt{Quick Start Tutorial} under the drop down \texttt{Tutorials} on the upper right of
LieART's documentation home.

This loads the package:
\begin{mathin}
<<\:LieART`
\end{mathin}\par
\stepcounter{outcount}

\subsection{Entering Irreducible Representations}

Irreps are internally described by their Dynkin
label with a combined head of \com{Irrep} and the Lie algebra.
\definition{
    \com{Irrep[\args{algebraClass}][\args{label}]} & irrep described by its \args{algebraClass} and Dynkin \args{label}.
}{Entering irreps by Dynkin label.}

The \args{algebraClass} follows the Dynkin classification of simple Lie algebras
and can only be \com{A}, \com{B}, \com{C}, \com{D} for the classical algebras
and \com{E6}, \com{E7}, \com{E8}, \com{F4} and \com{G2} for the exceptional
algebras. The precise classical algebra is determined by the length of the
Dynkin label.

Entering the \irrepbar{10} of \SU5 by its Dynkin label and algebra class:
\begin{mathin}
Irrep[A][0,0,1,0]//FullForm
\end{mathin}
\begin{mathout}
Irrep[A][0,0,1,0]
\end{mathout}

In \com{StandardForm} the irrep is displayed in the textbook notation of Dynkin labels:
\begin{mathin}
Irrep[A][0,0,1,0]//StandardForm
\end{mathin}
\begin{mathout}
\dynkin{0,0,1,0}
\end{mathout}

In \com{TraditionalForm} (default) the irrep is displayed by its dimensional name:
\begin{mathin}
Irrep[A][0,0,1,0]
\end{mathin}
\begin{mathout}
\irrepbar{10}
\end{mathout}
The default output format type of LieART is \com{TraditionalForm}. The
associated user setting is overwritten for the front-end session LieART is loaded in. For
\com{StandardForm} as default output format type evaluate:
{\small\ttfamily\bfseries CurrentValue[\$FrontEndSession,\,\{CommonDefaultFormatTypes,\,"Output"\}]\,=\,StandardForm}

As an example for entering an irrep of an exceptional algebra, consider the \irrep{27} of \E6:
\begin{mathin}
Irrep[E6][1,0,0,0,0,0]
\end{mathin}
\begin{mathout}
\irrep{27}
\end{mathout}

Irreps may also be entered by their dimensional name. The package
transforms the irrep into its Dynkin label. Since the algebra of an irrep of a classical Lie algebra
becomes ambiguous with only the dimensional name, it has to be specified.
\definition{
    \com{Irrep[\args{algebra}][\args{dimname}]} & irrep entered by its \args{algebra} and dimensional name \args{dimname}.
}{Entering irreps by dimensional name.}
\pagebreak 

Entering the \irrepbar{10} of \SU5 by its dimensional name specifying the algebra by its Dynkin classification \A4:
\begin{mathin}
Irrep[A4][Bar[10]]//InputForm
\end{mathin}
\begin{mathout}
Irrep[A][0,0,1,0]
\end{mathout}

The traditional name of the algebra \SU{5} may also be used:
\begin{mathin}
Irrep[SU5][Bar[10]]//InputForm
\end{mathin}
\begin{mathout}
Irrep[A][0,0,1,0]
\end{mathout}

Irreps of product algebras like $\SU3{\otimes}\SU2{\otimes}\U1$ are specified by
\com{ProductIrrep} with the individual irreps of simple Lie algebras as arguments.
\definition{
    \com{ProductIrrep[\args{irreps}]} & head of product \args{irreps}, gathering irreps of simple Lie algebras.
}{Product irreps.}

The product irrep $(\irrep{3}, \irrepbar{3})$ of $\SU3{\otimes}\SU3$:
\begin{mathin}
ProductIrrep[Irrep[SU3][3],Irrep[SU3][Bar[3]]]
\end{mathin}
\begin{mathout}
(\irrep{3},\irrepbar{3})
\end{mathout}
\begin{mathin}
\%//InputForm
\end{mathin}
\begin{mathout}
ProductIrrep[Irrep[A][1,0],Irrep[A][0,1]]
\end{mathout}
\begin{mathin}
ProductIrrep[Irrep[A][1,0],Irrep[A][0,1]]
\end{mathin}
\begin{mathout}
(\irrep{3},\irrepbar{3})
\end{mathout}

Take for example the left-handed quark doublet in the Standard-Model gauge group
$\SU3{\otimes}\SU2{\otimes}\U1$ (The \U1 charge is not typeset in bold face):
\begin{mathin}
ProductIrrep[Irrep[SU3][3],Irrep[SU2][2],Irrep[U1][1/3]]
\end{mathin}
\begin{mathout}
(\irrep{3},\irrep{2})\!($\mathtt{1/3}$)
\end{mathout}
\begin{mathin}
 \%//InputForm
\end{mathin}
\begin{mathout}
ProductIrrep[Irrep[A][1,0],Irrep[A][1],Irrep[U][1/3]]
\end{mathout}

\subsection{Decomposing Tensor Products}

\definition{
    \com{DecomposeProduct[\args{irreps}]} & decomposes the tensor product of several \args{irreps}.
}{Tensor product decomposition.}

Decompose the tensor product $\irrep{3}{\otimes}\irrepbar{3}$ of \SU3:
\begin{mathin}
DecomposeProduct[Irrep[SU3][3],Irrep[SU3][Bar[3]]]
\end{mathin}
\begin{mathout}
$\irrep{1}+\irrep{8}$
\end{mathout}

Decompose the tensor product $\irrep{27}{\otimes}\irrepbar{27}$ of \E6:
\begin{mathin}
DecomposeProduct[Irrep[E6][27],Irrep[E6][Bar[27]]]
\end{mathin}
\begin{mathout}
$\irrep{1}+\irrep{78}+\irrep{650}$
\end{mathout}

Decompose the tensor product $\irrep{3}{\otimes}\irrep{3}{\otimes}\irrep{3}$ of \SU3:
\begin{mathin}
DecomposeProduct[Irrep[SU3][3],Irrep[SU3][3],Irrep[SU3][3]]
\end{mathin}
\begin{mathout}
$\irrep{1}+2(\irrep{8})+\irrep{10}$
\end{mathout}

Decompose the tensor product $\irrep{8}{\otimes}\irrep{8}$ of \SU3:
\begin{mathin}
DecomposeProduct[Irrep[SU3][8],Irrep[SU3][8]]
\end{mathin}
\begin{mathout}
$\irrep{1}+2(\irrep{8})+\irrep{10}+\irrepbar{10}+\irrep{27}$
\end{mathout}
Internally a sum of irreps is represented by \com{IrrepPlus} and \com{IrrepTimes}, an analog of the built-in functions \com{Plus} and \com{Times}:
\begin{mathin}
\%//InputForm
\end{mathin}
\par
\medskip
\begin{mathout}
IrrepPlus[Irrep[A][0,0],\:IrrepTimes[2,\:Irrep[A][1,1]],\linebreak  Irrep[A][3,\,0],\:Irrep[A][0,3],\:Irrep[A][2,2]]
\end{mathout}
The multiplicity of irreps in the sum can be extracted with \com{IrrepMultiplicity}:
\begin{mathin}
IrrepMultiplicity[\%,Irrep[SU3][8]]
\end{mathin}
\begin{mathout}
2
\end{mathout}
Results can also be transformed into a list of irreps with \com{IrrepList}, suitable for further processing with Mathematica built-in functions like \com{Select} or \com{Cases}:
\begin{mathin}
\%\%//IrrepList
\end{mathin}
\begin{mathout}
\{\irrep{1},\irrep{8},\irrep{8},\irrep{10},\irrepbar{10},\irrep{27}\}
\end{mathout}
Decompose the tensor product $\irrep{4}{\otimes}\irrep{4}{\otimes}\irrep{6}{\otimes}\irrep{15}$ of \SU4:
\begin{mathin}
DecomposeProduct[Irrep[SU4][4],Irrep[SU4][4],Irrep[SU4][6],Irrep[SU4][15]]
\end{mathin}
\begin{mathout}
$2(\irrep{1})+7(\irrep{15})+4(\irrep[1]{20})+\irrep{35}+5(\irrep{45})+3(\irrepbar{45})+3(\irrep{84})+2(\irrep{175})+\irrep{256}$
\end{mathout}

The Mathematica built-in command \com{Times} for products is replaced by 
\com{DecomposeProduct} for irreps as arguments. E.g., decompose the tensor 
product $\irrepbar{10}{\otimes}\irrep{24}{\otimes}\irrep{45}$ of \SU5:
\begin{mathin}
	Irrep[SU5][Bar[10]]*Irrep[SU5][24]*Irrep[SU5][45]
\end{mathin}
\begin{mathout}
$3(\irrepbar{5})+6(\irrepbar{45})+3(\irrepbar{50})+5(\irrepbar{70})+2(\irrepbar{105})+\irrepbar[2]{175}+6(\irrepbar{280})+2(\irrepbar[1]{280})+\irrepbar{420}+\irrepbar[1]{450}+3(\irrepbar{480})+2(\irrepbar{720})+\irrepbar{1120}+\irrepbar{2520}$
\end{mathout}

For powers of irreps the Mathematica built-in command \com{Power} may be used. 
E.g., decompose the tensor product 
$\irrep{27}{\otimes}\irrep{27}{\otimes}\irrep{27}$ of \E6:
\begin{mathin}
Irrep[E6][27]\textasciicircum3
\end{mathin}
\begin{mathout}
$\irrep{1}+2(\irrep{78})+3(\irrep{650})+\irrep{2925}+\irrep{3003}+2(\irrep{5824})$
\end{mathout}

Decompose tensor products of product irreps $(\irrep{3},\,\irrepbar{3},\,\irrep{1}){\otimes}(\irrepbar{3},\,\irrep{3},\,\irrep{1})$ of $\SU3{\otimes}\SU3{\otimes}\SU3$:
\begin{mathin}
DecomposeProduct[\linebreak ProductIrrep[Irrep[SU3][3],Irrep[SU3][Bar[3]],Irrep[SU3][1]],\linebreak ProductIrrep[Irrep[SU3][Bar[3]],Irrep[SU3][3],Irrep[SU3][1]]]
\end{mathin}
\begin{mathout}
$(\irrep{1},\irrep{1},\irrep{1})+(\irrep{8},\irrep{1},\irrep{1})+(\irrep{1},\irrep{8},\irrep{1})+(\irrep{8},\irrep{8},\irrep{1})$
\end{mathout}

\subsection{Decomposition to Subalgebras}

\definition{
    \com{DecomposeIrrep[\args{irrep},\,\args{subalgebra}]} & decomposes \args{irrep} to the specified \args{subalgebra}.\\
    \com{DecomposeIrrep[\args{pirrep},\,\args{subalgebra},\,\args{pos}]} & decomposes the product irrep \args{pirrep} at position \args{pos}.\\
}{Decompose irreps and product irreps.}

Decompose the \irrepbar{10} of \SU5 to $\SU3{\otimes}\SU2{\otimes}\U1$:
\begin{mathin}
DecomposeIrrep[Irrep[SU5][Bar[10]],ProductAlgebra[SU3,SU2,U1]]
\end{mathin}
\begin{mathout}
$(\irrep{1},\irrep{1})(6)+(\irrep{3},\irrep{1})(-4)+(\irrepbar{3},\irrep{2})(1)$
\end{mathout}

\pagebreak

Decompose the \irrep{10} and the \irrepbar{5} of \SU5 to $\SU3{\otimes}\SU2{\otimes}\U1$ (\com{DecomposeIrrep} is \com{Listable}):
\begin{mathin}
DecomposeIrrep[\{Irrep[SU5][10],Irrep[SU5][Bar[5]]\},ProductAlgebra[SU3,SU2,U1]]
\end{mathin}
\begin{mathout}
$\{(\irrepbar{3},\irrep{1})(4)+(\irrep{3},\irrep{2})(-1)+(\irrep{1},\irrep{1})(-6),\,(\irrepbar{3},\irrep{1})(-2)+(\irrep{1},\irrep{2})(3)\}$
\end{mathout}

Decompose the \irrep{16} of \SO{10} to $\SU5{\otimes}\U1$:
\begin{mathin}
DecomposeIrrep[Irrep[SO10][16],ProductAlgebra[SU5,U1]]
\end{mathin}
\begin{mathout}
$(\irrep{1})(-5)+(\irrepbar{5})(3)+(\irrep{10})(-1)$
\end{mathout}

Decompose the \irrep{27} of \E6 to $\SU3{\otimes}\SU3{\otimes}\SU3$:
\begin{mathin}
DecomposeIrrep[Irrep[E6][27],ProductAlgebra[SU3,SU3,SU3]]
\end{mathin}
\begin{mathout}
$(\irrep{3},\irrep{1},\irrep{3})+(\irrep{1},\irrep{3},\irrepbar{3})+(\irrepbar{3},\irrepbar{3},\irrep{1})$
\end{mathout}

Decompose the \SU3 irrep \irrep{3} in $(\irrep{24},\irrep{3})(-3)$ of $\SU5{\otimes}\SU3{\otimes}\U1$ to
$\SU2{\otimes}\text{U}'(1)$,\newline i.e., $\SU5{\otimes}\SU3{\otimes}\U1\to\SU5{\otimes}\SU2{\otimes}\text{U}'(1){\otimes}\U1$:
\begin{mathin}
DecomposeIrrep[ProductIrrep[Irrep[SU5][24],Irrep[SU3][3],Irrep[U1][-3]], ProductAlgebra[SU2,U1],2]
\end{mathin}
\begin{mathout}
$(\irrep{24},\irrep{1})(-2)(-3)+(\irrep{24},\irrep{2})(1)(-3)$
\end{mathout}

The same decomposition as above displayed as branching rule:
\begin{mathin}
IrrepRule[\slot,DecomposeIrrep[\slot,ProductAlgebra[SU2,U1],2]]\&@
 ProductIrrep[Irrep[SU5][24],Irrep[SU3][3],Irrep[U1][-3]]
\end{mathin}
\begin{mathout}
$(\irrep{24},\irrep{3})(-3)\to(\irrep{24},\irrep{1})(-2)(-3)+(\irrep{24},\irrep{2})(1)(-3)$
\end{mathout}

Branching rules for all totally antisymmetric irreps, so-called basic irreps, of \SU6 to $\SU3{\otimes}\SU3{\otimes}\U1$:
\begin{mathin}
IrrepRule[\slot,DecomposeIrrep[\slot,ProductAlgebra[SU3,SU3,U1]]]\&/@\,\newline BasicIrreps[SU6]//TableForm
\end{mathin}
\begin{mathout}\hangindent=0ex%
$\irrep{6}\rightarrow(\irrep{3},\irrep{1})(1)+(\irrep{1},\irrep{3})(-1)$\newline
$\irrep{15}\rightarrow(\irrepbar{3},\irrep{1})(2)+(\irrep{1},\irrepbar{3})(-2)+(\irrep{3},\irrep{3})(0)$\newline
$\irrep{20}\rightarrow(\irrep{1},\irrep{1})(3)+(\irrep{1},\irrep{1})(-3)+(\irrep{3},\irrepbar{3})(-1)+(\irrepbar{3},\irrep{3})(1)$\newline
$\irrepbar{15}\rightarrow(\irrep{3},\irrep{1})(-2)+(\irrep{1},\irrep{3})(2)+(\irrepbar{3},\irrepbar{3})(0)$\newline
$\irrepbar{6}\rightarrow(\irrepbar{3},\irrep{1})(-1)+(\irrep{1},\irrepbar{3})(1)$\newline
\end{mathout}
\vspace{-15pt}

\subsection{Young Tableaux}
\Yautoscale0
\Yboxdim13pt

The irreps of \SU{N} have a correspondence to Young tableaux, which can be displayed by \com{YoungTableau}.
\definition{
    \com{YoungTableau[\args{irrep}]} & Displays the Young tableau associated with an \SU{N} \args{irrep}.\\
}{Young tableaux.}

Young tableau of the \irrep{720} of \SU5:
\begin{mathin}
YoungTableau[Irrep[A][1,2,0,1]]
\end{mathin}
\begin{mathout}\label{out:YoungTableau720SU5}
\large\yng(4,3,1,1)
\end{mathout}

Display Young tableaux of \SU4 irreps with a maximum of one column per box count:
\begin{mathin}
Row[Row[\{\textcolor{DarkGreen}{\#},"{:}\ ",YoungTableau[\textcolor{DarkGreen}{\#}]\}]\&/@\newline
SortBy[Irrep[A]@@@Tuples[\{0,1\},3],Dim],Spacer[10]]
\end{mathin}
\newcommand{\irrepandtableau}[2]{\irrep{#1}{:}\:#2\quad}
\begin{mathout}
$\irrepandtableau{\irrep{1}}{\bullet}
\irrepandtableau{\irrepbar{4}}{\yng(1,1,1)}
\irrepandtableau{\irrep{4}}{\yng(1)}
\irrepandtableau{\irrep{6}}{\yng(1,1)}
\irrepandtableau{\irrep{15}}{\yng(2,1,1)}
\irrepandtableau{\irrep{20}}{\yng(2,2,1)}
\irrepandtableau{\irrepbar{20}}{\yng(2,1)}
\irrepandtableau{\irrep{64}}{\yng(3,2,1)}$
\end{mathout}

\subsection{Tables of LieART Commands}
\label{subsec:TablesOfLieARTCommands}

While various LieART commands are scattered throughout the text, we provide the complete set here for easy future reference.
    \subsubsection{Basic Algebra Properties}
	\definition{
		\com{Algebra[\args{algebraClass}][\args{rank}]}     & represents a classical algebra of the type \args{algebraClass}, which can only be \com{A}, \com{B}, \com{C} or \com{D}, with \args{rank}.\\
		\com{Algebra[\args{expr}]}                          & algebra (classical or exceptional) of \args{expr}, which may be an irrep, a weight or a root in any basis.\\
		\com{Rank[\args{expr}]}                             & rank of the algebra of \args{expr}, which can be an irrep, a weight a root or an algebra itself.\\
		\com{OrthogonalSimpleRoots[\args{algebra}]}         & simple roots of \args{algebra} in the orthogonal basis.\\
		\com{CartanMatrix[\args{algebra}]}                  & Cartan matrix of \args{algebra}.\\
		\com{OmegaMatrix[\args{algebra}]}                   & matrix of fundamental weights of \args{algebra} as rows.\\
		\com{OrthogonalFundamentalWeights[\args{algebra}]}  & fundamental weights of \args{algebra} in the orthogonal basis.\\
		\com{OrthogonalBasis[\args{expr}]}                  & transforms \args{expr} from any basis into the orthogonal basis.\\
		\com{OmegaBasis[\args{expr}]}                       & transforms \args{expr} from any basis into the $\omega$-basis. \\
		\com{AlphaBasis[\args{expr}]}                       & transforms \args{expr} from any basis into the $\alpha$-basis.\\
		\com{DMatrix[\args{algebra}]}                       & matrix with inverse length factors of simple roots on the main diagonal.\\
		\com{ScalarProduct[\args{weight1},\args{weight2}]}  & scalar product of \args{expr1} and \args{expr2} in any basis. \args{expr1} and \args{expr2} may be weights or roots.\\
		\com{MetricTensor[\args{algebra}]}                  & metric tensor or quadratic-form matrix of \args{algebra}.\\
		}{\ }
    
    \subsubsection{Weyl Group Orbits}
	\definition{
		\com{Reflect[\args{weightOrRoot},\args{simpleroots}]}       & reflects \args{weightOrRoot} at the hyperplanes orthogonal to the specified \args{simpleroots}.\\
		\com{Reflect[\args{weightOrRoot}]}                          & reflects \args{weightOrRoot} at the hyperplanes orthogonal to all simple roots of the Lie algebra of \args{weightOrRoot}.\\
		\com{ReflectionMatrices[\args{algebra}]}                    & reflection matrices of the Weyl group of \args{algebra}.\\
		\com{Orbit[\args{weightOrRoot},\args{simpleroots}]}         & generates the Weyl group orbit of \args{weightOrRoot} using only the specified \args{simpleroots}.\\
		\com{Orbit[\args{weightOrRoot}]}                            & generates the full Weyl group orbit of \args{weightOrRoot} using all simple roots of the Lie algebra of \args{weightOrRoot}.\\
		\com{DimOrbit[\args{weightOrRoot},\args{simpleroots}]}      & size of the orbit of \args{weightOrRoot} using only the \args{simpleroots}.\\
		\com{DimOrbit[\args{weightOrRoot}]}                         & size of the orbit of \args{weightOrRoot} using all simple roots of the Lie algebra of \args{weightOrRoot}.\\
	}{\ }
\pagebreak
    \subsubsection{Roots}
	\definition{
		\com{RootSystem[\args{algebra}]}            & root system of \args{algebra}\\
		\com{ZeroRoots[\args{algebra}]}             & zero roots associated with the Cartan subalgebra of \args{algebra}\\
		\com{Height[\args{root}]}                   & height of a \args{root} within the root system\\
		\com{HighestRoot[\args{algebra}]}           & highest root of the root system of \args{algebra}\\
		\com{PositiveRoots[\args{algebra}]}         & only the positive roots of \args{algebra}\\
		\com{PositiveRootQ[\args{root}]}            & gives \com{True} if \args{root} is a positive root and \com{False} otherwise\\
		\com{NumberOfPositiveRoots[\args{algebra}]} & number of positive roots of \args{algebra}\\
		\com{Delta[\args{algebra}]}                     & half the sum of positive roots of \args{algebra} ($\delta{=}\dynkincomma{1,1,\ldots}$)\\
	}{\ }
	
    \subsubsection{Weights}
	\definition{
		\com{Weight[\args{algebraClass}][\args{label}]}  & weight in the $\omega$-basis defined by its \args{algebraClass} and Dynkin \args{label}\\
		\com{WeightOrthogonal[\args{algebraClass}][\args{label}]}  & weight in the orthogonal basis defined by its \args{algebraClass} and Dynkin \args{label}\\
		\com{WeightAlpha[\args{algebraClass}][\args{label}]}  & weight in the $\alpha$-basis defined by its \args{algebraClass} and Dynkin \args{label}\\
	}{\ }
	
	\subsubsection{Basic Properties of Irreps}
	\definition{
		\com{Irrep[\args{algebraClass}][\args{label}]}  & irrep described by its \args{algebraClass} and Dynkin \args{label}\\
		\com{Irrep[\args{algebra}][\args{dimname}]}     & irrep entered by its \args{algebra} and \args{dimname}\\
		\com{ProductIrrep[\args{irreps}]}               & head of product \args{irreps}\\
		\com{DynkinLabel[\args{irrep}]}  & gives the Dynkin label of \args{irrep}\\
		\com{WeightLevel[\args{weight},\args{irrep}]}          & level of the \args{weight} within the \args{irrep}\\
		\com{Height[\args{irrep}]}                      & height of \args{irrep}\\
		\com{SingleDominantWeightSystem[\args{irrep}]}          & dominant weights of \args{irrep} without their multiplicities\\
		\com{WeightMultiplicity[\args{weight},\args{irrep}]}   & computes the multiplicity of \args{weight} within \args{irrep}\\
		\com{DominantWeightSystem[\args{irrep}]}                & dominant weights of \args{irrep} with their multiplicities\\
		\com{WeightSystem[\args{irrep}]}                        & full weight system of \args{irrep}\\
		\com{WeylDimensionFormula[\args{algebra}]}      & explicit Weyl dimension formula for \args{algebra}\\
		\com{Dim[\args{irrep}]}                         & dimension of \args{irrep}\\
		\com{DimName[\args{irrep}]}                     & dimensional name of \args{irrep}\\
		\com{Index[\args{irrep}]}                       & index of \args{irrep}\\
		\com{CasimirInvariant[\args{irrep}]}&  (new in LieART 2.0) Casimir invariant of an irrep\\
		\com{CongruencyClass[\args{irrep}]}             & congruency class number of \args{irrep}\\
	}{\ }
	
	\pagebreak
	
	\subsubsection{Tensor Product Decomposition}
	\definition{
		\com{DecomposeProduct[\args{irreps}]} & decomposes the tensor product of \args{irreps}\\
		\com{DominantWeightsAndMul[\args{weights}]} & filters and tallies dominant weights of \args{weights} by multiplicities\\
		\com{SortOutIrrep[\args{dominantWeightsAndMul}]} & sorts out the irrep of largest height from the collection of dominant weights \args{dominantWeightsAndMul}\\
		\com{WeightSystemWithMul[\args{irrep}]} & weight system of \args{irrep} with multiplicities\\
		\com{TrivialStabilizerWeights[\args{weights}]} & drops weights that lie on a chamber wall\\
		\com{ReflectToDominantWeightWithMul[\args{weightAndMul}]} & reflects \args{weightAndMul} to the dominant chamber and multiplies the parity of the reflection to the multiplicity\\
		\com{IrrepMultiplicity[\args{decomp},\args{irrep}]}&  (new in LieART 2.0) gives the multiplicity of an irrep, e.g., in a decomposition
	}{\ }

% \pagebreak
	\subsubsection{Subalgebra Decomposition of Irreps and Product Algebra Irreps}
	\definition{
		\com{DecomposeIrrep[\args{irrep},\,\args{subalgebra}]} & decomposes \args{irrep} to \args{subalgebra}.\\
		\com{DecomposeIrrep[\args{irrep},\,\args{subalgebra},\,\args{index}]} & decomposes \args{irrep} to the specified \args{subalgebra}. \args{index} specifies which branching rule to use, for algebra--subalgebra pairs with multiple branching rules. Only applicable for \SU{15} $\to$ \SU3, \E7 $\to$ \SU2, and \linebreak \E8 $\to$ \SU2.\\
		\com{DecomposeIrrep[\args{productIrrep},\,\args{subalgebra},\,\args{pos}]} & decomposes \args{productIrrep} at position \args{pos}.\\
		\com{DecomposeIrrep[\args{productIrrep},\,\args{subalgebra},\,\args{pos},\,\args{index}]} & decomposes \args{productIrrep} at position \args{pos}. \args{index} specifies which branching rule to use, for algebra--subalgebra pairs with multiple branching rules. Only applicable for ...$\times$\SU{15}$\times$... $\to$ ...$\times$\SU3$\times$..., ...$\times$\E7$\times$... $\to$ ...$\times$\SU2$\times$..., and ...$\times$\E8$\times$... $\to$ ...$\times$\SU2$\times$....
	}{\ }
		
\pagebreak

	\subsubsection{Subalgebra Projection}
	\definition{
		\com{ProjectionMatrix[\args{origin},\args{target}]} & defines the projection matrix for the algebra--subalgebra pair specified by \args{origin} and \args{target}\\
		\com{Project[\args{projectionMatrix},\args{weights}]} & applies \args{projectionMatrix} to \args{weights}\\
		\com{GroupProjectedWeights[\args{projectedWeights},\args{target}]} & groups the projected weights according to the subalgebra specified in \args{target}\\
		\com{NonSemiSimpleSubalgebra[\args{origin},\args{simpleRootToDrop}]} & computes the projection matrix of a maximal non-semi-simple subalgebra by dropping one dot of the Dynkin diagram \args{simpleRootToDrop} and turning it into a \U1 charge\\
		\com{SemiSimpleSubalgebra[\args{origin},\args{simpleRootToDrop}]} & computes the projection matrix of a maximal semi-simple subalgebra by dropping one dot from the extended Dynkin diagram.\\
		\com{ExtendedWeightScheme[\args{algebra},\args{simpleRootToDrop}]} & adds the Dynkin label associated with the extended simple root ${-}\gamma$ to each weight of the lowest orbit of \args{algebra} and drops the simple root \args{simpleRootToDrop}\\
		\com{SpecialSubalgebra[\args{origin},\args{targetirreps}]} & computes the projection matrix of a maximal special subalgebra of \args{origin} by specifying the branching rule of the generating irrep with \args{targetirreps}.
	}{\ }
	
% 	\newpage
	
	\subsubsection{Branching Rules Tables}
	(load with \com{<<\,LieART`Tables`})
	\definition{
		\com{BranchingRulesTable[\args{algebra},\,\args{subalgebras}]} & constructs a table of branching rules for the decomposition of representations of \args{algebra} to \args{subalgebras}.\\
		\com{BranchingRulesTable[\args{algebra},\,\args{subalgebras},\args{index}]} & constructs a table of branching rules for the decomposition of representations of \args{algebra} to \args{subalgebras}; \args{index} specifies which branching rule to use, for algebra--subalgebra pairs with multiple branching rules. Only applicable for \SU{15} $\to$ \SU3, \E7 $\to$ \SU2, and \E8 $\to$ \SU2.
	}{\ }
	
	\subsubsection{Young Tableaux}
	\definition{
		\com{YoungTableau[\args{irrep}]} & Displays the Young tableau associated with an \SU{N} \args{irrep}.\\
	}{}
\pagebreak

\section{Implementation: Extension of the Subalgebra Branching}
\label{sec:Implementation}

The maximal subalgebras of simple Lie algebras are thoroughly categorized by Dynkin \cite{Dynkin:1957um,Dynkin:1957dm} and Yamatsu \cite{Yamatsu:2015npn}, hence the main body of this work has been on computing the projection matrices for these branchings. These projection matrices are different for each algebra--maximal-subalgebra pair and are not unique. However, once a projection matrix is known it can be used for the decomposition of all irreps of the algebra--subalgebra pair.

The previous version of LieART already came with some functions for calculating projection matrices: \com{NonSemiSimpleSubalgebra[\args{algebra},\args{simpleRootToDrop}]} computes the projection matrix of a maximal non-semisimple subalgebra by dropping one dot of the Dynkin diagram, \args{simpleRootToDrop}, and turning it into a U(1) charge. \com{SemiSimpleSubalgebra[\args{algebra}, \args{simpleRootToDrop}]} computes the projection matrix of a maximal semisimple subalgebra by dropping one dot, \args{simpleRootToDrop}, from the extended Dynkin diagram. \com{SpecialSubalgebra[\args{algebra}, \args{targetirreps}]} computes the projection matrix of a maximal special subalgebra of \args{algebra} by specifying the branching rule of the generating irrep with \args{targetirreps}\cite{Feger:2015bs}. However, \linebreak\com{NonSemiSimpleSubalgebra} and \com{SemiSimpleSubalgebra} are only applicable to regular subalgebras, and \com{SpecialSubalgebra} requires exact knowledge of the branching rules for the generating irrep. Thus, many projection matrices, or rather their general form, were explicitly put into LieART, with the form of the projection matrices being found in \cite{Yamatsu:2015npn} and \cite{Kim:1982jg}. One issue, however, was that some of these projection matrices do not properly map between weight systems in LieART's conventions; this was also a problem already present for \com{NonSemiSimpleSubalgebra} and \com{SemiSimpleSubalgebra}. However, this issue can be (and was) solved by permuting some of the rows of the projection matrices to ensure that the irrep decomposition works properly.

Additionally, LieART already had some functionality for finding branching rules. Using projection matrices, the LieART function \com{DecomposeIrrep[\args{irrep}, \args{subalgebra}]} decomposes an \args{irrep} of a simple Lie algebra into a maximal \args{subalgebra}, which can be simple, semisimple, or non-semisimple. To decompose an irrep of a semisimple or non-semisimple irrep, a third argument, \args{pos}, allows one to specify which part of \args{productIrrep} should be decomposed into the subalgebra, i.e., \com{DecomposeIrrep[\args{productIrrep},\,\args{subalgebra},\,\args{pos}]} \cite{Feger:2015bs}. However, some subalgebra branchings have multiple branching rules, namely, $\E7 \to \SU2$ has two different sets of branching rules, $\E8 \to \SU2$ has three different sets of branching rules, and $\SU{15} \to \SU3$ has two different sets of branching rules. Thus, we modified \com{DecomposeIrrep} to take an \args{index} parameter, which determines the branching rules to use: \com{DecomposeIrrep[\args{productIrrep},\,\args{subalgebra},\,\args{pos},\,\args{index}]}. If no \args{index} is specified, \com{DecomposeIrrep} defaults to an \args{index} of 1. An example is shown below. 

%decompose irrep example
\begin{mathin}
	DecomposeIrrep[Irrep[E7][56], SU2]
\end{mathin}
\begin{mathout}
	\irrep{10}+\irrep{18}+\irrep{28}
\end{mathout}
\begin{mathin}
	DecomposeIrrep[Irrep[E7][56], SU2, 1]
\end{mathin}
\begin{mathout}
	\irrep{10}+\irrep{18}+\irrep{28}
\end{mathout}
\begin{mathin}
	DecomposeIrrep[Irrep[E7][56], SU2, 2]
\end{mathin}
\begin{mathout}
	\irrep{6}+\irrep{12}+\irrep{16}+\irrep{22}
\end{mathout}
\begin{mathin}
	DecomposeIrrep[ProductIrrep[Irrep[E7][56], Irrep[SU2][1]], SU2, 1, 1]
\end{mathin}
\begin{mathout}
	(\irrep{10},\irrep{1})+(\irrep{18},\irrep{1})+(\irrep{28},\irrep{1})
\end{mathout}
\begin{mathin}
	DecomposeIrrep[ProductIrrep[Irrep[E7][56], Irrep[SU2][1]], SU2, 1, 2]
\end{mathin}
\begin{mathout}
	(\irrep{6},\irrep{1})+(\irrep{12},\irrep{1})+(\irrep{16},\irrep{1})+(\irrep{22},\irrep{1})
\end{mathout}

The command \com{BranchingRulesTable[\args{algebra}, \args{subalgebras}]} in the subpackage \com{LieART`Tables`} (load with \com{<<\,LieART`Tables`}) of the previous LieART version constructs a series of tables of the branching rules from \args{algebra} to each element of \args{subalgebras} \cite{Feger:2015bs}. However, because of the aforementioned multiple branching rules for some algebra--subalgebra pairs, we modified the command to take an \args{index} parameter as well, which determines the branching rules to display: \com{BranchingRulesTable[\args{algebra},\,\args{subalgebras},\,\args{index}]}. If no \args{index} is specified it displays branching rules tables for all of the branchings. An example is shown below.

%branching rules table example
{
\setlength\extrarowheight{1.3pt}
\renewcommand{\tabcolsep}{2pt}
\rowcolors{2}{tablerowcolor}{}
\begin{mathin}
<< LieART`Tables`
\end{mathin}
\begin{mathin}
	BranchingRulesTable[E8, \{SU2\}, 1, MaxDim -> 147250]
\end{mathin}
\begin{mathout}
    \begin{tabularx}{\textwidth}{rcX}
		\toprule
		\rowcolor{tableheadcolor}
		\E8& $\to$ &\SU2\\
		\midrule 
		\irrep{248} & = & $\irrep{3}+\irrep{11}+\irrep{15}+\irrep{19}+\irrep{23}+\irrep{27}+\irrep{29}+\irrep{35}+\irrep{39}+\irrep{47}$ \\
		\irrep{3875} & = & $2(\irrep{1})+3(\irrep{5})+\irrep{7}+4(\irrep{9})+2(\irrep{11})+6(\irrep{13})+3(\irrep{15})+6(\irrep{17})+4(\irrep{19})+7(\irrep{21})+4(\irrep{23})+7(\irrep{25})+5(\irrep{27})+7(\irrep{29})+5(\irrep{31})+6(\irrep{33})+4(\irrep{35})+7(\irrep{37})+4(\irrep{39})+5(\irrep{41})+3(\irrep{43})+5(\irrep{45})+3(\irrep{47})+4(\irrep{49})+2(\irrep{51})+3(\irrep{53})+2(\irrep{55})+2(\irrep{57})+\irrep{59}+2(\irrep{61})+\irrep{63}+\irrep{65}+\irrep{69}+\irrep{73}$ \\
		\irrep{27000} & = & $7(\irrep{1})+\irrep{3}+13(\irrep{5})+7(\irrep{7})+19(\irrep{9})+14(\irrep{11})+25(\irrep{13})+19(\irrep{15})+29(\irrep{17})+23(\irrep{19})+33(\irrep{21})+26(\irrep{23})+35(\irrep{25})+28(\irrep{27})+36(\irrep{29})+28(\irrep{31})+35(\irrep{33})+28(\irrep{35})+34(\irrep{37})+27(\irrep{39})+31(\irrep{41})+24(\irrep{43})+28(\irrep{45})+22(\irrep{47})+25(\irrep{49})+18(\irrep{51})+21(\irrep{53})+15(\irrep{55})+18(\irrep{57})+12(\irrep{59})+14(\irrep{61})+9(\irrep{63})+11(\irrep{65})+7(\irrep{67})+8(\irrep{69})+5(\irrep{71})+6(\irrep{73})+3(\irrep{75})+4(\irrep{77})+2(\irrep{79})+3(\irrep{81})+\irrep{83}+2(\irrep{85})+\irrep{89}+\irrep{93}$ \\
		\irrep{30380} & = & $10(\irrep{3})+6(\irrep{5})+17(\irrep{7})+14(\irrep{9})+24(\irrep{11})+22(\irrep{13})+30(\irrep{15})+26(\irrep{17})+35(\irrep{19})+31(\irrep{21})+37(\irrep{23})+34(\irrep{25})+40(\irrep{27})+34(\irrep{29})+40(\irrep{31})+34(\irrep{33})+38(\irrep{35})+34(\irrep{37})+36(\irrep{39})+30(\irrep{41})+33(\irrep{43})+27(\irrep{45})+29(\irrep{47})+24(\irrep{49})+25(\irrep{51})+19(\irrep{53})+21(\irrep{55})+16(\irrep{57})+16(\irrep{59})+13(\irrep{61})+13(\irrep{63})+9(\irrep{65})+10(\irrep{67})+6(\irrep{69})+7(\irrep{71})+5(\irrep{73})+5(\irrep{75})+2(\irrep{77})+3(\irrep{79})+2(\irrep{81})+2(\irrep{83})+\irrep{85}+\irrep{87}+\irrep{91}$ \\
		\irrep{147250} & = & $8(\irrep{1})+22(\irrep{3})+41(\irrep{5})+49(\irrep{7})+69(\irrep{9})+80(\irrep{11})+93(\irrep{13})+102(\irrep{15})+118(\irrep{17})+121(\irrep{19})+133(\irrep{21})+138(\irrep{23})+144(\irrep{25})+147(\irrep{27})+153(\irrep{29})+149(\irrep{31})+153(\irrep{33})+151(\irrep{35})+149(\irrep{37})+144(\irrep{39})+144(\irrep{41})+134(\irrep{43})+132(\irrep{45})+124(\irrep{47})+118(\irrep{49})+110(\irrep{51})+105(\irrep{53})+94(\irrep{55})+89(\irrep{57})+81(\irrep{59})+73(\irrep{61})+66(\irrep{63})+61(\irrep{65})+51(\irrep{67})+47(\irrep{69})+41(\irrep{71})+36(\irrep{73})+30(\irrep{75})+27(\irrep{77})+21(\irrep{79})+19(\irrep{81})+16(\irrep{83})+12(\irrep{85})+10(\irrep{87})+9(\irrep{89})+6(\irrep{91})+5(\irrep{93})+4(\irrep{95})+3(\irrep{97})+2(\irrep{99})+2(\irrep{101})+\irrep{105}+\irrep{107}$ \\
		\bottomrule
	\end{tabularx}
    % 	\caption{\label{tab:E8BranchingRulesIndex1}\E8 Branching Rules}
\end{mathout}
}

{
\setlength\extrarowheight{1.3pt}
\renewcommand{\tabcolsep}{2pt}
\rowcolors{2}{tablerowcolor}{}
\begin{mathin}
	BranchingRulesTable[E8, \{SU2\}, 2, MaxDim -> 147250]
\end{mathin}
\begin{mathout}
    \begin{tabularx}{\textwidth}{rcX}
		\toprule
		\rowcolor{tableheadcolor}
		\E8& $\to$ &\SU2\\
		\midrule 
		\irrep{248} & = & $\irrep{3}+\irrep{7}+\irrep{11}+\irrep{15}+\irrep{17}+\irrep{19}+2(\irrep{23})+\irrep{27}+\irrep{29}+\irrep{35}+\irrep{39}$ \\
\irrep{3875} & = & $3(\irrep{1})+5(\irrep{5})+3(\irrep{7})+6(\irrep{9})+4(\irrep{11})+9(\irrep{13})+6(\irrep{15})+9(\irrep{17})+7(\irrep{19})+10(\irrep{21})+7(\irrep{23})+10(\irrep{25})+7(\irrep{27})+9(\irrep{29})+6(\irrep{31})+8(\irrep{33})+5(\irrep{35})+7(\irrep{37})+5(\irrep{39})+5(\irrep{41})+3(\irrep{43})+4(\irrep{45})+2(\irrep{47})+3(\irrep{49})+\irrep{51}+2(\irrep{53})+\irrep{55}+\irrep{57}+\irrep{61}$ \\
\irrep{27000} & = & $9(\irrep{1})+4(\irrep{3})+20(\irrep{5})+16(\irrep{7})+31(\irrep{9})+25(\irrep{11})+40(\irrep{13})+33(\irrep{15})+47(\irrep{17})+39(\irrep{19})+50(\irrep{21})+43(\irrep{23})+51(\irrep{25})+42(\irrep{27})+50(\irrep{29})+40(\irrep{31})+46(\irrep{33})+37(\irrep{35})+41(\irrep{37})+32(\irrep{39})+35(\irrep{41})+26(\irrep{43})+29(\irrep{45})+20(\irrep{47})+22(\irrep{49})+16(\irrep{51})+16(\irrep{53})+11(\irrep{55})+12(\irrep{57})+7(\irrep{59})+8(\irrep{61})+4(\irrep{63})+5(\irrep{65})+2(\irrep{67})+3(\irrep{69})+\irrep{71}+2(\irrep{73})+\irrep{77}$ \\
\irrep{30380} & = & $\irrep{1}+15(\irrep{3})+13(\irrep{5})+29(\irrep{7})+26(\irrep{9})+39(\irrep{11})+38(\irrep{13})+48(\irrep{15})+45(\irrep{17})+55(\irrep{19})+50(\irrep{21})+57(\irrep{23})+52(\irrep{25})+57(\irrep{27})+49(\irrep{29})+54(\irrep{31})+46(\irrep{33})+48(\irrep{35})+41(\irrep{37})+42(\irrep{39})+34(\irrep{41})+35(\irrep{43})+27(\irrep{45})+26(\irrep{47})+21(\irrep{49})+21(\irrep{51})+14(\irrep{53})+15(\irrep{55})+10(\irrep{57})+9(\irrep{59})+7(\irrep{61})+6(\irrep{63})+3(\irrep{65})+4(\irrep{67})+\irrep{69}+2(\irrep{71})+\irrep{73}+\irrep{75}$ \\
\irrep{147250} & = & $14(\irrep{1})+39(\irrep{3})+71(\irrep{5})+88(\irrep{7})+118(\irrep{9})+138(\irrep{11})+158(\irrep{13})+173(\irrep{15})+194(\irrep{17})+199(\irrep{19})+213(\irrep{21})+217(\irrep{23})+222(\irrep{25})+221(\irrep{27})+223(\irrep{29})+213(\irrep{31})+212(\irrep{33})+202(\irrep{35})+192(\irrep{37})+180(\irrep{39})+172(\irrep{41})+154(\irrep{43})+144(\irrep{45})+130(\irrep{47})+117(\irrep{49})+103(\irrep{51})+93(\irrep{53})+78(\irrep{55})+69(\irrep{57})+59(\irrep{59})+49(\irrep{61})+40(\irrep{63})+35(\irrep{65})+26(\irrep{67})+22(\irrep{69})+17(\irrep{71})+13(\irrep{73})+10(\irrep{75})+8(\irrep{77})+4(\irrep{79})+4(\irrep{81})+3(\irrep{83})+\irrep{85}+\irrep{87}+\irrep{89}$ \\
		\bottomrule
	\end{tabularx}
    % 	\caption{\label{tab:E8BranchingRulesIndex1}\E8 Branching Rules}
\end{mathout}
}
\pagebreak

\section{Review of Theoretical Background and Implementation}
\label{sec:TheoryAndImplementation}

In this section we review the Lie algebra theory used
and implemented in LieART \cite{Feger:2015bs}, specifically, the basic
properties of Lie algebras, roots, weights, Weyl orbits, representations and decompositions.
Relevant LieART functions can be found in section \ref{subsec:TablesOfLieARTCommands}. 
See also  \cite{Feger:2015bs} for a more detailed discussion of their use.
For a pedagogical introduction to Lie algebras and we refer the
reader to the  literature  \cite{Slansky:1981yr,
Georgi:1982jb, cahn1984semi}.

\subsection{Algebras}
\label{ssec:Algebras}

\newcommand{\liebracket}[2]{\left[#1, #2\right]}

\subsubsection{Definition}

A \emph{Lie Algebra} is a vector space $g$ over a field $F$ with an associated \emph{Lie
bracket} $\left[\cdot,\cdot\right]$, which is a binary operation that is bilinear,
alternating and fulfills the Jacobi identity (e.g., commutators, cross products).   The Lie brackets of the generators $t_i$ of the Lie
algebra are
\begin{equation}\label{eq:DefinitionStructureConstants}
    \liebracket{t_i}{t_j} = f_{ijk} t_k
\end{equation}
with the  \emph{structure constants} $f_{ijk}$,  which fully determine
the algebra. A Lie algebra is called \emph{simple} if it contains no
non-trivial ideals and \emph{semi-simple}  if it is a sum of simple Lie algebras (or, equivalently, if it contains no non-trivial Abelian ideals).

\subsubsection{Roots}

The generators $t_i$   fall into
two sets: One set is called  the  \emph{Cartan subalgebra}, $H$, which contains all
simultaneously diagonalizable generators $h_i$ where
\begin{equation}
    h_i=h_i^\dagger, \qquad\liebracket{h_i}{h_j} = 0, \qquad  i,j=1,\ldots,n
\end{equation}
and where $n$ is called the
\emph{rank} of the algebra, which is determined by the function
\com{Rank[\args{expr}]} in LieART. All other generators, $e_\alpha$
satisfy  eigenvalue equations with  $h_i$ of the form
\begin{equation}\label{eq:DefinitionRootVector}
    \liebracket{h_i}{e_\alpha} = \alpha_i\,e_\alpha, \qquad  i=1,\ldots,n,
\end{equation}
which is a subset of \eqref{eq:DefinitionStructureConstants}. Thus, the
$\alpha_i$ are real structure constants,  due to the
hermiticity of the $h_i$'s. The vectors
$\alpha{=}(\alpha_1,\ldots,\alpha_n)$ are called the \emph{root vectors}, which
lie in   \emph{root space}.
\emph{Roots} are functionals mapping the Cartan subalgebra $H$ onto the real
numbers, for all generators $t_i$.

A zero root with an $n$-fold degeneracy is associated with the Cartan
subalgebra. In the Cartan--Weyl basis the $e_\alpha$ generators come in conjugated
pairs $e_\alpha^\dagger{=}e_{{-}\alpha}$ and correspond to  ladder operators,
i.e.,  raising operator  \emph{positive roots} 
$e_\alpha$ and  lowering operators  \emph{negative roots} $e_{{-}\alpha}$.  

Linearly independent positive roots   are called \emph{simple roots} and a Lie algebra has as many
simple roots as its rank, i.e.,  the simple roots fully
determines a Lie algebra.  

\subsubsection{Classification of Lie Algebras}

\begin{figure}[t]
    \begin{center}
        \includegraphics{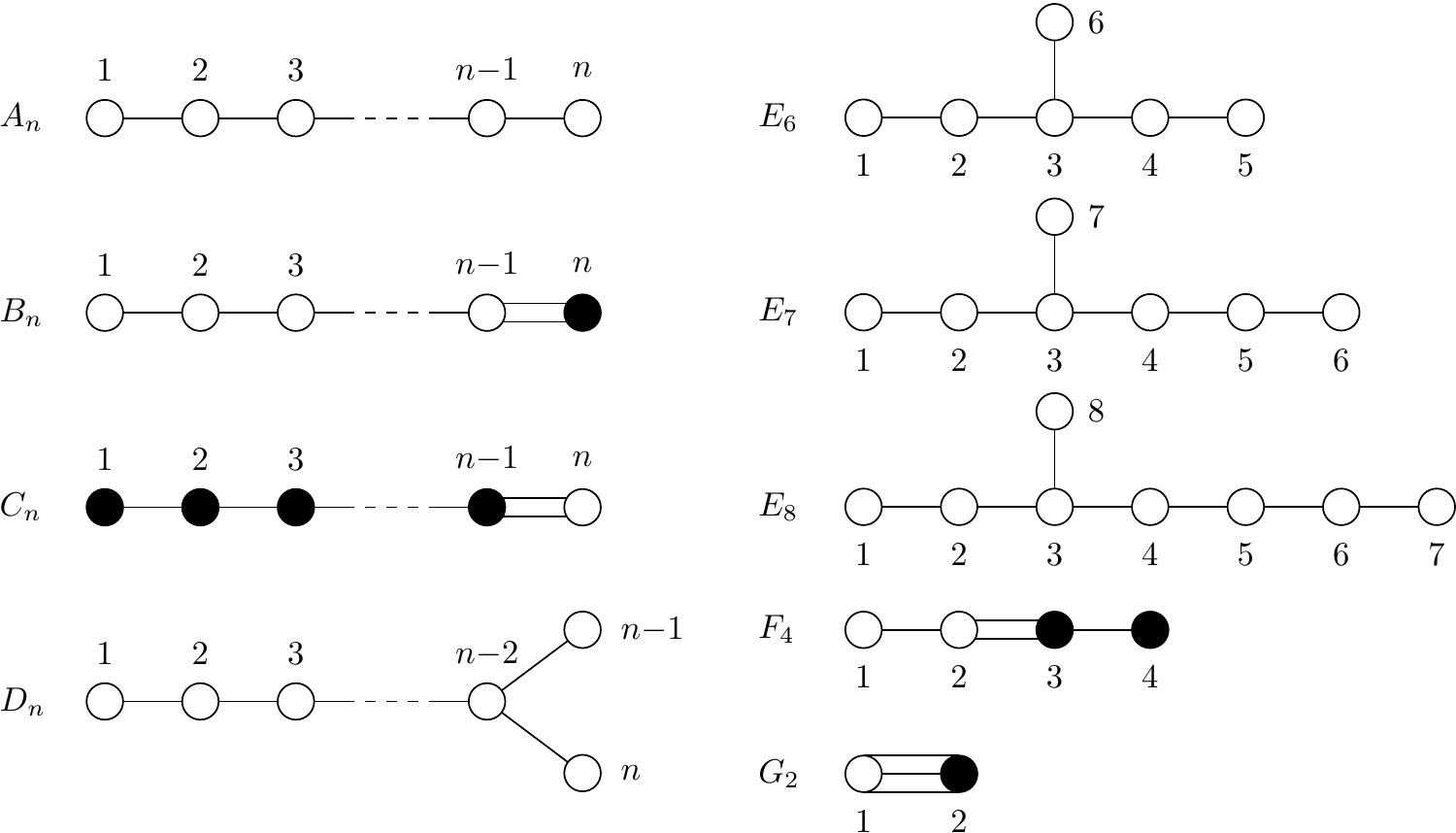}
        \caption{\label{fig:DynkinDiagrams} Dynkin Diagrams of classical and exceptional simple Lie algebras.}
    \end{center}
\end{figure}
The commutation relations and the Jacobi identity can be used to find all possible
root systems, which is equivalent to identifying all allowable Lie algebras.
Simple roots are, in general, not  orthogonal and come in at most two lengths 
 and   four different angles between  pairs.   \emph{Dynkin diagrams}   depict these relations: 
 simple longer (or for all roots if
they only come in one length) roots are represented by open dots,
 {\Large $\Circle$},  while filled dots, {\Large $\CIRCLE$}, represent the shorter
roots. Angles between two simple roots are represented by lines connecting the
dots: no line for an angle of 90\textdegree, one line for 120\textdegree, two
lines for 135\textdegree\ and three for 150\textdegree. Figure
\ref{fig:DynkinDiagrams} shows the Dynkin diagrams for all simple Lie algebras.

The simple Lie algebras fall into two types: four  infinite series
\A{n}, \B{n}, \C{n} and \D{n},  called the \emph{classical Lie
algebras} and five   \emph{exceptional algebras}, \E6, \E7, \E8, \F4 and
\G2, with their rank as subscript (see Table
\ref{tab:LieAlgebrasClassification}).

\begin{table}[h]
    \begin{center}
    \rowcolors{2}{tablerowcolor}{}
    \begin{tabular}{lllll}\toprule\rowcolor{tableheadcolor}
    \textbf{Type}   & \textbf{Cartan} & \textbf{Name} & \textbf{Rank}   & \textbf{Description}\\\midrule
    classical       & \A{n}           & \SU{n{+}1}    & $n\geq1$        & Special unitary algebras of $n{+}1$ complex dimension\\
                    & \B{n}           & \SO{2n{+}1}   & $n\geq3$        & Special orthogonal algebras of odd ($2n{+}1$) real dimension\\
                    & \C{n}           & \Sp{2n}       & $n\geq2$        & Symplectic algebras of even ($2n$) complex dimension\\
                    & \D{n}           & \SO{2n}       & $n\geq4$        & Special orthogonal algebras of even ($2n$) real dimension\\\midrule
    exceptional     & \E6             & \E6           & 6               & Exceptional algebra of rank 6\\
                    & \E7             & \E7           & 7               & Exceptional algebra of rank 7\\
                    & \E8             & \E8           & 8               & Exceptional algebra of rank 8\\
                    & \F4             & \F4           & 4               & Exceptional algebra of rank 4\\
                    & \G2             & \G2           & 2               & Exceptional algebra of rank 2\\
    \bottomrule
    \end{tabular}
    \caption{\label{tab:LieAlgebrasClassification} Classification of simple Lie algebras.}
    \end{center}
\end{table}
This classification is due to Cartan. A few of the lower dimensional algebras (but not necessarily the associated Lie group) are isomorphic; see \cite{Gilmore} for a discussion. 

The classical Lie algebras are internally represented (i.e., in \com{FullForm}) by \com{Algebra[\args{algebraClass}][\args{n}]}, with \args{algebraClass} being either \com{A}, \com{B}, \com{C} or \com{D} and \args{n} being the rank. The exceptional algebras are defined in LieART as \com{E6}, \com{E7}, \com{E8}, \com{F4} and \com{G2} and short forms of the classical algebras are predefined up to rank 30 for the Cartan classification, i.e., \com{A1}$,\ldots,$\com{A30}, \com{B3}$,\ldots,$\com{B30}, \com{C2}$,\ldots,$\com{C30}, \com{D4}$,\ldots,$\com{D30}, 
and up to dimension 30 for the conventional name, i.e., \com{SU2}$,\ldots,$\com{SU30}, 
\com{SO7}$,\ldots,$\com{SO30} and \com{Sp4}, \com{Sp6}$,\ldots,$\com{Sp30}. In \com{StandardForm} the Cartan classification is explicitly displayed and in \com{TraditionalForm} the Lie algebra is written by its conventional name.    

\subsubsection{Bases}
\label{ssec:Bases}

It is convenient to express the root space in an
orthogonal coordinate system.  For \A{n} it is a subspace of
$\mathbb{R}^{n+1}$, where the coordinates sum to one. As the simple roots define the Lie
algebra, they are explicitly specified in LieART using orthogonal coordinates
and can be retrieved by \com{OrthogonalSimpleRoots[\args{algebra}]}.\linebreak
\pagebreak

E.g., the four simple roots of \A4, i.e., \SU5, in orthogonal coordinates are: 
\begin{mathin}
OrthogonalSimpleRoots[A4]//Column
\end{mathin}
\begin{mathout}
\nohangingindent
\rootorthogonal{1, {-}1, 0, 0, 0}\newline
\rootorthogonal{0, 1, {-}1, 0, 0}\newline
\rootorthogonal{0, 0, 1, {-}1, 0}\newline
\rootorthogonal{0, 0, 0, 1, {-}1}
\end{mathout}

The   \emph{Cartan matrix} 
\begin{equation}\label{eq:CartanMatrix}
    A_{ij} = \frac{2 \scalarproduct{\alpha_i}{\alpha_j}}{\scalarproduct{\alpha_j}{\alpha_j}}\qquad i,j=1,\ldots,n
\end{equation}
exhibits the non-orthogonality of the simple
roots, where   $\scalarproduct{\cdot}{\cdot}$ is the ordinary scalar
product in $\mathbb{R}^{n+1}$ for \A{n}. The rows of the Cartan matrix are the simple roots in the so-called
\emph{$\omega$-basis}, which is the bases of \emph{fundamental weights}, also
called the \emph{Dynkin basis}.   The Cartan matrix is implemented in LieART as the function
\com{CartanMatrix[\args{algebra}]} following the definition of
\eqref{eq:CartanMatrix}. The Cartan matrix for \A4 reads:
\begin{mathin}
CartanMatrix[A4]
\end{mathin}
\begin{mathout}
$\begin{pmatrix}
 2 & {-}1 & 0 & 0 \\
 {-}1 & 2 & {-}1 & 0 \\
 0 & {-}1 & 2 & {-}1 \\
 0 & 0 & {-}1 & 2 \\
\end{pmatrix}$
\end{mathout}
Besides the orthogonal basis, and the $\omega$-basis, the $\alpha$-basis is also
useful.   It is the basis of simple roots and it  
shows how, e.g., a root is composed  of simple roots.   The Cartan matrix
mediates between the $\omega$- and $\alpha$-bases:
\begin{equation}
    \alpha_i
    = \sum_{j=1}^n A_{ij}\omega_j, \qquad \omega_i
    = \sum_{j=1}^n (A^{{-}1})_{ij}\alpha_j.
\end{equation}
where the $\omega_i$ are called the fundamental weights.
These bases can be scaled to be dual by defining the $\emph{coroot}$  $\alpha_i^\vee$ of the $\alpha_i$:
\begin{equation}\label{eq:AlphaOmegaDual}
    \frac{2\scalarproduct{\alpha_i}{\omega_j}}{\scalarproduct{\alpha_i}{\alpha_i}}
    \equiv \scalarproduct{\alpha_i^\vee}{\omega_j}
    = \delta_{ij},\qquad i,j=1,\ldots,n
\end{equation}
where  
\begin{equation}
    \alpha^\vee = \frac{2\alpha}{\scalarproduct{\alpha}{\alpha}}.
\end{equation}
The transformation to the orthogonal basis expressing $\alpha_i$ and $\omega_j$ in orthogonal
coordinates as $\hat\alpha_i$ and $\hat\omega_j$ equation \eqref{eq:AlphaOmegaDual} reads
\begin{equation}\label{eq:AlphaOmegaDualOrthogonal}
    \frac{2 \hat\alpha_i\cdot\hat\omega_j}{\hat\alpha_i\cdot\hat\alpha_i}
    \equiv \hat\alpha_i^\vee\cdot\hat\omega_j
    = \delta_{ij},\qquad i,j=1,\ldots,n.
\end{equation}
In matrix form we can write  
\eqref{eq:AlphaOmegaDualOrthogonal} as:
\begin{equation}\label{eq:AlphaOmegaDualOrthogonalMatrices}
    \matrixhat{A}\transpose{\matrixhat{\Omega}} = I_n,
\end{equation}
where both $\matrixhat{A}$ and $\matrixhat\Omega$ are $n{\times}m$ matrices.
The dimension of the orthogonal space $m$ is not necessarily
the same as the rank of the algebra $n$. These exceptions are: \A{n} with
$m{=}n+1$, \E6 with $m{=}8$,  \E7 with $m{=}8$ and \G2 with $m{=}3$.  
The matrix of the simple coroots in the orthogonal basis
$\matrixhat{A}$ is easily calculated from the simple roots given in LieART, but
the matrix of fundamental weights in the orthogonal basis $\matrixhat{\Omega}$
must be determined by \eqref{eq:AlphaOmegaDualOrthogonalMatrices}. In the cases
where $\matrixhat{A}$ is not a square matrix its inverse does not exist.  
The Mathematica built-in function \com{PseudoInverse[\args{matrix}]} yields the
right-inverse for our case of a \args{matrix} with linear independent rows. 
The matrix of the fundamental weights $\matrixhat{\Omega}$ is implemented as
\com{OmegaMatrix[\args{algebra}]}, e.g., for \A4:
 
\begin{mathin}
    OmegaMatrix[A4]
\end{mathin}
\begin{mathout}
\setlength\extrarowheight{2pt}
    $\begin{pmatrix}
        \frac{4}{5} & {-}\frac{1}{5} & {-}\frac{1}{5} & {-}\frac{1}{5} & {-}\frac{1}{5} \\
        \frac{3}{5} & \frac{3}{5} & {-}\frac{2}{5} & {-}\frac{2}{5} & {-}\frac{2}{5} \\
        \frac{2}{5} & \frac{2}{5} & \frac{2}{5} & {-}\frac{3}{5} & {-}\frac{3}{5} \\
        \frac{1}{5} & \frac{1}{5} & \frac{1}{5} & \frac{1}{5} & {-}\frac{4}{5}
    \end{pmatrix}$
\end{mathout}

\enlargethispage{2ex}

The LieART commands \com{OrthogonalBasis[\args{weightOrRoot}]},  \com{OmegaBasis[\args{weightOrRoot}]} and \linebreak \com{AlphaBasis[\args{weightOrRoot}]}
transform \args{weightOrRoot} from
any basis into the orthogonal basis, the $\omega$-basis and the
 $\alpha$-basis, respectively.

\subsubsection{Scalar Product}

The standard choice for the length factors $\scalarproduct{\alpha_j}{\alpha_j}$
in \eqref{eq:CartanMatrix} is 2 for the longer roots, if there are two root
lengths. The factors $2/\scalarproduct{\alpha_j}{\alpha_j}$ can only take three
different values which are: 1 for all roots of \A{n}, \D{n}, \E6, \E7, \E8 and
for the long roots of \B{n}, \C{n}, \F4 and \G2; 2 for the short roots of \B{n},
\C{n} and \F4 and 3 for the short root of \G2. Their implementation in LieART
is in the form of diagonal matrices $D$ with the inverse factors for the simple roots
corresponding to the row on the main diagonal, i.e.,
\begin{equation}
D=\text{diag}\left(\frac{1}{2}\scalarproduct{\alpha_1}{\alpha_1},\ldots, \frac{1}{2}\scalarproduct{\alpha_n}{\alpha_n}\right)
\end{equation}
as defined in \cite{klimyk_orbit_2006}.   The matrix
\begin{equation}
    G_{ij} = (A^{{-}1})_{ij} \frac{\scalarproduct{\alpha_j}{\alpha_j}}{2} = (A^{{-}1})_{ij} D_j
\end{equation}
is called the \emph{quadratic-form matrix} or \emph{metric tensor} of the Lie
algebra.   The LieART function for the metric tensor $G$ is
\com{MetricTensor[\args{algebra}]}, e.g., for \A4:
\begin{mathin}
MetricTensor[A4]
\end{mathin}
\begin{mathout}
\setlength\extrarowheight{2pt}
$\begin{pmatrix}
 \frac{4}{5} & \frac{3}{5} & \frac{2}{5} & \frac{1}{5} \\
 \frac{3}{5} & \frac{6}{5} & \frac{4}{5} & \frac{2}{5} \\
 \frac{2}{5} & \frac{4}{5} & \frac{6}{5} & \frac{3}{5} \\
 \frac{1}{5} & \frac{2}{5} & \frac{3}{5} & \frac{4}{5} \\
\end{pmatrix}$
\end{mathout}

\subsubsection{Definition of a Representation}
\label{ssec:Representations}

Simple Lie algebras have an infinite number of representations. If the set of  matrices that make up the representation can not be
simultaneously diagonalized, the representations are irreducible, or an  \emph{irrep}.
A \emph{representation} is a  homomorphism that maps the generators $t_i$ onto invertible matrices
$T_i$, that satisfy the same ``commutation'' relations as the Lie algebra,
namely
\begin{equation}\label{eq:DefinitionRepresentation}
    \liebracket{T_i}{T_j} = f_{ijk} T_k.
\end{equation}

% Analogously to roots, functionals mapping the Cartan subalgebra to eigenvalues
% can be defined, which are called \emph{weights}. They are defined in
% \emph{weight space}, which is the the same as the roots space and are labeled by
% \emph{weight vectors}.

An eigenvalue of a matrix representing a generator of the Cartan
subalgebra  is called a \emph{weight vector}, and the
associated functional \emph{weight}, denoted by $\lambda$. 
Weights can be expressed as rational linear combinations of
roots, and eventually by simple roots.  
The structure functions themselves form an irreducible representation of the
algebra: the \emph{adjoint representation}, which has the same dimension as the
algebra.

\subsection{Weyl Group Orbits}

The finite group $W(L)$, called the Weyl group of the Lie algebra $L$, is a
reflection group inherent in the root systems of all simple Lie algebras. 
The Dynkin diagram describes the Weyl group of the Lie algebra.

The transformations $r_i$ generating the Weyl group are reflections of a vector $x$
in root space at the hyperplanes orthogonal to the simple roots $\alpha_i$
of the Lie algebra.  

While we expect most LieART users to be primarily interested in irrep products and decompositions, some may want to delve into the details of properties of individual  irreps.
In that case LieART can be used to obtain a wealth of information about roots, weights, Weyl orbits and much more.
All this information can be found using the LieART commands listed in section \ref{subsec:TablesOfLieARTCommands} and by consulting their full descriptions in \cite{Feger:2015bs}.

\subsection{Properties of Representations}

As explained in section \ref{ssec:Representations} a \emph{representation} is a set of
matrices that satisfies the same commutation relations as the algebra. Each of
the matrices can be labeled by a \emph{weight vector} (or simply a \emph{weight}) which has eigenvalues  
corresponding to the generators of the Cartan subalgebra. The weight vector has
the dimension of the Cartan subalgebra, not
the dimension of the space the matrices act on.  

The weights $\lambda$ can be written as linear combination of simple roots and 
 the   \emph{Dynkin labels} $a_i$ defined by
\begin{equation}
    a_i = \frac{2\scalarproduct{\lambda}{\alpha_i}}{\scalarproduct{\alpha_i}{\alpha_i}},\qquad i=1,\ldots,n,
\end{equation}
are integers for all simple roots $\alpha_i$.   The smallest
non-zero weights with $a_i\geq 0$ are called the \emph{fundamental weights}
$\omega_i$. They define the $\omega$-basis or Dynkin basis, which has already been introduced.
They are implemented in LieART  
  in the orthogonal basis.
The $a_i$ are the coordinates in the
$\omega$-basis.

Weights are linear combinations of roots,  and the Weyl group also applies to weights, where the weight space is
divided into Weyl chambers. A weight with only positive coordinates lies in
the dominant Weyl chamber and is called a \emph{dominant weight}. In analogy
with the highest root, every irrep has a
non-degenerate \emph{highest weight}, denoted as $\Lambda$, which is also a
dominant weight, but not necessarily the only dominant weight of the irrep. The
weight system of the irrep can be computed from the highest weight $\Lambda$ by
subtracting simple roots. Thus, a highest weight $\Lambda$ uniquely defines the
irrep, and  it also serves
as a label for the irrep.

In LieART an irrep is represented by
\com{Irrep[\args{algebraClass}][\args{label}]}, where \args{algebraClass}
defines the Lie algebra class in the same manner as for weights and roots, and
\args{label} is the comma-separated label of the highest weight of the irrep.

 \subsubsection{Weight System}

The conventional approach to computing all weights of an irrep is to subtract
simple roots from the highest weight $\Lambda$ that defines the irrep.  
 The \emph{level} of a weight is the number of simple
root that need to be subtracted from the highest weight to obtain it. 
The highest level of an irrep is called its \emph{height}. 

The algorithm to compute the weight system used in LieART is an implementation 
of the scheme developed in \cite{Moody:1982}. It deviates from the traditional 
procedure but provides better performance by exploiting the Weyl group in both the weight and the
root system.

\subsubsection{Properties of Irreducible Representations}

\paragraph{Dimension}
The dimension of an irrep, i.e., the number of its weights, can be calculated
with the \emph{Weyl dimension
formula}, which gives the
dimension of an irrep in terms of its highest weight $\Lambda$, positive roots
$\alpha\in\Delta^{+}$ and $\delta=\dynkincomma{1,1,\ldots,1}$, which is half the sum of the positive roots:
\begin{equation}
\dim(\Lambda) = \prod_{\alpha\in\Delta^{\!+}\!}\frac{\scalarproduct{\alpha}{\Lambda + \delta}}{\scalarproduct{\alpha}{\delta}}.
\end{equation}
  By using
variables the simple structure of the formula becomes explicit, e.g., for a
general irrep of \A4:
\begin{mathin}
    Dim[Irrep[A][a,b,c,d]]
\end{mathin}
\begin{mathout}
    $\displaystyle\frac{1}{288}(a+1)(b+1)(c+1)(d+1)(a+b+2)(b+c+2)(c+d+2)(a+b+c+3)(b+c+d+3)(a+b+c+d+4)$
\end{mathout}
Internally, the Weyl dimension formula is computed by
\com{WeylDimensionFormula[\args{algebra}]} as a pure function with the digits of
the Dynkin label as parameters. E.g. for \A4:
\begin{mathin}
    WeylDimensionFormula[A4]//InputForm
\end{mathin}
\begin{mathout}
    Function[\{a1,a2,a3,a4\},((1{+}a1){*}(1{+}a2){*}(1{+}a3){*}(1{+}a4){*}(2{+}a1{+}a2){*}(2{+}a2{+}a3)\newline
    {*}(2{+}a3{+}a4){*}(3{+}a1{+}a2{+}a3){*}(3{+}a2{+}a3{+}a4){*}(4{+}a1{+}a2{+}a3{+}a4))/288]
\end{mathout}

\paragraph{Index}
Another important property of an irrep $\Lambda$ is its \emph{Dynkin index}, denoted as
$l(\Lambda)$, which is an eigenvalue of the \emph{Casimir invariant} normalized
to be an integer:
\begin{equation}
    l(\Lambda) = \frac{\text{dim}(\Lambda)}{\text{ord}(L)}\scalarproduct{\Lambda}{\Lambda+2\delta},
\end{equation}
where $\text{ord}(L)$ is  the dimension of the adjoint irrep. The index is related
to the length of the weights and has applications in renormalization group
equations and elsewhere \cite{Fonseca:2011sy}. The corresponding LieART function is
\com{Index[\args{irrep}]}, while the eigenvalue of the Casimir invariant can be obtained by \com{CasimirInvariant[\args{irrep}]}.  

\paragraph{Casimir Invariant}
Casimir invariants can be constructed from structure constants \cite{Wybourne}. There are a set of $N$ independent Casimir invariants for a rank $N$ Lie algebra. The Quadratic Casimir invariant shows up in renormalization group calculations in gauge theories. The higher order Casimirs are important in chiral anomaly and elsewhere in higher dimensional gauge theories and superstrings. The Quadratic Casimir invariant is delivered by the command \com{CasimirInvariant[\args{irrep}]}. LieART does not yet provide the higher order Casimirs. 

\paragraph{Congruency Class}
The \emph{congruency class} generalizes the concept of $n$-ality of \SU{N} to all other simple Lie
algebras. LieART uses congruency classes to characterize irreps, especially for
the distinguishing irreps with degenerate dimension and  index. Again, for details see \cite{Feger:2015bs}. 
 
 \subsubsection{Examples of Simple Explorations with LieART}
 LieART makes explorations in representation theory easy and convenient. It allows one to find relations
 not easily found by other methods. Take for instance the above form of the dimension of the representation
 $[a,b,c,d]$ for \A4. Setting $a=b=c=d=1$, LieART delivers the dimension $2^{10}$. In fact the dimension of 
  $[a,a,a,a]$ is $(a+1)^{10}$. Furthermore one easily finds that the dimension of any irrep 
  with equal entries in its Dynkin label, $[a,a,a,...]$, is always
  $(a+1)^p$ for any classical or exceptional Lie algebra. The power $p$  depends on the algebra, and  
 for \A{n} it is $p =1,3,6,10,...$ for $n = 1,2,3,4,...$, i.e., a binomial coefficient.
For \B{n} and for \C{n} the power is $p=4,9,16,25,36,...$ for $n = 2,3,4,5,...$, i.e., $p=n^2$.
For \D{n} the power is $p=2, 6, 10, 20, ...$ for $n=2,3,4,5...$
i.e., the power is 2 times a binomial coefficient. And for the exceptional cases we find
$p=6,\, 24,\, 36,\, 63$ and $ 120$ for \G2, \F4, \E6, \E7 and \E8 respectively.
Hence we  find exact results like \com{Irrep[E8][7, 7, 7, 7, 7, 7, 7, 7]} $=2^{360}$ which is somewhat slow to obtain when
calculated directly from the Weyl dimension formula, but easy to prove as follows:

First we rewrite the Weyl's dimension formula for a general irrep $\Lambda$ in the form
\begin{equation}
\dim(\Lambda)=\frac{\prod_{\alpha\in \Delta^+} (\Lambda+\delta,\alpha)}{\prod_{\alpha\in \Delta^{^+}} (\delta,\alpha)}
\end{equation}
where $\Delta^+$ is the set of positive roots $\alpha$. Recall that in the Dynkin bases 
one has $\delta=(1,1,1,...,1)$. Now for the specific irrep $\Lambda=(a,a,a,...,a)=a\delta$ that means we have 
\begin{equation}
\dim(\Lambda)=\frac{\prod_{\alpha\in \Delta^+} (a+1)(\delta,\alpha)}{\prod_{\alpha\in \Delta^{^+}} (\delta,\alpha)}=\frac{\prod_{\alpha\in \Delta^+} (a+1)}{\prod_{\alpha\in \Delta^{^+}} (1)}=(a+1)^p
\end{equation}
where $p$ is the number of positive roots, i.e.,  the number of elements in the set $\Delta^+$, which agrees with the numbers in the examples above. Hence LieART has lead us an to the asymptotic form for an interesting class if irreps of all classical and exceptional groups.
  
 Some simple related results are that for \SU3
 \begin{equation}
 (0,a)\otimes(a,0)= (0,0)+(1,1)+(2,2)+...+(a,a),
 \end{equation}
for example
\begin{mathin}
DecomposeProduct[Irrep[A][9, 0], Irrep[A][0, 9]]
\end{mathin}
\begin{mathout}
$\irrep{1}+\irrep{8}+\irrep{27}+\irrep{64}+\irrep{125}+\irrep{216}+\irrep{343}+\irrep{512}+\irrep{729}+\irrep{1000}$
\end{mathout}
Another interesting fact is that for any classical or exceptional algebra, the product $(b_1,b_2,...,b_n)\otimes(a-b_1,a-b_2,...a-b_n)$ always contains $(a,a,...a)$ as the largest irrep in the decomposition. 
 
\subsubsection{Representation Names}
\label{ssec:irrepNames}

The Dynkin label of an irreducible representation together with its Lie algebra
uniquely specifies it.  
However, it is common practice to name representations by their dimension, the
\emph{dimensional name}, which is often  shorter. The dimension of a
representation is not necessarily unique. If
there is an accidental dimension degeneracy, then irreps with the same dimension have primes ($\text{\bf
dim}^\prime$) in their dimensional name. Irreps can be related by conjugation, when they
are complex. In that case one of the irreps is written with an overbar ($\irrepbar{dim}$).
Due to the
high symmetry of \SO8 irrep, more than two related irreps of the same dimension
exist. In the case of \SO8 subscripts specify the irreps completely.\footnote{One should be aware that   different choices of notation can be found in the literature for the decomposition
$\SO8 \rightarrow \SO7$ due to the special properties of the \SO8 Dynkin diagram.
Because of triality (the symmetry of the \SO8 Dynkin diagram) we could choose any one of \irrepsub{8}{v}, \irrepsub{8}{s} or \irrepsub{8}{c}, to
reduce to $\irrep{1} + \irrep{7}$ under the decomposition $\SO8\to\SO7$ while the other two \irrep{8}s of \SO8 both go to
the \irrep{8} of \SO7. Our choice here is the same as Slansky and Yamatsu and is also the same as we used in LieART 1.} For a complete description of LieART's algorithm to determine the dimensional name refer to \cite{Feger:2015bs}.

\subsection{Tensor Product Decomposition}

Tensor products of irreps can be decomposed into a direct sum of irreps. The product of two irreps $R_1$ and $R_2$ can be decomposed as
\begin{equation}\label{eq:TensorProduct}
    R_1\otimes R_2 = \sum_i m_i R_i
\end{equation}
with the following dimension and index sum rules:
\begin{align}
    \dim(R_1\otimes R_2)&= \dim(R_1)\cdot\dim(R_2)=\sum_i m_i \dim(R_i)\\
    l(R_1\otimes R_2)&= l(R_1)\dim(R_2)+\dim(R_1)l(R_2)=\sum_i m_i\, l(R_i).
\end{align}
% \vspace{-13pt}

\subsubsection{Generic Algorithm}
\label{ssec:TensorProductGeneric}

To compute the right-side of \eqref{eq:TensorProduct} 
add all weights of $R_2$ to each weight of $R_1$. The 
resulting $\dim(R_1)\cdot\dim(R_2)$ weights belong to the different irreps 
$R_i$, which must be sorted out. Instead of all weights, one can consider just 
the dominant weights in the product, as each of the dominant weights represents 
an orbit in the irreps $R_i$.
There is a unique dominant weight that represents the irrep of 
largest height in the decomposition. Starting with 
this dominant weight viewed as the highest weight of an irrep,
the dominant weight system of the corresponding irrep,   should then be subtracted from the 
combined dominant weights. The same procedure is 
applied recursively to the remaining set of dominant weights until it is empty.

LieART provides the function \com{DecomposeProduct[\args{irreps}]} for the 
decomposition of the tensor product of arbitrary many \args{irreps} of any 
classical or exceptional Lie algebra as argument.

\subsubsection{Algorithm Based on Klimyk's Formula}

Adding all weights of $R_2$ to each weight of $R_1$ is costly for large irreps. 
LieART's algorithm to decompose tensor-product implements 
Klimyk's formula~\cite{klimyk_1967,klimyk_orbit_2006,Humphreys:1980dw}, which improves the runtime 
of tensor product decompositions considerably: Let $\lambda_1$ and $\lambda_2$ 
be weights and $\Lambda_1$ and $\Lambda_2$ the highest weights of $R_1$ and 
$R_2$, respectively. Instead of adding all weights $\lambda_1$ to each weight 
$\lambda_2$, the weights $\lambda_1$ are added only to the highest weight 
$\Lambda_2$ of $R_2$ together with half the sum of positive simple roots, 
$\delta{=}\dynkincomma{1,\ldots,1}$, building the set of weights 
\begin{equation}
	\mu = \lambda_1 + \Lambda_2 + \delta.
\end{equation}
Each $\mu$ is reflected to the dominant chamber, yielding a highest weight 
denoted as $\{\mu\}$, with $\sgn(\mu)$ as the parity of the reflection. Of these 
dominant weights all that lie on a chamber wall are dropped. The irreps in the 
decomposition are $R(\{\mu\}-\delta)$. Klimyk's formula reads 
\begin{equation}
	R_1(\Lambda_1)\otimes R_2(\Lambda_2) = \sum\limits_{\lambda_1} m_{1\lambda_1} t(\lambda_1+ \Lambda_2 + \delta) R(\{\lambda_1+ \Lambda_2 + \delta\}-\delta),
\end{equation}
where $m_{1\lambda_1}$ denotes the multiplicity of $\lambda_1$ in $R_1$. We 
define $t(\mu)$ to be $\sgn(\mu)$ if $\mu$ has a trivial-stabilizer subgroup $\text{Stab}(\mu) = \{1\}$,
 and zero if the stabilizer subgroup 
is non-trivial, i.e. the weight lies on a chamber wall:
\begin{equation}
	t(\mu) = \left\{
		\begin{array}{rcl}
			\sgn(\mu) & : & \text{Stab}(\mu) = \{1\}\\
			          0 & : & \text{else}
		\end{array}
		\right..
\end{equation}

\subsubsection{\SU{N} Decomposition via Young Tableaux}

The correspondence of \SU{N} irreps with Young tableaux is very useful for the 
calculation of tensor products and subalgebra decomposition by hand. We have 
found that the algorithm for the \SU{N} tensor product decomposition via Young 
tableaux also performs better on the computer, with respect to CPU time and 
memory consumption, than the procedure described in the previous section. Thus, 
LieART uses the Young tableaux algorithm for the tensor-product decomposition of 
\SU{N} irreps but the procedure of adding weights and filtering out irreps for 
all other classical and exceptional Lie algebras.

A \emph{Young tableau} is a left-aligned set of boxes, with successive rows 
having an equal or smaller number of boxes. Young tableaux correspond to the 
symmetry of the tensors of \SU{N} irreps, by first writing each index of the 
tensor into one box of a Young tableau and the prescription that they ought to 
be first symmetrized in the rows and then antisymmetrized in the columns. Please 
see \outref{out:YoungTableau720SU5} in Section~\ref{sec:QuickStart} for a 
non-trivial example for a Young tableau displayed by LieART.  

\paragraph{IrrepMultiplicity} The new command \com{IrrepMultiplicity[\args{decomp}, \args{irrep}]} extracts the number of times an irrep appears in a product decomposition. For example:

\begin{mathin}
    IrrepMultiplicity[DecomposeProduct[Irrep[G2][14, 0], Irrep[G2][0, 14]], Irrep[G2][7, 8]]
\end{mathin}
\begin{mathout}
    11
\end{mathout}
%**********************************

\subsection{Subalgebra Decomposition}
\label{ssec:SubalgebraDecomposition}

LieART  decomposes an irrep of a simple Lie algebra into a maximal subalgebra, which can be simple, semi-simple or non-semi-simple.

\subsubsection{Branching Rules and Maximal Subalgebras}

 Subalgebras fall into two classes: \emph{regular} and \emph{special} subalgebras. The derivation  of  maximal subalgebras  \cite{Dynkin:1957um,Dynkin:1957dm}  which are either regular non-semisimple, regular semisimple or special subalgebras,
 is utilized by LieART to determine the projection matrices.

\paragraph{Non-Semisimple Subalgebras}

A non-semisimple subalgebra is a semisimple subalgebra times a \U1 factor, obtained by removing a dot from the Dynkin diagram. The resulting two or more disconnected
Dynkin diagrams symbolize the semisimple subalgebra and the removed dot,  becomes the \U1 generator. 

Dropping a dot from the Dynkin diagram corresponds to
dropping the associated digit from the Dynkin label.  
Writing the weights of an irrep  as \emph{columns} of a matrix $\matrixhat{W}$ and the weights with the removed digit expressed in non-normalized $\alpha$-basis coordinates moved to the end as rows of a matrix $\matrixhat{W}'$,
the projection matrix $\matrixhat{P}$ can be determined from
\begin{equation}
    \matrixhat{P}\matrixhat{W} = \matrixhat{W}'
\end{equation}
with the right-inverse $\matrixhat{W}^{\!+}$ of $\matrixhat{W}$ defined in section \ref{ssec:Bases}. Since $\matrixhat{W}$ is in general not a rectangular matrix:
\begin{equation}
    \matrixhat{P} = \matrixhat{W}'\matrixhat{W}^{\!+}.
\end{equation}

The projection matrix found by this procedure can  be used to decompose any irrep of the initial algebra to the subalgebra.  The fundamental irrep must be used for the determination of the projection matrices.
The generating irreps of representative Lie algebras are listed in Table \ref{tab:GeneratingIrreps}.
\begin{table}[t]
    \begin{center}
        \rowcolors{2}{}{tablerowcolor}
        \begin{tabular}{lll}
            \toprule\rowcolor{tableheadcolor}
            \textbf{Algebra} & \textbf{Irrep}    & \textbf{Irrep}\\\rowcolor{tableheadcolor}
                             & \textbf{(Dynkin)} & \textbf{(Name)} \\
            \midrule%\\[-10pt]
            \A4 (\SU5) & \dynkin{1,0,0,0}          & \irrep{5}       \\
            \B4 (\SO9) & \dynkin{0,0,0,1}          & \irrep{16}      \\
            \C4 (\Sp8) & \dynkin{1,0,0,0}          & \irrep{8}       \\
            \D4 (\SO8) & \dynkin{0,0,0,1}          & \irrepsub{8}{s} \\
            \E6 & \dynkin{1,0,0,0,0,0}      & \irrep{27}      \\
            \E7 & \dynkin{0,0,0,0,0,1,0}    & \irrep{56}      \\
            \E8 & \dynkin{0,0,0,0,0,0,1,0}  & \irrep{248}     \\
            \F4 & \dynkin{0,0,0,1}          & \irrep{26}      \\
            \G2 & \dynkin{1,0}              & \irrep{7}       \\
            \bottomrule
        \end{tabular}
        \caption{\label{tab:GeneratingIrreps} Generating Irreps of representative Lie algebras}
    \end{center}
    % \vspace{-15pt}
\end{table}
Only the lowest non-trivial orbit is needed for the determination of the projection matrices.

\paragraph{Semisimple Subalgebras}
\enlargethispage{2ex}

\begin{figure}[t]
    \begin{center}
        \includegraphics[scale=0.95]{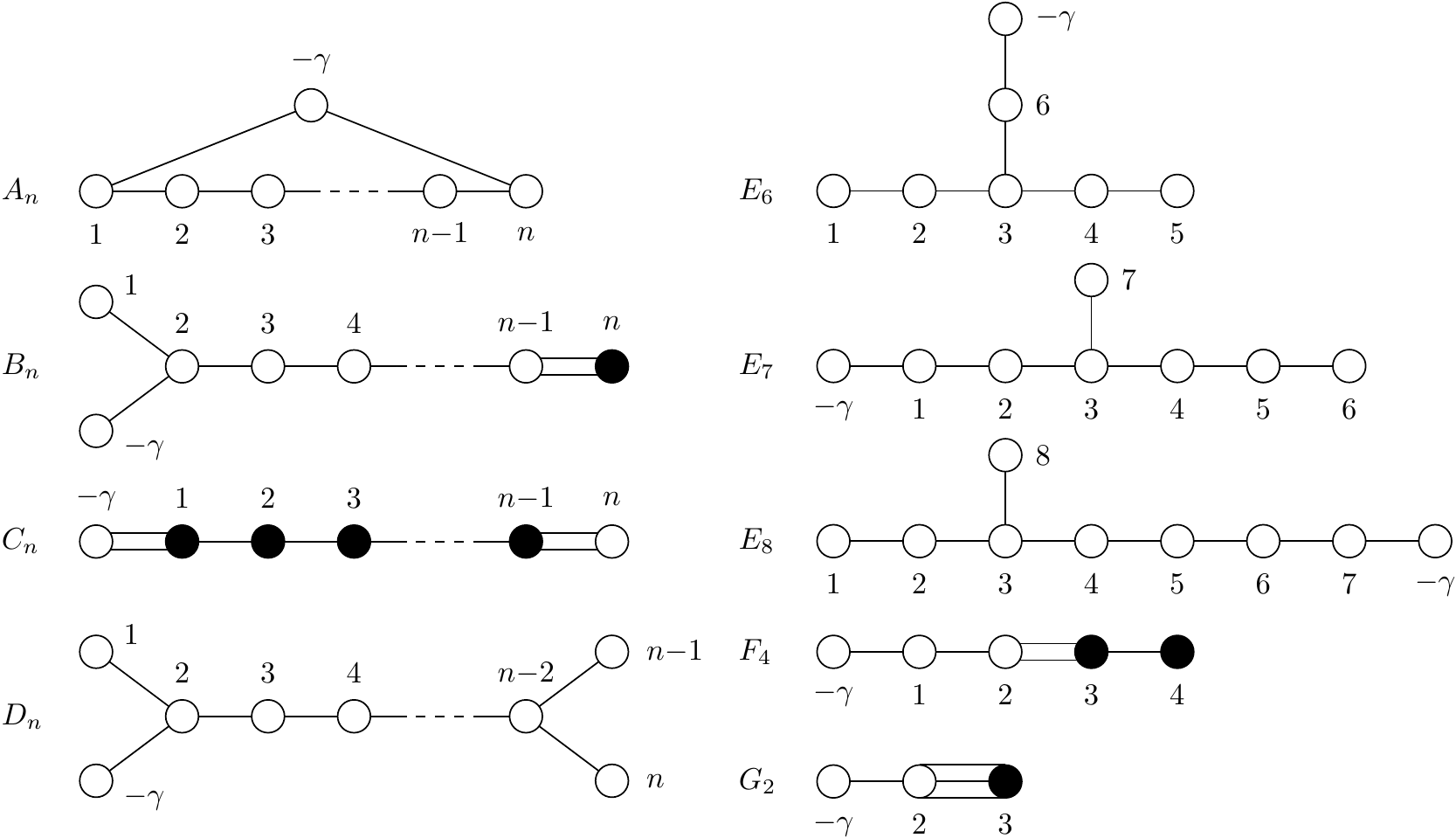}
        \caption{\label{fig:ExtendedDynkinDiagrams} Extended Dynkin Diagrams of classical and exceptional simple Lie algebras.}
    \end{center}
\end{figure}

To obtain a semisimple subalgebra without a \U1 generator, a root from the
 \emph{extended Dynkin diagram} is removed. The extended Dynkin diagram
is constructed by adding the most negative root $-\gamma$ to the set of simple roots.  
Removing one root restores the linear independence yielding a
valid system of simple roots of a subalgebra, which in general is semisimple. 
  The extended Dynkin diagrams for all classical and exceptional Lie 
Algebras are shown in Figure~\ref{fig:ExtendedDynkinDiagrams}.  
%**********************************
%**********************************

With the weight of the generating irrep as columns of the matrix
$\matrixhat{W}$ and the weights in the  decomposition   as columns of $\matrixhat{W}'$ the
projection matrix $\matrixhat{P}$ is computed as described for non-semisimple
regular subalgebras as $\matrixhat{P}=\matrixhat{W}'\matrixhat{W}^{+}$ with
the right-inverse $\matrixhat{W}^{+}\!$ of $\matrixhat{W}$, which is analogous to the semisimple case.

\paragraph{Special Subalgebras}

Special maximal subalgebras cannot be derived from the root system. The embedding of a special subalgebra does not follow a general pattern and must be derived for every algebra--subalgebra pair individually. (See \cite{Kim:1982jg} for a discussion and useful examples.)
Generating irreps are  used to derive the special subalgebra embedding, which may be simple or semisimple and can involve more than one irrep of the subalgebra. LieART 2.0 contains all projection matrices for decompositions to special maximal subalgebras through rank 15, and for many patterns beyond rank 15. Examples of these decompositions are tabulated in the appended supplementary material, and examples for the branchings $\SO{14}\rightarrow \G2$ and $\SO{26}\rightarrow \F4$ are shown below in Tables \ref{tab:SO14BranchingRulesExample} and \ref{tab:SO26BranchingRulesExample}.

\begin{table}[!h]
    \setlength\extrarowheight{1.3pt}
    \renewcommand{\tabcolsep}{2pt}
    \rowcolors{2}{tablerowcolor}{}
    \begin{tabularx}{\textwidth}{rcX}
        \toprule
            \rowcolor{tableheadcolor}
            \SO{14}& $\to$ &\G2\\
        \midrule 
            \irrep{14} & = & $\irrep{14}$\\
            \irrep{64} & = & $\irrep{64}$\\
            \irrep{91} & = & $\irrep{14}+\irrep{77}$\\
            \irrep{104} & = & $\irrep{27}+\irrep[1]{77}$\\
            \irrep{364} & = & $\irrep{1}+\irrep{27}+\irrep{77}+\irrep[1]{77}+\irrep{182}$\\
            \irrep{546} & = & $\irrep{7}+\irrep{77}+\irrep{189}+\irrep{273}$\\
            \irrep{832} & = & $\irrep{7}+\irrep{27}+\irrep{64}+\irrep{77}+\irrep{182}+\irrep{189}+\irrep{286}$\\
            \irrep{896} & = & $\irrep{14}+\irrep{27}+\irrep{64}+\irrep{77}+\irrep[1]{77}+\irrep{189}+\irrep{448}$\\
            \irrep{1001} & = & $\irrep{14}+\irrep{27}+\irrep{64}+\irrep[1]{77}+\irrep{182}+\irrep{189}+\irrep{448}$\\
            \irrep{1716} & = & $\irrep{27}+\irrep{64}+\irrep[1]{77}+\irrep{182}+\irrep{189}+\irrep{448}+\irrep{729}$\\
            \irrep{2002} & = & $\irrep{7}+\irrep{14}+\irrep{64}+2(\irrep{77})+2(\irrep{189})+\irrep{273}+\irrep{286}+\irrep{378}+\irrep{448}$\\
            \irrep{2275} & = & $\irrep{27}+\irrep{64}+\irrep[1]{77}+\irrep{182}+\irrep{448}+\irrep{729}+\irrep{748}$\\
            \irrep{3003} & = & $\irrep{7}+\irrep{27}+\irrep{64}+2(\irrep{77})+\irrep[1]{77}+\irrep{182}+2(\irrep{189})+\irrep{273}+\irrep{286}+\irrep{378}+\irrep{448}+\irrep{729}$\\
            \irrep{3080} & = & $\irrep{1}+2(\irrep{27})+\irrep{64}+\irrep{77}+2(\irrep[1]{77})+2(\irrep{182})+\irrep{189}+\irrep{286}+\irrep{448}+\irrep{714}+\irrep{729}$\\
            \irrep{4004} & = & $\irrep{7}+2(\irrep{14})+\irrep{27}+2(\irrep{64})+3(\irrep{77})+\irrep[1]{77}+\irrep{182}+3(\irrep{189})+\irrep{273}+\irrep{286}+\irrep{378}+2(\irrep{448})+\irrep{924}$\\
            % \irrep{4928} & = & $\irrep{7}+\irrep{14}+2(\irrep{27})+3(\irrep{64})+2(\irrep{77})+\irrep[1]{77}+2(\irrep{182})+3(\irrep{189})+2(\irrep{286})+\irrep{378}+2(\irrep{448})+\irrep{729}+\irrep{924}$\\
            % \irrep{5265} & = & $\irrep{7}+\irrep{14}+\irrep{27}+2(\irrep{64})+2(\irrep{77})+\irrep[1]{77}+\irrep{182}+3(\irrep{189})+\irrep{273}+\irrep{286}+\irrep{378}+2(\irrep{448})+\irrep{729}+\irrep{1547}$\\
            % \irrep{5824} & = & $\irrep{7}+\irrep{14}+2(\irrep{27})+3(\irrep{64})+2(\irrep{77})+\irrep[1]{77}+2(\irrep{182})+3(\irrep{189})+2(\irrep{286})+\irrep{378}+2(\irrep{448})+\irrep{729}+\irrep{896}+\irrep{924}$\\
        \bottomrule
    \end{tabularx}
    \caption{\label{tab:SO14BranchingRulesExample}$\SO{14}\rightarrow \G2$ Branching Rules Example}
\end{table}

\begin{table}[!h]
    \setlength\extrarowheight{1.3pt}
    \renewcommand{\tabcolsep}{2pt}
    \rowcolors{2}{tablerowcolor}{}
    \begin{tabularx}{\textwidth}{rcX}
        \toprule
            \rowcolor{tableheadcolor}
            \SO{26}& $\to$ &\F4\\
        \midrule 
        \irrep{26} & = & $\irrep{26}$ \\
\irrep{325} & = & $\irrep{52}+\irrep{273}$ \\
\irrep{2600} & = & $\irrep{273}+\irrep{1053}+\irrep{1274}$ \\
\irrep{4096} & = & $\irrep{4096}$ \\
\irrep{5824} & = & $\irrep{26}+\irrep{52}+\irrep{273}+\irrep{324}+\irrep{1053}+\irrep{4096}$ \\
\irrep{14950} & = & $\irrep{324}+\irrep{1053}+\irrep[1]{1053}+\irrep{4096}+\irrep{8424}$ \\
\irrep{52325} & = & $\irrep{26}+\irrep{52}+2(\irrep{273})+\irrep{324}+2(\irrep{1053})+2(\irrep{1274})+2(\irrep{4096})+\irrep{8424}+\irrep{10829}+\irrep{19278}$ \\
\irrep{65780} & = & $\irrep{324}+\irrep{1053}+\irrep[1]{1053}+\irrep{2652}+\irrep{4096}+\irrep{8424}+\irrep{10829}+\irrep{17901}+\irrep{19448}$ \\
\irrep{102400} & = & $\irrep{273}+\irrep{324}+\irrep{1053}+\irrep{1274}+\irrep{2652}+\irrep{4096}+\irrep{8424}+\irrep{10829}+\irrep{19278}+\irrep{19448}+\irrep{34749}$ \\
\irrep{230230} & = & $\irrep{273}+\irrep{1053}+\irrep{1274}+\irrep{2652}+\irrep{4096}+\irrep{8424}+2(\irrep{10829})+\irrep{12376}+\irrep{17901}+\irrep{19278}+\irrep{34749}+\irrep{106496}$ \\
\irrep{320320} & = & $\irrep{26}+\irrep{52}+2(\irrep{273})+2(\irrep{324})+3(\irrep{1053})+\irrep[1]{1053}+2(\irrep{1274})+\irrep{2652}+4(\irrep{4096})+3(\irrep{8424})+2(\irrep{10829})+\irrep{17901}+2(\irrep{19278})+\irrep{19448}+\irrep{29172}+\irrep{34749}+\irrep{106496}$ \\
\irrep{450450} & = & $\irrep{1}+2(\irrep{26})+\irrep{52}+3(\irrep{273})+3(\irrep{324})+3(\irrep{1053})+\irrep[1]{1053}+2(\irrep{1274})+2(\irrep{2652})+4(\irrep{4096})+3(\irrep{8424})+2(\irrep{10829})+\irrep{17901}+2(\irrep{19278})+2(\irrep{19448})+\irrep{29172}+\irrep{34749}+\irrep{106496}+\irrep{107406}$ \\
\irrep{657800} & = & $\irrep{52}+\irrep{273}+\irrep{1053}+\irrep{1274}+2(\irrep{4096})+\irrep{8424}+2(\irrep{10829})+\irrep{17901}+2(\irrep{19278})+\irrep{29172}+2(\irrep{34749})+\irrep{76076}+\irrep{106496}+\irrep{119119}+\irrep{160056}$ \\
        \bottomrule
    \end{tabularx}
    \caption{\label{tab:SO26BranchingRulesExample}$\SO{26}\rightarrow \F4$ Branching Rules Example}
\end{table}

%A few additional interesting cases beyond rank 15, like %$\Sp{56}\rightarrow \E7$ are also included.

\pagebreak
\ 
\newpage

\colorlet{tableoverheadcolor}{gray!37.5}
\colorlet{tableheadcolor}{gray!40}
\colorlet{tablerowcolor}{gray!20}

\section{Results}
\label{sec:Results}

LieART now has additional functionality for finding the branching rules of regular and special maximal subalgebras of Lie algebras. We have tested that LieART can now correctly reproduce the branching rules for all maximal subalgebras, both regular and special, of Lie algebras up to rank 15. Indeed, LieART is now able to find all branching rules for regular subalgebras and all branching rules for the special subalgebra patterns given in the tables below, regardless of rank. While special subalgebras cannot be simply obtained from manipulation of an algebra's Dynkin diagram, they follow the patterns given in Tables \ref{Table:Maximal-S-sub-classical-1} and \ref{Table:Maximal-S-sub-classical-2} below, with some abnormal exceptions. Table \ref{Table:Maximal-subalgebra} gives all regular and special subalgebras up to rank 15, labeled (R) and (S), respectively. LieART 2.0 can compute every branching rule in Tables \ref{Table:Maximal-S-sub-classical-1}, \ref{Table:Maximal-S-sub-classical-2}, and \ref{Table:Maximal-subalgebra}. Examples of branching rules tables are given in the Appendix.

\begin{table}[!h]
    \vspace{2ex}
    \begin{center}
        \renewcommand{\arraystretch}{1.3}
        \rowcolors{2}{tablerowcolor}{}
        \begin{tabular}{lll}
            \toprule\rowcolor{tableheadcolor}
            \textbf{Rank} & \textbf{Algebra}    & \textbf{Subalgebra}\\
            \midrule%\\[-10pt]
            $mn-1$ & \SU{mn} & $\SU{m}\otimes \SU{n}$\\
            $\left[\frac{mn}{2}\right]$ & \SO{mn} & $\SO{m}\otimes \SO{n}$\\
        	$2mn$  &\SO{4mn}&$\Sp{2m}\otimes\Sp{2n}$\\
        	$mn$   &\Sp{2mn}&$\SO{m}\otimes\Sp{2n}$\\
            \bottomrule
        \end{tabular}
        \caption{\label{Table:Maximal-S-sub-classical-1} Maximal Special Subalgebras of Classical Algebras}
    \end{center}
    % \vspace{-15pt}
\end{table}

% \newpage
\begin{table}[!h]
    \begin{center}
        \renewcommand{\arraystretch}{1.3}
        \rowcolors{2}{tablerowcolor}{}
        \begin{tabular}{llll}
            \toprule\rowcolor{tableheadcolor}
            \textbf{Rank} & \textbf{Algebra}    & \textbf{Subalgebra} &\\
            \midrule%\\[-10pt]
            $mn-1$ &$\A{mn-1}$&$\A{m-1}\otimes \A{n-1}$&$m,n\geq2$\\
        	$2n$   &$\A{2n}$  &$\B{n}$ &$n\geq 2$\\
        	$2n-1$ &$\A{2n-1}$&$\C{n}$ &$n\geq 2$\\
        	$2n-1$ &$\A{2n-1}$&$\D{n}$ &$n\geq 2$\\
        	$\frac{(n-1)(n+2)}{2}$ 
        	&$\A{\frac{(n-1)(n+2)}{2}}$  &$\A{n}$ &$n\geq 3$\\
        	$\frac{n(n+3)}{2}$&$\A{\frac{n(n+3)}{2}}$  &$\A{n}$ &$n\geq 2$\\
        	$n$ &$\B{n}$ &$\A1$ &$n\geq 4$\\
        	$n$ &$\C{n}$ &$\A1$ &$n\geq 2$\\
        	$n+1$ &$\D{n+1}$ &$\B{n}$ &$n\geq 3$\\
        	$n+2$&$\D{n+2}$&$\A1\otimes\B{n}$ &$n\geq 4$\\
        	$n+3$&$\D{n+3}$&$\C2\otimes\B{n}$ &$n\geq 4$\\
        	$m+n+1$&$\D{m+n+1}$&$\B{m}\otimes\B{n}$ &$m+n\geq 4$\\
        	$2mn$ &$\D{2mn}$&$\C{m}\otimes \C{n}$&\\
            \bottomrule
        \end{tabular}
        \caption{\label{Table:Maximal-S-sub-classical-2} Maximal Special Subalgebras of Classical Algebras (Dynkin Classification)}
    \end{center}
    % \vspace{-15pt}
\end{table}

\pagebreak
\enlargethispage{1ex}
\newcommand{\x}{$\otimes$}
\rowcolors{2}{tablerowcolor}{}
{
\renewcommand{\arraystretch}{1.3}
\begin{longtable}{crc>{\raggedright\arraybackslash}p{10cm}c}
	\rowcolor{white}
	\caption{\label{Table:Maximal-subalgebra}Maximal Subalgebras}\\
	\toprule
	\rowcolor{tableheadcolor}
	\textbf{Rank}&\multicolumn{2}{l}{\textbf{Algebra}}&\multicolumn{1}{l}{\textbf{Maximal subalgebras}} &\textbf{Type}\\
	\midrule
	\endfirsthead
	\rowcolor{white}
	\caption[]{Maximal Subalgebras (continued)}\\
	\toprule
	\rowcolor{tableheadcolor}
	\textbf{Rank}&\multicolumn{2}{l}{\textbf{Algebra}}&\multicolumn{1}{l}{\textbf{Maximal subalgebras}}&\textbf{Type}\\
	\midrule
	\endhead
	\endfoot
	\bottomrule
    \endlastfoot
	$1$ &\SU2&$\supset$&\U1&$(R)$\\
	&\multicolumn{4}{l}{(The algebras \SU2, \SO3, and \Sp2 are all isomorphic.)}\\
	\hline
	$2$ &\SU3&$\supset$&\SU2\x \U1&$(R)$\\
	&       &$\supset$&\SU2&$(S)$\\
	&\Sp4&$\supset$&\SU2\x \SU2; \SU2\x \U1&$(R)$\\
	&       &$\supset$&\SU2&$(S)$\\
	&\multicolumn{4}{l}
	{(\SO5 is isomorphic to \Sp4, and \SO4 is isomorphic to \SU2\x \SU2.)}\\
	&\G2  &$\supset$&\SU3; \SU2\x \SU2&$(R)$\\
	&       &$\supset$&\SU2&$(S)$\\
	\hline
	$3$ &\SU4&$\supset$
	&\SU3\x \U1; \SU2\x \SU2\x \U1
	&$(R)$\\
	&       &$\supset$&\Sp4; \SU2\x \SU2&$(S)$\\
	&\SO7&$\supset$
	&\SU4; \SU2\x \SU2\x \SU2; \Sp4\x \U1&$(R)$\\
	&       &$\supset$&$\G2$&$(S)$\\
	&\Sp6&$\supset$
	&\SU3\x \U1; \SU2\x \Sp4
	&$(R)$\\
	&       &$\supset$
	&\SU2; \SU2\x \SU2&$(S)$\\
	&\multicolumn{4}{l}{(\SO6 is isomorphic to \SU4.)}\\
	\hline
	$4$ &\SU5&$\supset$
	&\SU4\x \U1; \SU3\x \SU2\x \U1
	&$(R)$\\
	&       &$\supset$&\Sp4&$(S)$\\
	&\SO9&$\supset$
	&\SO8; \SU2 \x \SU2\x \Sp4; \SU4\x \SU2; \SO7\x \U1
	&$(R)$\\
	&       &$\supset$&\SU2; \SU2\x \SU2&$(S)$\\
	&\Sp8&$\supset$
	&\SU4\x \U1; \SU2\x \Sp6; \Sp4\x \Sp4&$(R)$\\
	&       &$\supset$&\SU2; \SU2\x \SU2\x \SU2&$(S)$\\
	&\SO8&$\supset$
	&\SU2\x \SU2\x \SU2\x \SU2; \SU4\x \U1&$(R)$\\
	&       &$\supset$
	&\SU3; \SO7; \SU2\x \Sp4&$(S)$\\
	&\F4  &$\supset$
	&\SO9; \SU3\x \SU3; \SU2\x \Sp6&$(R)$\\
	&       &$\supset$
	&\SU2; \SU2\x \G2&$(S)$\\
	\hline
	$5$ &\SU6&$\supset$
	&\SU5\x \U1; \SU4\x \SU2\x \U1; \SU3\x \SU3\x \U1
	&$(R)$\\
	&       &$\supset$
	&\SU3; \SU4; \Sp6; \SU3\x \SU2&$(S)$\\
	&\SO{11}&$\supset$
	&\SO{10}; \SO8\x \SU2; \SU4\x \Sp4; \SU2\x \SU2\x \SO7; \SO9\x \U1&$(R)$\\
	&       &$\supset$&\SU2&$(S)$\\
	&\Sp{10}&$\supset$
	&\SU5\x \U1; \SU2\x \Sp8; \Sp4\x \Sp6&$(R)$\\
	&       &$\supset$
	&\SU2; \SU2\x \Sp4&$(S)$\\
	&\SO{10}&$\supset$
	&\SU5\x \U1; \SU2\x \SU2\x \SU4; \SO8\x \U1&$(R)$\\
	&       &$\supset$
	&\Sp4; \SO9; \SU2\x \SO7; \Sp4\x \Sp4&$(S)$\\
	\hline
	$6$ &\SU7&$\supset$
	&\SU6\x \U1; \SU5\x \SU2\x \U1; \SU4\x \SU3\x \U1
	&$(R)$\\
	&       &$\supset$&\SO7&$(S)$\\
	&\SO{13}&$\supset$
	&\SO{12}; \SO{10}\x \SU2; \SO8\x \Sp4; \SU4\x \SO7; \SU2\x \SU2\x \SO9; \SO{11}\x \U1
	&$(R)$\\
	&       &$\supset$&\SU2&$(S)$\\
	&\Sp{12}&$\supset$
	&\SU6\x \U1; \SU2\x \Sp{10}; \Sp4\x \Sp8; \Sp6\x \Sp6&$(R)$\\
	&       &$\supset$
	&\SU2; \SU2\x \SU4; \SU2\x \Sp4&$(S)$\\
	&\SO{12}&$\supset$
	&\SU6\x \U1; \SU2\x \SU2\x \SO8; \SU4\x \SU4; \SO{10}\x \U1&$(R)$\\
	&       &$\supset$
	&\SU2\x \Sp6; \SO{11}; 
	\SU2\x \SO9;
	\Sp4\x \SO7&$(S)$\\
	&\E6  &$\supset$
	&\SO{10}\x \U1; \SU6\x \SU2; \SU3\x \SU3\x \SU3
	&$(R)$\\
	&       &$\supset$
	&\F4; \SU3\x \G2; \Sp8; \G2; \SU3
	&$(S)$\\
	\hline
	$7$ &\SU8&$\supset$
	&\SU7\x \U1; \SU6\x \SU2\x \U1; \SU5\x \SU3\x \U1;
	\SU4\x \SU4\x \U1
	&$(R)$\\
	&       &$\supset$
	&\SO8; \Sp8; \SU4\x \SU2&$(S)$\\
	&\SO{15}&$\supset$
	&\SO{14}; \SO{12}\x \SU2; \SO{10}\x \Sp4; \SO8\x \SO7; \SU4\x \SO9;
	\SU2\x \SU2\x \SO{11}; \SO{13}\x \U1&$(R)$\\
	&       &$\supset$&\SU2; \SU4;
	\SU2\x \Sp4&$(S)$\\
	&\Sp{14}&$\supset$
	&\SU7\x \U1; \SU2\x \Sp{12}; \Sp4\x \Sp{10}; \Sp6\x \Sp8&$(R)$\\
	&       &$\supset$&\SU2; \SU2\x \SO7; \Sp6&$(S)$\\
	&\SO{14}&$\supset$
	&\SU7\x \U1; \SU2\x \SU2\x \SO{10}; \SU4\x \SO8; \SO{12}\x \U1&$(R)$\\
	&       &$\supset$&
	\Sp4; \Sp6; \G2; \SO{13}; \SU2\x \SO{11}; \Sp4\x \SO9; \SO7\x \SO7&$(S)$\\
	&\E7  &$\supset$&\E6\x \U1; \SU8; \SO{12}\x \SU2; \SU6\x \SU3
	&$(R)$\\
	&       &$\supset$&\SU2\x \F4; \G2\x \Sp6; \SU2\x\G2; \SU3; \SU2\x \SU2; \SU2; \SU2&$(S)$\\
	\hline
	$8$ &\SU9&$\supset$
	&\SU8\x \U1; \SU7\x \SU2\x \U1; \SU6\x \SU3\x \U1;
	\SU5\x \SU4\x \U1&$(R)$\\
	&       &$\supset$&\SO9; \SU3\x \SU3&$(S)$\\
	&\SO{17}&$\supset$
	&\SO{16}; \SO{14}\x \SU2; \SO{12}\x \Sp4; \SO{10}\x \SO7;
	\SO8\x \SO9; \SU4\x \SO{11}; \SU2\x \SU2\x \SO{13}; \SO{15}\x \U1&$(R)$\\
	&       &$\supset$&\SU2&$(S)$\\
	&\Sp{16}&$\supset$
	&\SU8\x \U1; \SU2\x \Sp{14}; \Sp4\x \Sp{12}; \Sp6\x \Sp{10}; \Sp8\x \Sp8&$(R)$\\
	&       &$\supset$&\SU2; \Sp4; \SU2\x \SO8&$(S)$\\
	&\SO{16}&$\supset$
	&\SU8\x \U1; \SU2\x \SU2\x \SO{12}; \SU4\x \SO{10}; \SO8\x \SO8;
	\SO{14}\x \U1&$(R)$\\
	&       &$\supset$&
	\SO9; \SU2\x \Sp8; \Sp4\x \Sp4; \SO{15}; \SU2\x \SO{13};
	\Sp4\x \SO{11}; \SO7\x \SO9&$(S)$\\
	&\E8  &$\supset$&\SO{16}; \SU5\x \SU5; \E6\x \SU3; \E7\x \SU2; \SU9
	&$(R)$\\
	&       &$\supset$&
	\G2\x\F4; \SU2\x \SU3; \Sp4; \SU2; \SU2; \SU2&$(S)$\\
	\hline
	\pagebreak
	\enlargethispage{1ex}
	$9$ &\SU{10}&$\supset$
	&\SU9\x \U1; \SU8\x \SU2\x \U1; \SU7\x \SU3\x \U1;
	\SU6\x \SU4\x \U1; \SU5\x \SU5\x \U1
	&$(R)$\\
	&       &$\supset$&
	\SU3; \SU4; \SU5; \SO{10}; \Sp{10}; \SU5\x \SU2&$(S)$\\
	&\SO{19}&$\supset$
	&\SO{18}; \SO{16}\x \SU2; \SO{14}\x \Sp4; \SO{12}\x \SO7;
	\SO{10}\x \SO9; \SO8\x \SO{11}; \SU4\x \SO{13};
	\SU2\x \SU2\x \SO{15}; \SO{17}\x \U1
	&$(R)$\\
	&       &$\supset$&\SU2&$(S)$\\
	&\Sp{18}&$\supset$
	&\SU9\x \U1; \SU2\x \Sp{16}; \Sp4\x \Sp{14}; \Sp6\x \Sp{12};
	\Sp8\x \Sp{10}&$(R)$\\
	&       &$\supset$&\SU2; \SU2\x \SO9; \SU2\x \Sp6&$(S)$\\
	&\SO{18}&$\supset$
	&\SU9\x \U1; \SU2\x \SU2\x \SO{14}; \SU4\x \SO{12}; \SO8\x \SO{10};
	\SO{16}\x \U1&$(R)$\\
	&       &$\supset$&
	\SU2\x \SU4; \SO{17}; \SU2\x \SO{15}; \Sp4\x \SO{13}; \SO7\x \SO{11};
	\SO9\x \SO9&$(S)$\\
	\hline
	$10$&\SU{11}&$\supset$
	&\SU{10}\x \U1; \SU9\x \SU2\x \U1; \SU8\x \SU3\x \U1;
	\SU7\x \SU4\x \U1; \SU6\x \SU5\x \U1&$(R)$\\
	&       &$\supset$&\SO{11}&$(S)$\\
	&\SO{21}&$\supset$
	&\SO{20}; \SO{18}\x \SU2; \SO{16}\x \Sp4; \SO{14}\x \SO7;
	\SO{12}\x \SO9; \SO{10}\x \SO{11}; \SO8\x \SO{13}; \SU4\x \SO{15};
	\SU2\x \SU2\x \SO{17}; \SO{19}\x \U1
	&$(R)$\\
	&       &$\supset$&\SU2; \SU2\x \SO7; \SO7; \Sp6&$(S)$\\
	&\Sp{20}&$\supset$
	&\SU{10}\x \U1; \SU2\x \Sp{18}; \Sp4\x \Sp{16}; \Sp6\x \Sp{14};
	\Sp8\x \Sp{12}; \Sp{10}\x \Sp{10}&$(R)$\\
	&       &$\supset$&
	\SU2; \Sp4\x \Sp4; \SU2\x \SO{10}; \SU6&$(S)$\\
	&\SO{20}&$\supset$&
	\SU{10}\x \U1; \SU2\x \SU2\x \SO{16}; \SU4\x \SO{14};
	\SO8\x \SO{12}; \SO{10}\x \SO{10}; \SO{18}\x \U1&$(R)$\\
	&       &$\supset$&
	\SU2\x \Sp{10}; \SO{19}; \SU2\x \SO{17}; \Sp4\x \SO{15}; \SO7\x \SO{13};
	\SO9\x \SO{11}; \SU4
	&$(S)$\\
 	\hline
	$11$&\SU{12}&$\supset$
	&\SU{11}\x \U1; \SU{10}\x \SU2\x \U1; \SU9\x \SU3\x \U1;
	\SU8\x \SU4\x \U1; \SU7\x \SU5\x \U1; \SU6\x \SU6\x \U1
	&$(R)$\\
	&       &$\supset$
	&\SO{12}; \Sp{12}; \SU6\x \SU2; \SU4\x \SU3
	&$(S)$\\
	&\SO{23}&$\supset$
	&\SO{22}; \SO{20}\x \SU2; \SO{18}\x \Sp4; \SO{16}\x \SO7;
	\SO{14}\x \SO9; \SO{12}\x \SO{11}; \SO{10}\x \SO{13}; \SO8\x \SO{15};
	\SU4\x \SO{17}; \SU2\x \SU2\x \SO{19}; \SO{21}\x \U1
	&$(R)$\\
	&       &$\supset$&\SU2&$(S)$\\
	&\Sp{22}&$\supset$
	&\SU{11}\x \U1; \SU2\x \Sp{20}; \Sp4\x \Sp{18}; \Sp6\x \Sp{16};
	\Sp8\x \Sp{14}; \Sp{10}\x \Sp{12}&$(R)$\\
	&       &$\supset$&
	\SU2; \SU2\x \SO{11}&$(S)$\\
	&\SO{22}&$\supset$&
	\SU{11}\x \U1; \SU2\x \SU2\x \SO{18}; \SU4\x \SO{16};
	\SO8\x \SO{14}; \SO{10}\x \SO{12}; \SO{20}\x \U1&$(R)$\\
	&       &$\supset$&
	\SO{21}; \SU2\x \SO{19}; \Sp4\x \SO{17}; \SO7\x \SO{15};
	\SO9\x \SO{13}; \SO{11}\x \SO{11}&$(S)$\\
 	\hline
	\pagebreak
	$12$&\SU{13}&$\supset$
	&\SU{12}\x \U1; \SU{11}\x \SU2\x \U1; \SU{10}\x \SU3\x \U1;
	\SU9\x \SU4\x \U1; \SU8\x \SU5\x \U1; \SU7\x \SU6\x \U1
	&$(R)$\\
	&       &$\supset$
	&\SO{13}&$(S)$\\
	&\SO{25}&$\supset$
	&\SO{24}; \SO{22}\x \SU2; \SO{20}\x \Sp4; \SO{18}\x \SO7;
	\SO{16}\x \SO9; \SO{14}\x \SO{11}; \SO{12}\x \SO{13}; \SO{10}\x \SO{15};
	\SO8\x \SO{17}; \SU4\x \SO{19}; \SU2\x \SU2\x \SO{21};
	\SO{23}\x \U1
	&$(R)$\\
	&       &$\supset$
	&\SU2; \Sp4\x \Sp4&$(S)$\\
	&\Sp{24}&$\supset$
	&\SU{12}\x \U1; \SU2\x \Sp{22}; \Sp4\x \Sp{20}; \Sp6\x \Sp{18};
	\Sp8\x \Sp{16}; \Sp{10}\x \Sp{14}; \Sp{12}\x \Sp{12}
	&$(R)$\\
	&       &$\supset$&
	\SU2; \SU2\x \Sp8; \SU4\x \Sp4; \SU2\x \SO{12}
	&$(S)$\\
	&\SO{24}&$\supset$&
	\SU{12}\x \U1; \SU2\x \SU2\x \SO{20}; \SU4\x \SO{18};
	\SO8\x \SO{16}; \SO{10}\x \SO{14}; \SO{12}\x \SO{12}; \SO{22}\x \U1
	&$(R)$\\
	&       &$\supset$&
	\SO{23}; \SU2\x \SO{21}; \Sp4\x \SO{19}; \SO7\x \SO{17}; \SO9\x \SO{15};
	\SO{11}\x \SO{13}; \Sp6\x \Sp4; \SU2\x \SO8; \SU5; \SU2\x \Sp{12}
	&$(S)$\\
 	\hline
	$13$&\SU{14}&$\supset$
	&\SU{13}\x \U1; \SU{12}\x \SU2\x \U1; \SU{11}\x \SU3\x \U1;
	\SU{10}\x \SU4\x \U1; \SU9\x \SU5\x \U1; \SU8\x \SU6\x \U1;
	\SU7\x \SU7\x \U1
	&$(R)$\\
	&       &$\supset$
	&\SO{14}; \Sp{14}; \SU7\x \SU2
	&$(S)$\\
	&\SO{27}&$\supset$
	&\SO{26}; \SO{24}\x \SU2; \SO{22}\x \Sp4; \SO{20}\x \SO7;
	\SO{18}\x \SO9; \SO{16}\x \SO{11}; \SO{14}\x \SO{13}; \SO{12}\x \SO{15};
	\SO{10}\x \SO{17}; \SO8\x \SO{19}; \SU4\x \SO{21};
	\SU2\x \SU2\x \SO{23}; \SO{25}\x \U1
	&$(R)$\\
	&       &$\supset$
	&\SU2; \SU3; \SO7; \SU2\x \SO9; \Sp8
	&$(S)$\\
	&\Sp{26}&$\supset$
	&\SU{13}\x \U1; \SU2\x \Sp{24}; \Sp4\x \Sp{22}; \Sp6\x \Sp{20};
	\Sp8\x \Sp{18}; \Sp{10}\x \Sp{16}; \Sp{12}\x \Sp{14}
	&$(R)$\\
	&       &$\supset$&
	\SU2; \SU2\x \SO{13}&$(S)$\\
	&\SO{26}&$\supset$&
	\SU{13}\x \U1; \SU2\x \SU2\x \SO{22}; \SU4\x \SO{20};
	\SO8\x \SO{18}; \SO{10}\x \SO{16}; \SO{12}\x \SO{14};\SO{24}\x \U1
	&$(R)$\\
	&       &$\supset$&
	\SO{25}; \SU2\x \SO{23}; \Sp4\x \SO{21}; \SO7\x \SO{19}; \SO9\x \SO{17};
	\SO{11}\x \SO{15}; \SO{13}\x \SO{13}; \F4
	&$(S)$\\
 	\hline
	$14$&\SU{15}&$\supset$
	&\SU{14}\x \U1; \SU{13}\x \SU2\x \U1; \SU{12}\x \SU3\x \U1;
	\SU{11}\x \SU4\x \U1; \SU{10}\x \SU5\x \U1; \SU9\x \SU6\x \U1;
	\SU8\x \SU7\x \U1
	&$(R)$\\
	&       &$\supset$
	&\SO{15}; \SU5\x \SU3; \SU3; \SU3; \SU5; \SU6
	&$(S)$\\
	&\SO{29}&$\supset$
	&\SO{28}; \SO{26}\x \SU2; \SO{24}\x \Sp4; \SO{22}\x \SO7;
	\SO{20}\x \SO9; \SO{18}\x \SO{11}; \SO{16}\x \SO{13}; \SO{14}\x \SO{15};
	\SO{12}\x \SO{17}; \SO{10}\x \SO{19}; \SO8\x \SO{21}; \SU4\x \SO{23};
	\SU2\x \SU2\x \SO{25}; \SO{27}\x \U1
	&$(R)$\\
	&       &$\supset$
	&\SU2&$(S)$\\
	&\Sp{28}&$\supset$
	&\SU{14}\x \U1; \SU2\x \Sp{26}; \Sp4\x \Sp{24}; \Sp6\x \Sp{22};
	\Sp8\x \Sp{20}; \Sp{10}\x \Sp{18}; \Sp{12}\x \Sp{16}; \Sp{14}\x \Sp{14}
	&$(R)$\\
	&       &$\supset$&
	\SU2; \SO7\x \Sp4; \SU2\x \SO{14}
	&$(S)$\\
	&\SO{28}&$\supset$&
	\SU{14}\x \U1; \SU2\x \SU2\x \SO{24}; \SU4\x \SO{22};
	\SO8\x \SO{20}; \SO{10}\x \SO{18}; \SO{12}\x \SO{16}; \SO{14}\x \SO{14};
	\SO{26}\x \U1
	&$(R)$\\
	&       &$\supset$&
	\SO{27}; \SU2\x \SO{25}; \Sp4\x \SO{23}; \SO7\x \SO{21}; \SO9\x \SO{19};
	\SO{11}\x \SO{17}; \SO{13}\x \SO{15}; \SO8; \SU2\x \Sp{14}
	&$(S)$\\
	\hline
	$15$&\SU{16}&$\supset$
	&\SU{15}\x \U1; \SU{14}\x \SU2\x \U1; \SU{13}\x \SU3\x \U1;
	\SU{12}\x \SU4\x \U1; \SU{11}\x \SU5\x \U1; \SU{10}\x \SU6\x \U1;
	\SU9\x \SU7\x \U1; \SU8\x \SU8\x \U1
	&$(R)$\\
	&       &$\supset$
	&\SO{16}; \Sp{16}; \SO{10}; \SU8\x \SU2; \SU4\x \SU4
	&$(S)$\\
	&\SO{31}&$\supset$
	&\SO{30}; \SO{28}\x \SU2; \SO{26}\x \Sp4; \SO{24}\x \SO7;
	\SO{22}\x \SO9; \SO{20}\x \SO{11}; \SO{18}\x \SO{13}; \SO{16}\x \SO{15};
	\SO{14}\x \SO{17}; \SO{12}\x \SO{19}; \SO{10}\x \SO{21}; \SO8\x \SO{23};
	\SU4\x \SO{25}; \SU2\x \SU2\x \SO{27}; \SO{29}\x \U1
	&$(R)$\\
	&       &$\supset$
	&\SU2&$(S)$\\
	&\Sp{30}&$\supset$
	&\SU{15}\x \U1; \SU2\x \Sp{28}; \Sp4\x \Sp{26}; \Sp6\x \Sp{24};
	\Sp8\x \Sp{22}; \Sp{10}\x \Sp{20}; \Sp{12}\x \Sp{18}; \Sp{14}\x \Sp{16}
	&$(R)$\\
	&       &$\supset$&
	\SU2; \SU2\x \Sp{10}; \Sp4\x \Sp6; \SU2\x \SO{15}
	&$(S)$\\
	&\SO{30}&$\supset$&
	\SU{15}\x \U1; \SU2\x \SU2\x \SO{26}; \SU4\x \SO{24};
	\SO8\x \SO{22}; \SO{10}\x \SO{20}; \SO{12}\x \SO{18}; \SO{14}\x \SO{16};
	\SO{28}\x \U1
	&$(R)$\\
	&       &$\supset$&
	\SO{29}; \SU2\x \SO{27}; \Sp4\x \SO{25}; \SO7\x \SO{23}; \SO9\x \SO{21};
	\SO{11}\x \SO{19}; \SO{13}\x \SO{17}; \SO{15}\x \SO{15}; \SU2\x \SO{10};
	\Sp4\x \SU4
	&$(S)$\\
\end{longtable}
}

We fixed a bug wherein \texttt{DecomposeIrrep[Irrep[SO10][16],ProductAlgebra[SO8,U1]]} would fail to compute the branching rules and throw an error. Similar bugs for other $\SO{m}\to\SO{m-2}\otimes\U1$ branchings were found and corrected. These functions now correctly compute the branching rules without throwing errors.

Over the course of assembling the modified program, several patterns in subalgebra branchings were observed. The first of these is that for $\SU{mn}\to\SU{m}\otimes\SU{n}$, the \irrep{mn} irrep of $\SU{mn}$ branches to the $(\irrep{m},\irrep{n})$ irrep of $\SU{m}\otimes\SU{n}$. Likewise, for $\SO{mn}\to\SO{m}\otimes\SO{n}$, the \irrep{mn} irrep of \SO{mn} branches to the $(\irrep{m},\irrep{n})$ irrep of $\SO{m}\otimes\SO{n}$. Similarly, for $\SO{m+n}\to\SO{m}\otimes\SO{n}$, the $\irrep{m + n}$ irrep of \SO{m+n} branches to the $(\irrep{m},\irrep{1})+(\irrep{1},\irrep{n})$ irrep of $\SO{m}\otimes\SO{n}$. Lastly, for $\Sp{2m+2n}\to \Sp{2m}\otimes\Sp{2n}$, the \irrep{2m+2n} irrep of \Sp{2m+2n} branches to the $(\irrep{2m},\irrep{1})+(\irrep{1},\irrep{2n})$ irrep of $\Sp{2m}\otimes\Sp{2n}$.

\pagebreak

\section{Benchmarks}
\label{sec:Benchmarks}
\enlargethispage{2ex}

In this section, we give runtime benchmarks for subalgebra decomposition. Since exceptional 
algebras have rather complicated Weyl reflection groups of high orders, computations involving them are significantly more CPU and memory demanding than with classical algebras of equal rank. In the 
following discussion, we give runtime benchmarks for the subalgebra decomposition of irreps of exceptional algebras. 

We use the Mathematica command 
\texttt{Timing[]}, which gives the CPU time spent in the Mathematica kernel in 
seconds. It does not include the time needed for the display of results in the 
front end. Note that each of the following timings are taken with a newly launched Kernel 
to avoid speedup due to caching of intermediate results from previous 
computations.

The following timings were taken with Mathematica 11.1.1.\ on a HP\textsuperscript{\textregistered}
Omen\textsuperscript{\texttrademark} with an Intel\textsuperscript{\textregistered} Core\textsuperscript{\texttrademark}
i7-6700HQ (2.60\,GHz) processor and 16\,GB RAM.

As an example for subalgebra decomposition of a large irrep we decompose
the $\irrep{6696000}$ of \E8 to $\G2\otimes\F4:$
\begin{mathin}
	Timing[DecomposeIrrep[Irrep[E8][6696000], ProductAlgebra[G2, F4]]]
\end{mathin}
\begin{mathout}
	$\{1066.14,2(\irrep{7},\irrep{1})+2(\irrep{14},\irrep{1})+(\irrep{1},\irrep{26})+(\irrep{27},\irrep{1})+6(\irrep{7},\irrep{26})+5(\irrep{14},\irrep{26})+2(\irrep{1},\irrep{52})+6(\irrep{27},\irrep{26})+3(\irrep{7},\irrep{52})+2(\irrep{64},\irrep{1})+3(\irrep{14},\irrep{52})+2(\irrep{77},\irrep{1})+5(\irrep{27},\irrep{52})+5(\irrep{64},\irrep{26})+4(\irrep{77},\irrep{26})+(\irrep[1]{77},\irrep{26})+3(\irrep{64},\irrep{52})+2(\irrep{77},\irrep{52})+(\irrep[1]{77},\irrep{52})+(\irrep{189},\irrep{1})+(\irrep{182},\irrep{26})+2(\irrep{189},\irrep{26})+(\irrep{182},\irrep{52})+(\irrep{189},\irrep{52})+3(\irrep{1},\irrep{273})+6(\irrep{7},\irrep{273})+4(\irrep{14},\irrep{273})+8(\irrep{27},\irrep{273})+(\irrep{1},\irrep{324})+6(\irrep{7},\irrep{324})+5(\irrep{64},\irrep{273})+5(\irrep{14},\irrep{324})+3(\irrep{77},\irrep{273})+(\irrep[1]{77},\irrep{273})+4(\irrep{27},\irrep{324})+3(\irrep{64},\irrep{324})+3(\irrep{77},\irrep{324})+(\irrep{182},\irrep{273})+(\irrep{189},\irrep{273})+(\irrep{189},\irrep{324})+2(\irrep{1},\irrep{1053})+5(\irrep{7},\irrep{1053})+(\irrep{7},\irrep[1]{1053})+3(\irrep{14},\irrep{1053})+(\irrep{14},\irrep[1]{1053})+5(\irrep{27},\irrep{1053})+3(\irrep{64},\irrep{1053})+2(\irrep{77},\irrep{1053})+(\irrep{77},\irrep[1]{1053})+(\irrep{189},\irrep{1053})+2(\irrep{1},\irrep{1274})+2(\irrep{7},\irrep{1274})+2(\irrep{14},\irrep{1274})+3(\irrep{27},\irrep{1274})+(\irrep{64},\irrep{1274})+(\irrep{77},\irrep{1274})+(\irrep[1]{77},\irrep{1274})+2(\irrep{7},\irrep{2652})+(\irrep{14},\irrep{2652})+(\irrep{27},\irrep{2652})+2(\irrep{1},\irrep{4096})+5(\irrep{7},\irrep{4096})+3(\irrep{14},\irrep{4096})+4(\irrep{27},\irrep{4096})+2(\irrep{64},\irrep{4096})+(\irrep{77},\irrep{4096})+2(\irrep{7},\irrep{8424})+(\irrep{14},\irrep{8424})+(\irrep{27},\irrep{8424})+(\irrep{64},\irrep{8424})+(\irrep{1},\irrep{10829})+(\irrep{7},\irrep{10829})+(\irrep{14},\irrep{10829})+(\irrep{27},\irrep{10829})+(\irrep{1},\irrep{19278})+(\irrep{7},\irrep{19278})+(\irrep{27},\irrep{19278})+(\irrep{7},\irrep{19448})+(\irrep{14},\irrep{19448})+(\irrep{1},\irrep{34749})+(\irrep{7},\irrep{34749})\}$
\end{mathout}
or we can decompose to $\SU2\otimes \SU3:$
\begin{mathin}
	Timing[DecomposeIrrep[Irrep[E8][6696000], ProductAlgebra[SU2, SU3]]]
\end{mathin}
\begin{mathout}
	$\{569.313,19(\irrep{1},\irrep{1})+66(\irrep{3},\irrep{1})+78(\irrep{5},\irrep{1})+87(\irrep{7},\irrep{1})+121(\irrep{1},\irrep{8})+64(\irrep{9},\irrep{1})+322(\irrep{3},\irrep{8})+103(\irrep{1},\irrep{10})+103(\irrep{1},\irrepbar{10})+48(\irrep{11},\irrep{1})+253(\irrep{3},\irrep{10})+253(\irrep{3},\irrepbar{10})+428(\irrep{5},\irrep{8})+24(\irrep{13},\irrep{1})+349(\irrep{5},\irrep{10})+349(\irrep{5},\irrepbar{10})+426(\irrep{7},\irrep{8})+13(\irrep{15},\irrep{1})+333(\irrep{7},\irrep{10})+333(\irrep{7},\irrepbar{10})+342(\irrep{9},\irrep{8})+4(\irrep{17},\irrep{1})+272(\irrep{9},\irrep{10})+272(\irrep{9},\irrepbar{10})+228(\irrep{11},\irrep{8})+2(\irrep{19},\irrep{1})+173(\irrep{11},\irrep{10})+173(\irrep{11},\irrepbar{10})+126(\irrep{13},\irrep{8})+97(\irrep{13},\irrep{10})+97(\irrep{13},\irrepbar{10})+57(\irrep{15},\irrep{8})+40(\irrep{15},\irrep{10})+40(\irrep{15},\irrepbar{10})+20(\irrep{17},\irrep{8})+15(\irrep{17},\irrep{10})+15(\irrep{17},\irrepbar{10})+5(\irrep{19},\irrep{8})+185(\irrep{1},\irrep{27})+36(\irrep{1},\irrep{28})+36(\irrep{1},\irrepbar{28})+3(\irrep{19},\irrep{10})+3(\irrep{19},\irrepbar{10})+(\irrep{21},\irrep{8})+526(\irrep{3},\irrep{27})+115(\irrep{3},\irrep{28})+115(\irrep{3},\irrepbar{28})+(\irrep{21},\irrep{10})+(\irrep{21},\irrepbar{10})+675(\irrep{5},\irrep{27})+137(\irrep{5},\irrep{28})+137(\irrep{5},\irrepbar{28})+673(\irrep{7},\irrep{27})+136(\irrep{7},\irrep{28})+136(\irrep{7},\irrepbar{28})+138(\irrep{1},\irrep{35})+138(\irrep{1},\irrepbar{35})+520(\irrep{9},\irrep{27})+95(\irrep{9},\irrep{28})+95(\irrep{9},\irrepbar{28})+361(\irrep{3},\irrep{35})+361(\irrep{3},\irrepbar{35})+342(\irrep{11},\irrep{27})+61(\irrep{11},\irrep{28})+61(\irrep{11},\irrepbar{28})+476(\irrep{5},\irrep{35})+476(\irrep{5},\irrepbar{35})+176(\irrep{13},\irrep{27})+25(\irrep{13},\irrep{28})+25(\irrep{13},\irrepbar{28})+455(\irrep{7},\irrep{35})+455(\irrep{7},\irrepbar{35})+77(\irrep{15},\irrep{27})+10(\irrep{15},\irrep{28})+10(\irrep{15},\irrepbar{28})+353(\irrep{9},\irrep{35})+353(\irrep{9},\irrepbar{35})+23(\irrep{17},\irrep{27})+2(\irrep{17},\irrep{28})+2(\irrep{17},\irrepbar{28})+220(\irrep{11},\irrep{35})+220(\irrep{11},\irrepbar{35})+6(\irrep{19},\irrep{27})+112(\irrep{13},\irrep{35})+112(\irrep{13},\irrepbar{35})+44(\irrep{15},\irrep{35})+44(\irrep{15},\irrepbar{35})+13(\irrep{17},\irrep{35})+13(\irrep{17},\irrepbar{35})+2(\irrep{19},\irrep{35})+2(\irrep{19},\irrepbar{35})+5(\irrep{1},\irrep{55})+5(\irrep{1},\irrepbar{55})+6(\irrep{3},\irrep{55})+6(\irrep{3},\irrepbar{55})+10(\irrep{5},\irrep{55})+10(\irrep{5},\irrepbar{55})+7(\irrep{7},\irrep{55})+7(\irrep{7},\irrepbar{55})+5(\irrep{9},\irrep{55})+5(\irrep{9},\irrepbar{55})+156(\irrep{1},\irrep{64})+(\irrep{11},\irrep{55})+(\irrep{11},\irrepbar{55})+412(\irrep{3},\irrep{64})+(\irrep{13},\irrep{55})+(\irrep{13},\irrepbar{55})+533(\irrep{5},\irrep{64})+509(\irrep{7},\irrep{64})+385(\irrep{9},\irrep{64})+235(\irrep{11},\irrep{64})+116(\irrep{13},\irrep{64})+43(\irrep{15},\irrep{64})+25(\irrep{1},\irrep{80})+25(\irrep{1},\irrepbar{80})+11(\irrep{17},\irrep{64})+90(\irrep{1},\irrep{81})+90(\irrep{1},\irrepbar{81})+65(\irrep{3},\irrep{80})+65(\irrep{3},\irrepbar{80})+2(\irrep{19},\irrep{64})+224(\irrep{3},\irrep{81})+224(\irrep{3},\irrepbar{81})+81(\irrep{5},\irrep{80})+81(\irrep{5},\irrepbar{80})+294(\irrep{5},\irrep{81})+294(\irrep{5},\irrepbar{81})+72(\irrep{7},\irrep{80})+72(\irrep{7},\irrepbar{80})+269(\irrep{7},\irrep{81})+269(\irrep{7},\irrepbar{81})+49(\irrep{9},\irrep{80})+49(\irrep{9},\irrepbar{80})+200(\irrep{9},\irrep{81})+200(\irrep{9},\irrepbar{81})+26(\irrep{11},\irrep{80})+26(\irrep{11},\irrepbar{80})+114(\irrep{11},\irrep{81})+114(\irrep{11},\irrepbar{81})+10(\irrep{13},\irrep{80})+10(\irrep{13},\irrepbar{80})+53(\irrep{13},\irrep{81})+53(\irrep{13},\irrepbar{81})+2(\irrep{15},\irrep{80})+2(\irrep{15},\irrepbar{80})+17(\irrep{15},\irrep{81})+17(\irrep{15},\irrepbar{81})+4(\irrep{17},\irrep{81})+4(\irrep{17},\irrepbar{81})+63(\irrep{1},\irrep{125})+179(\irrep{3},\irrep{125})+216(\irrep{5},\irrep{125})+204(\irrep{7},\irrep{125})+140(\irrep{9},\irrep{125})+80(\irrep{11},\irrep{125})+32(\irrep{13},\irrep{125})+10(\irrep{15},\irrep{125})+(\irrep{17},\irrep{125})+(\irrep{1},\irrep{143})+(\irrep{1},\irrepbar{143})+(\irrep{3},\irrep{143})+(\irrep{3},\irrepbar{143})+(\irrep{5},\irrep{143})+(\irrep{5},\irrepbar{143})+(\irrep{7},\irrep{143})+(\irrep{7},\irrepbar{143})+28(\irrep{1},\irrep{154})+28(\irrep{1},\irrepbar{154})+70(\irrep{3},\irrep{154})+70(\irrep{3},\irrepbar{154})+88(\irrep{5},\irrep{154})+88(\irrep{5},\irrepbar{154})+75(\irrep{7},\irrep{154})+75(\irrep{7},\irrepbar{154})+51(\irrep{9},\irrep{154})+51(\irrep{9},\irrepbar{154})+4(\irrep{1},\irrep{162})+4(\irrep{1},\irrepbar{162})+25(\irrep{11},\irrep{154})+25(\irrep{11},\irrepbar{154})+16(\irrep{3},\irrep{162})+16(\irrep{3},\irrepbar{162})+9(\irrep{13},\irrep{154})+9(\irrep{13},\irrepbar{154})+15(\irrep{5},\irrep{162})+15(\irrep{5},\irrepbar{162})+14(\irrep{7},\irrep{162})+14(\irrep{7},\irrepbar{162})+2(\irrep{15},\irrep{154})+2(\irrep{15},\irrepbar{154})+7(\irrep{9},\irrep{162})+7(\irrep{9},\irrepbar{162})+3(\irrep{11},\irrep{162})+3(\irrep{11},\irrepbar{162})+14(\irrep{1},\irrep{216})+36(\irrep{3},\irrep{216})+43(\irrep{5},\irrep{216})+36(\irrep{7},\irrep{216})+22(\irrep{9},\irrep{216})+10(\irrep{11},\irrep{216})+3(\irrep{13},\irrep{216})+4(\irrep{1},\irrep{260})+4(\irrep{1},\irrepbar{260})+8(\irrep{3},\irrep{260})+8(\irrep{3},\irrepbar{260})+11(\irrep{5},\irrep{260})+11(\irrep{5},\irrepbar{260})+7(\irrep{7},\irrep{260})+7(\irrep{7},\irrepbar{260})+4(\irrep{9},\irrep{260})+4(\irrep{9},\irrepbar{260})+(\irrep{11},\irrep{260})+(\irrep{11},\irrepbar{260})+(\irrep{3},\irrep{280})+(\irrep{3},\irrepbar{280})+(\irrep{5},\irrep{280})+(\irrep{5},\irrepbar{280})+(\irrep{1},\irrep{343})+3(\irrep{3},\irrep{343})+(\irrep{5},\irrep{343})+2(\irrep{7},\irrep{343})\}$
\end{mathout}
\newpage
or we can decompose to $\Sp4:$
\begin{mathin}
	Timing[DecomposeIrrep[Irrep[E8][6696000], Sp4]]
\end{mathin}
\begin{mathout}
	$\{1218.64,4(\irrep{1})+27(\irrep{5})+64(\irrep{10})+61(\irrep{14})+104(\irrep{30})+161(\irrep{35})+127(\irrep[1]{35})+127(\irrep{55})+278(\irrep{81})+201(\irrep{84})+134(\irrep{91})+297(\irrep{105})+116(\irrep[2]{140})+355(\irrep{154})+175(\irrep{165})+86(\irrep{204})+417(\irrep{220})+355(\irrep{231})+373(\irrep{260})+52(\irrep{285})+132(\irrep{286})+29(\irrep{385})+465(\irrep{390})+327(\irrep{405})+287(\irrep{429})+434(\irrep{455})+59(\irrep[1]{455})+10(\irrep{506})+244(\irrep{595})+412(\irrep{625})+3(\irrep{650})+24(\irrep{680})+169(\irrep{715})+406(\irrep{770})+292(\irrep{810})+\irrep{819}+154(\irrep{836})+314(\irrep{935})+3(\irrep{969})+69(\irrep{1105})+81(\irrep{1134})+313(\irrep{1190})+151(\irrep{1309})+238(\irrep{1326})+194(\irrep{1330})+\irrep[1]{1330}+35(\irrep{1495})+18(\irrep{1615})+197(\irrep{1729})+101(\irrep[1]{1820})+11(\irrep{1925})+46(\irrep{1976})+149(\irrep{1995})+98(\irrep{2090})+2(\irrep{2261})+104(\irrep{2401})+41(\irrep{2415})+3(\irrep{2430})+78(\irrep{2835})+10(\irrep[1]{2835})+50(\irrep[1]{3080})+25(\irrep{3094})+13(\irrep{3125})+42(\irrep{3220})+29(\irrep{3864})+\irrep{3960}+13(\irrep{4200})+19(\irrep{4301})+3(\irrep{4370})+8(\irrep{4485})+8(\irrep{5100})+3(\irrep{5355})+5(\irrep{5775})+2(\irrep{6175})+\irrep{6561}\}$
\end{mathout}

Another example is given by decomposing the $\irrep{1801371}$ of \F4 to $\SU2\otimes\G2:$
\begin{mathin}
	Timing[DecomposeIrrep[Irrep[F4][1801371],ProductAlgebra[SU2,G2]]]
\end{mathin}
\begin{mathout}
	$\{127.594,(\irrep{1},\irrep{1})+4(\irrep{3},\irrep{1})+6(\irrep{5},\irrep{1})+6(\irrep{1},\irrep{7})+6(\irrep{7},\irrep{1})+17(\irrep{3},\irrep{7})+5(\irrep{9},\irrep{1})+24(\irrep{5},\irrep{7})+4(\irrep{11},\irrep{1})+25(\irrep{7},\irrep{7})+2(\irrep{13},\irrep{1})+7(\irrep{1},\irrep{14})+23(\irrep{9},\irrep{7})+(\irrep{15},\irrep{1})+20(\irrep{3},\irrep{14})+17(\irrep{11},\irrep{7})+27(\irrep{5},\irrep{14})+10(\irrep{13},\irrep{7})+30(\irrep{7},\irrep{14})+5(\irrep{15},\irrep{7})+26(\irrep{9},\irrep{14})+2(\irrep{17},\irrep{7})+20(\irrep{11},\irrep{14})+13(\irrep{13},\irrep{14})+11(\irrep{1},\irrep{27})+7(\irrep{15},\irrep{14})+31(\irrep{3},\irrep{27})+2(\irrep{17},\irrep{14})+43(\irrep{5},\irrep{27})+(\irrep{19},\irrep{14})+47(\irrep{7},\irrep{27})+41(\irrep{9},\irrep{27})+32(\irrep{11},\irrep{27})+20(\irrep{13},\irrep{27})+11(\irrep{15},\irrep{27})+4(\irrep{17},\irrep{27})+(\irrep{19},\irrep{27})+13(\irrep{1},\irrep{64})+37(\irrep{3},\irrep{64})+50(\irrep{5},\irrep{64})+55(\irrep{7},\irrep{64})+48(\irrep{9},\irrep{64})+37(\irrep{11},\irrep{64})+23(\irrep{13},\irrep{64})+11(\irrep{1},\irrep{77})+6(\irrep{1},\irrep[1]{77})+13(\irrep{15},\irrep{64})+31(\irrep{3},\irrep{77})+17(\irrep{3},\irrep[1]{77})+5(\irrep{17},\irrep{64})+43(\irrep{5},\irrep{77})+23(\irrep{5},\irrep[1]{77})+2(\irrep{19},\irrep{64})+46(\irrep{7},\irrep{77})+25(\irrep{7},\irrep[1]{77})+40(\irrep{9},\irrep{77})+21(\irrep{9},\irrep[1]{77})+31(\irrep{11},\irrep{77})+16(\irrep{11},\irrep[1]{77})+19(\irrep{13},\irrep{77})+10(\irrep{13},\irrep[1]{77})+10(\irrep{15},\irrep{77})+5(\irrep{15},\irrep[1]{77})+4(\irrep{17},\irrep{77})+2(\irrep{17},\irrep[1]{77})+(\irrep{19},\irrep{77})+(\irrep{19},\irrep[1]{77})+6(\irrep{1},\irrep{182})+18(\irrep{3},\irrep{182})+24(\irrep{5},\irrep{182})+25(\irrep{7},\irrep{182})+11(\irrep{1},\irrep{189})+21(\irrep{9},\irrep{182})+32(\irrep{3},\irrep{189})+15(\irrep{11},\irrep{182})+43(\irrep{5},\irrep{189})+8(\irrep{13},\irrep{182})+46(\irrep{7},\irrep{189})+4(\irrep{15},\irrep{182})+39(\irrep{9},\irrep{189})+(\irrep{17},\irrep{182})+29(\irrep{11},\irrep{189})+17(\irrep{13},\irrep{189})+9(\irrep{15},\irrep{189})+3(\irrep{17},\irrep{189})+(\irrep{19},\irrep{189})+(\irrep{1},\irrep{273})+3(\irrep{3},\irrep{273})+4(\irrep{5},\irrep{273})+4(\irrep{7},\irrep{273})+3(\irrep{9},\irrep{273})+2(\irrep{11},\irrep{273})+(\irrep{13},\irrep{273})+5(\irrep{1},\irrep{286})+15(\irrep{3},\irrep{286})+20(\irrep{5},\irrep{286})+21(\irrep{7},\irrep{286})+17(\irrep{9},\irrep{286})+12(\irrep{11},\irrep{286})+6(\irrep{13},\irrep{286})+3(\irrep{15},\irrep{286})+(\irrep{17},\irrep{286})+2(\irrep{1},\irrep{378})+6(\irrep{3},\irrep{378})+7(\irrep{5},\irrep{378})+7(\irrep{7},\irrep{378})+5(\irrep{9},\irrep{378})+3(\irrep{11},\irrep{378})+(\irrep{13},\irrep{378})+5(\irrep{1},\irrep{448})+15(\irrep{3},\irrep{448})+19(\irrep{5},\irrep{448})+20(\irrep{7},\irrep{448})+15(\irrep{9},\irrep{448})+10(\irrep{11},\irrep{448})+5(\irrep{13},\irrep{448})+2(\irrep{15},\irrep{448})+(\irrep{3},\irrep{714})+(\irrep{5},\irrep{714})+(\irrep{7},\irrep{714})+2(\irrep{1},\irrep{729})+6(\irrep{3},\irrep{729})+7(\irrep{5},\irrep{729})+7(\irrep{7},\irrep{729})+5(\irrep{9},\irrep{729})+3(\irrep{11},\irrep{729})+(\irrep{13},\irrep{729})+(\irrep{3},\irrep{896})+(\irrep{5},\irrep{896})+(\irrep{7},\irrep{896})+(\irrep{9},\irrep{896})+(\irrep{1},\irrep{924})+4(\irrep{3},\irrep{924})+4(\irrep{5},\irrep{924})+4(\irrep{7},\irrep{924})+2(\irrep{9},\irrep{924})+(\irrep{11},\irrep{924})+(\irrep{3},\irrep{1547})+(\irrep{5},\irrep{1547})+(\irrep{7},\irrep{1547})+(\irrep{3},\irrep{1728})\}$
\end{mathout}

One last example is the decomposition of the $\irrep{2282280}$ of \E7 to $\SU2\otimes\G2:$
\begin{mathin}
	Timing[DecomposeIrrep[Irrep[E7][2282280],ProductAlgebra[SU2,G2]]]
\end{mathin}
\begin{mathout}
	$\{367.625,15(\irrep{2},\irrep{1})+21(\irrep{4},\irrep{1})+21(\irrep{6},\irrep{1})+54(\irrep{2},\irrep{7})+15(\irrep{8},\irrep{1})+80(\irrep{4},\irrep{7})+7(\irrep{10},\irrep{1})+75(\irrep{6},\irrep{7})+3(\irrep{12},\irrep{1})+50(\irrep{8},\irrep{7})+(\irrep{14},\irrep{1})+57(\irrep{2},\irrep{14})+24(\irrep{10},\irrep{7})+87(\irrep{4},\irrep{14})+8(\irrep{12},\irrep{7})+78(\irrep{6},\irrep{14})+2(\irrep{14},\irrep{7})+50(\irrep{8},\irrep{14})+23(\irrep{10},\irrep{14})+7(\irrep{12},\irrep{14})+(\irrep{14},\irrep{14})+92(\irrep{2},\irrep{27})+137(\irrep{4},\irrep{27})+124(\irrep{6},\irrep{27})+79(\irrep{8},\irrep{27})+36(\irrep{10},\irrep{27})+11(\irrep{12},\irrep{27})+2(\irrep{14},\irrep{27})+102(\irrep{2},\irrep{64})+151(\irrep{4},\irrep{64})+132(\irrep{6},\irrep{64})+80(\irrep{8},\irrep{64})+34(\irrep{10},\irrep{64})+9(\irrep{12},\irrep{64})+(\irrep{14},\irrep{64})+90(\irrep{2},\irrep{77})+45(\irrep{2},\irrep[1]{77})+129(\irrep{4},\irrep{77})+65(\irrep{4},\irrep[1]{77})+113(\irrep{6},\irrep{77})+55(\irrep{6},\irrep[1]{77})+67(\irrep{8},\irrep{77})+31(\irrep{8},\irrep[1]{77})+27(\irrep{10},\irrep{77})+12(\irrep{10},\irrep[1]{77})+7(\irrep{12},\irrep{77})+3(\irrep{12},\irrep[1]{77})+(\irrep{14},\irrep{77})+54(\irrep{2},\irrep{182})+76(\irrep{4},\irrep{182})+62(\irrep{6},\irrep{182})+34(\irrep{8},\irrep{182})+88(\irrep{2},\irrep{189})+11(\irrep{10},\irrep{182})+125(\irrep{4},\irrep{189})+2(\irrep{12},\irrep{182})+104(\irrep{6},\irrep{189})+58(\irrep{8},\irrep{189})+21(\irrep{10},\irrep{189})+4(\irrep{12},\irrep{189})+7(\irrep{2},\irrep{273})+9(\irrep{4},\irrep{273})+6(\irrep{6},\irrep{273})+3(\irrep{8},\irrep{273})+39(\irrep{2},\irrep{286})+53(\irrep{4},\irrep{286})+42(\irrep{6},\irrep{286})+21(\irrep{8},\irrep{286})+6(\irrep{10},\irrep{286})+(\irrep{12},\irrep{286})+19(\irrep{2},\irrep{378})+25(\irrep{4},\irrep{378})+19(\irrep{6},\irrep{378})+8(\irrep{8},\irrep{378})+2(\irrep{10},\irrep{378})+42(\irrep{2},\irrep{448})+58(\irrep{4},\irrep{448})+44(\irrep{6},\irrep{448})+21(\irrep{8},\irrep{448})+6(\irrep{10},\irrep{448})+4(\irrep{2},\irrep{714})+4(\irrep{4},\irrep{714})+3(\irrep{6},\irrep{714})+(\irrep{8},\irrep{714})+15(\irrep{2},\irrep{729})+19(\irrep{4},\irrep{729})+13(\irrep{6},\irrep{729})+5(\irrep{8},\irrep{729})+(\irrep{10},\irrep{729})+2(\irrep{2},\irrep{896})+2(\irrep{4},\irrep{896})+(\irrep{6},\irrep{896})+10(\irrep{2},\irrep{924})+14(\irrep{4},\irrep{924})+9(\irrep{6},\irrep{924})+3(\irrep{8},\irrep{924})+2(\irrep{2},\irrep{1547})+2(\irrep{4},\irrep{1547})+(\irrep{6},\irrep{1547})+(\irrep{2},\irrep{1728})+(\irrep{4},\irrep{1728})\}$
\end{mathout}
\pagebreak

\enlargethispage{2ex}
% \vspace*{-10ex}
\section{\LaTeX\ Package}
\label{LaTeXPackage}
LieART comes with a \LaTeX\ package (\texttt{lieart.sty} in the subdirectory \texttt{latex/}) that defines commands to display irreps, roots and weights properly (see Table~\ref{tab:LaTeXCommands}),
which are displayed by LieART using the \texttt{LaTeXForm} on an appropriate expression, e.g.:
\begin{mathin}
DecomposeProduct[Irrep[SU3][8],Irrep[SU3][8]]//LaTeXForm
\end{mathin}
\begin{mathout}
\verb#$\irrep{1}+2(\irrep{8})+\irrep{10}+\irrepbar{10}+\irrep{27}$#
\end{mathout}

\newcommand{\bsl}{\textbackslash}
\begin{table}[!h]
\begin{center}
\renewcommand{\arraystretch}{1.3}
\rowcolors{2}{tablerowcolor}{}
\begin{tabularx}{\textwidth}{llX}
    \toprule\rowcolor{tableheadcolor}
    \textbf{Command Example} & \textbf{Output} & \textbf{Description}\\
    \midrule
     \texttt{\bsl irrep\{10\}} & \irrep{10} & dimensional name of irrep\\
     \texttt{\bsl irrepbar\{10\}} & \irrepbar{10} & dimensional name of conjugated irrep\\
     \texttt{\bsl irrep[2]\{175\}} & \irrep[2]{175} & number of primes as optional parameter\\
     \texttt{\bsl irrepsub\{8\}\{s\}} & \irrepsub{8}{s} & irrep with subscript, e.g., irreps of \SO8 \\
     \texttt{\bsl irrepbarsub\{10\}\{a\}} & \irrepbarsub{10}{a} & conjugated irrep with subscript, e.g., for labeling antisymmetric product\\
     \texttt{\bsl dynkin\{0,1,0,0\}} & \dynkin{0,1,0,0} & Dynkin label of irrep\\
     \texttt{\bsl dynkincomma\{0,10,0,0\}} & \dynkincomma{0,10,0,0} & for Dynkin labels with at least one digit larger then 9\newline
                                                                      (also as \texttt{\bsl rootorthogonal}, \texttt{\bsl weightalpha}
                                                                      and \texttt{\bsl weightorthogonal} for negative integers) \\
     \texttt{\bsl weight\{0,1,0,{-}1\}} & \weight{0,1,0,{-1}} & weight in $\omega$-basis\\
     \texttt{\bsl rootomega\{{-}1,2,{-}1,0\}} & \rootomega{{-}1,2,{-}1,0} & root in $\omega$-basis\\
    \bottomrule
\end{tabularx}
\caption{\label{tab:LaTeXCommands}\LaTeX\ commands defined in supplemental style file \texttt{lieart.sty}}
\end{center}
\end{table}
% \vspace*{-7ex}

\section{Conclusions and Outlook}
\label{ConclusionsAndOutlook}

 The user friendly software package LieART, an application which brings Lie algebra and representation theory related computations to Mathematica, has been upgraded.
It provides functions for the decomposition of tensor products and branching rules of irreducible representations, which are of  interest
to many aspects of mathematical physics, including string theory, particle physics, and  unified model building. LieART exploits the Weyl reflection group in most of its applications, making it fast
and memory efficient. The user interface focuses on usability, allowing one to enter irreducible representations by their dimensional name and
giving results in textbook style. We have reproduced and extended existing tabulated data on irreducible representations, their tensor products
and branching rules. LieART 2.0 includes all products and decompositions of irreps for Lie algebras up to rank 15 including many special subgroup 
decompositions that were not implemented in earlier versions of LieART.

 We consider the tables given in the appendix as dynamical: They are included in LieART as Mathematica notebooks and can easily modified and extended by the user. Tables for algebras of high rank and/or higher dimensional irreducible representations have large CPU time and high memory consumption, so the quick look up tables can serve a useful purpose. Extended  tables will be available online in a standard format as supplemental material. 

\pagebreak

\section{Note Added in Proof}

Because of its importance for string theory we have included extended tables of 
irrep properties and tensor products for \SO{32}.  They were generated with
a ``minimal'' 2019 version iMac, but with more than the typical 4 or 8 GB of memory.
(Processor 3.6 GHz Quad-Core Intel\textsuperscript{\textregistered} Core\textsuperscript{\texttrademark} i3, Memory 32 GB 2400 MHz DDR4)
These tables demonstrate that with sufficient memory, LieART can easily be used for work with the classical
groups \A{n}, \B{n}, \C{n} and \D{n} when $n$ is large.

% \vspace{-10pt}
\section{Acknowledgments}

We thank the many users of LieART for their feedback and suggestions. 
We thank Georgios Papathanasiou for reporting a bug in the decomposition of $\SO{10}{\to}\SO8{\otimes}\U1$.
We thank Daniel Boer for suggestions leading to the implementation of the LieART commands \com{IrrepMultiplicity} and \com{CasimirInvariant}.
We also thank Itzhak Bars for extensive testing of LieART 2.0 under Mathematica 12.0, which enabled us to fix incompatibilities with this version as well as installation issues under Microsoft\textsuperscript{\textregistered} Windows\textsuperscript{\textregistered}.
 
The work of TWK was supported by US DOE grant DE-SC0019235.

\newpage

\bibliographystyle{elsarticle-num}
\bibliography{LieART}

\begin{thebibliography}{100}
\expandafter\ifx\csname url\endcsname\relax
  \def\url#1{\texttt{#1}}\fi
\expandafter\ifx\csname urlprefix\endcsname\relax\def\urlprefix{URL }\fi
\expandafter\ifx\csname href\endcsname\relax
  \def\href#1#2{#2} \def\path#1{#1}\fi

\bibitem{Dynkin:1957um}
E.~Dynkin, {Semisimple subalgebras of semisimple Lie algebras},
  Trans.Am.Math.Soc. 6 (1957) 111.

\bibitem{Dynkin:1957dm}
E.~Dynkin, {Maximal subgroups of the classical groups}, Trans.Am.Math.Soc. 6
  (1957) 245.

\bibitem{Feger:2015bs}
R.~Feger, T.~W. Kephart, {LieART -- A Mathematica application for Lie algebras
  and representation theory}, Comput. Phys. Commun. 192 (2015) 166--195.
\newblock \href {http://arxiv.org/abs/1206.6379} {\path{arXiv:1206.6379}},
  \href {http://dx.doi.org/10.1016/j.cpc.2014.12.023}
  {\path{doi:10.1016/j.cpc.2014.12.023}}.

\bibitem{Georgi:1974sy}
H.~Georgi, S.~Glashow, {Unity of All Elementary Particle Forces},
  Phys.Rev.Lett. 32 (1974) 438--441.
\newblock \href {http://dx.doi.org/10.1103/PhysRevLett.32.438}
  {\path{doi:10.1103/PhysRevLett.32.438}}.

\bibitem{Georgi:1974xy}
{H. Georgi. Particles And Fields: Williamsburg 1974. AIP Conference Proceedings
  No. 23 - C. E. Carlson (eds.)}.

\bibitem{Fritzsch:1974nn}
H.~Fritzsch, P.~Minkowski, {Unified Interactions of Leptons and Hadrons},
  Annals Phys. 93 (1975) 193--266.
\newblock \href {http://dx.doi.org/10.1016/0003-4916(75)90211-0}
  {\path{doi:10.1016/0003-4916(75)90211-0}}.

\bibitem{Gursey:1975ki}
F.~Gursey, P.~Ramond, P.~Sikivie, {A Universal Gauge Theory Model Based on E6},
  Phys.Lett. B60 (1976) 177.
\newblock \href {http://dx.doi.org/10.1016/0370-2693(76)90417-2}
  {\path{doi:10.1016/0370-2693(76)90417-2}}.

\bibitem{Slansky:1981yr}
R.~Slansky, {Group Theory for Unified Model Building}, Phys. Rept. 79 (1981)
  1--128.
\newblock \href {http://dx.doi.org/10.1016/0370-1573(81)90092-2}
  {\path{doi:10.1016/0370-1573(81)90092-2}}.

\bibitem{McKay:99021}
W.~G. McKay, J.~Patera, {Tables of dimensions, indices, and branching rules for
  representations of simple Lie algebras}, Lecture Notes in Pure and Applied
  Mathematics, Dekker, New York, NY, 1981.

\bibitem{Wybourne}
B.~Wybourne, {Classical Groups for Physicists}, Wiley-Interscience, New York,
  1974.

\bibitem{Georgi:1982jb}
H.~Georgi, {LIE ALGEBRAS IN PARTICLE PHYSICS. FROM ISOSPIN TO UNIFIED
  THEORIES}, Front.Phys. 54 (1982) 1--255.

\bibitem{Ramond:2010zz}
P.~Ramond, {Group theory: A physicist's survey}, Cambridge University Press,
  Cambridge, UK, 2010.

\bibitem{cahn1984semi}
R.~Cahn, {Semi-simple lie algebras and their representations}, Vol.~59 of
  Front.Phys., Benjamin Cummings, Menlo Park, CA, 1984.

\bibitem{Humphreys:1980dw}
J.~Humphreys, {Introduction to Lie Algebras and Representation Theory},
  Springer, New York, 1972.

\bibitem{Schur}
B.~G. Wybourne, \href{{http://smc.vnet.net/Schur.html}}{Schur} (2002).
\newline\urlprefix\url{{http://smc.vnet.net/Schur.html}}

\bibitem{Lie}
M.~van Leeuwen, A.~Cohen, B.~Lisser,
  \href{{http://young.sp2mi.univ-poitiers.fr/~marc/LiE/}}{{LiE, A Package for
  Lie Group Computations}}, Computer Algebra Nederland, Amsterdam, 1992.
\newline\urlprefix\url{{http://young.sp2mi.univ-poitiers.fr/~marc/LiE/}}

\bibitem{SimpLie}
T.~Nutma, \href{{http://code.google.com/p/simplie/}}{Simplie} (2009).
\newline\urlprefix\url{{http://code.google.com/p/simplie/}}

\bibitem{Nazarov:2011mv}
A.~Nazarov, {Affine.m - Mathematica package for computations in representation
  theory of finite-dimensional and affine Lie algebras}, Comput. Phys. Commun.
  183 (2012) 2480--2493.
\newblock \href {http://arxiv.org/abs/1107.4681} {\path{arXiv:1107.4681}},
  \href {http://dx.doi.org/10.1016/j.cpc.2012.06.014}
  {\path{doi:10.1016/j.cpc.2012.06.014}}.

\bibitem{Yamatsu:2015npn}
N.~Yamatsu, {Finite-Dimensional Lie Algebras and Their Representations for
  Unified Model Building }\href {http://arxiv.org/abs/1511.08771}
  {\path{arXiv:1511.08771}}.

\bibitem{Lyonnet:2016xiz}
F.~Lyonnet, I.~Schienbein, {PyR@TE 2: A Python tool for computing RGEs at
  two-loop}, Comput. Phys. Commun. 213 (2017) 181--196.
\newblock \href {http://arxiv.org/abs/1608.07274} {\path{arXiv:1608.07274}},
  \href {http://dx.doi.org/10.1016/j.cpc.2016.12.003}
  {\path{doi:10.1016/j.cpc.2016.12.003}}.

\bibitem{Apruzzi:2018xkw}
F.~Apruzzi, F.~Hassler, J.~J. Heckman, T.~B. Rochais, {Nilpotent Networks and
  4D RG Flows}, JHEP 05 (2019) 074.
\newblock \href {http://arxiv.org/abs/1808.10439} {\path{arXiv:1808.10439}},
  \href {http://dx.doi.org/10.1007/JHEP05(2019)074}
  {\path{doi:10.1007/JHEP05(2019)074}}.

\bibitem{Bednyakov:2018cmx}
A.~V. Bednyakov, {On three-loop RGE for the Higgs sector of
  2HDM}[JHEP11,154(2018)].
\newblock \href {http://arxiv.org/abs/1809.04527} {\path{arXiv:1809.04527}},
  \href {http://dx.doi.org/10.1007/JHEP11(2018)154}
  {\path{doi:10.1007/JHEP11(2018)154}}.

\bibitem{Ferretti:2013kya}
G.~Ferretti, D.~Karateev, {Fermionic UV completions of Composite Higgs models},
  JHEP 03 (2014) 077.
\newblock \href {http://arxiv.org/abs/1312.5330} {\path{arXiv:1312.5330}},
  \href {http://dx.doi.org/10.1007/JHEP03(2014)077}
  {\path{doi:10.1007/JHEP03(2014)077}}.

\bibitem{Mojaza:2014ppe}
M.~Mojaza, \href{http://cp3-origins.dk/a/12565}{{Conformality, Spontaneous
  Symmetry Breaking and Mass Hierarchies}}, Ph.D. thesis, Southern Denmark U.,
  CP3-Origins (2014).
\newline\urlprefix\url{http://cp3-origins.dk/a/12565}

\bibitem{Matsumoto:2014ila}
Y.~Matsumoto, Y.~Sakamura, {6D gauge-Higgs unification on T$^{2}$ /Z$_{N}$ with
  custodial symmetry}, JHEP 08 (2014) 175.
\newblock \href {http://arxiv.org/abs/1407.0133} {\path{arXiv:1407.0133}},
  \href {http://dx.doi.org/10.1007/JHEP08(2014)175}
  {\path{doi:10.1007/JHEP08(2014)175}}.

\bibitem{Barnard:2014tla}
J.~Barnard, T.~Gherghetta, T.~S. Ray, A.~Spray, {The Unnatural Composite
  Higgs}, JHEP 01 (2015) 067.
\newblock \href {http://arxiv.org/abs/1409.7391} {\path{arXiv:1409.7391}},
  \href {http://dx.doi.org/10.1007/JHEP01(2015)067}
  {\path{doi:10.1007/JHEP01(2015)067}}.

\bibitem{Lim:2014zfa}
K.~S. Lim, \href{http://archiv.ub.uni-heidelberg.de/volltextserver/17723/}{{New
  Aspects Of Scale And Discrete Flavor Symmetry Breaking}}, Ph.D. thesis,
  Heidelberg U. (2014).
\newline\urlprefix\url{http://archiv.ub.uni-heidelberg.de/volltextserver/17723/}

\bibitem{Hollands:2016kgm}
L.~Hollands, A.~Neitzke, {BPS states in the Minahan-Nemeschansky ${E_6}$
  theory}, Commun. Math. Phys. 353~(1) (2017) 317--351.
\newblock \href {http://arxiv.org/abs/1607.01743} {\path{arXiv:1607.01743}},
  \href {http://dx.doi.org/10.1007/s00220-016-2798-1}
  {\path{doi:10.1007/s00220-016-2798-1}}.

\bibitem{ElHedri:2016onc}
S.~El~Hedri, A.~Kaminska, M.~de~Vries, {A Sommerfeld Toolbox for Colored Dark
  Sectors}, Eur. Phys. J. C77~(9) (2017) 622.
\newblock \href {http://arxiv.org/abs/1612.02825} {\path{arXiv:1612.02825}},
  \href {http://dx.doi.org/10.1140/epjc/s10052-017-5168-z}
  {\path{doi:10.1140/epjc/s10052-017-5168-z}}.

\bibitem{Albright:2012zt}
C.~H. Albright, R.~P. Feger, T.~W. Kephart, {An explicit SU(12) family and
  flavor unification model with natural fermion masses and mixings}, Phys.Rev.
  D86 (2012) 015012.
\newblock \href {http://arxiv.org/abs/1204.5471} {\path{arXiv:1204.5471}},
  \href {http://dx.doi.org/10.1103/PhysRevD.86.015012}
  {\path{doi:10.1103/PhysRevD.86.015012}}.

\bibitem{Antipin:2015xia}
O.~Antipin, M.~Redi, A.~Strumia, E.~Vigiani, {Accidental Composite Dark
  Matter}, JHEP 07 (2015) 039.
\newblock \href {http://arxiv.org/abs/1503.08749} {\path{arXiv:1503.08749}},
  \href {http://dx.doi.org/10.1007/JHEP07(2015)039}
  {\path{doi:10.1007/JHEP07(2015)039}}.

\bibitem{Fonseca:2015aoa}
R.~M. Fonseca, {On the chirality of the SM and the fermion content of GUTs},
  Nucl. Phys. B897 (2015) 757--780.
\newblock \href {http://arxiv.org/abs/1504.03695} {\path{arXiv:1504.03695}},
  \href {http://dx.doi.org/10.1016/j.nuclphysb.2015.06.012}
  {\path{doi:10.1016/j.nuclphysb.2015.06.012}}.

\bibitem{Feger:2015xqa}
R.~P. Feger, T.~W. Kephart, {Grand Unification and Exotic Fermions}, Phys. Rev.
  D92~(3) (2015) 035005.
\newblock \href {http://arxiv.org/abs/1505.03403} {\path{arXiv:1505.03403}},
  \href {http://dx.doi.org/10.1103/PhysRevD.92.035005}
  {\path{doi:10.1103/PhysRevD.92.035005}}.

\bibitem{Maurer:2015zpa}
V.~K.~M. Maurer, \href{http://edoc.unibas.ch/diss/DissB_11268}{{Insights into
  Grand Unified Theories from Current Experimental Data}}, Ph.D. thesis, Basel
  U. (2015).
\newblock \href {http://dx.doi.org/10.5451/unibas-006387307}
  {\path{doi:10.5451/unibas-006387307}}.
\newline\urlprefix\url{http://edoc.unibas.ch/diss/DissB_11268}

\bibitem{GonzaloVelasco:2015hki}
T.~Gonzalo~Velasco, \href{http://discovery.ucl.ac.uk/id/eprint/1471160}{{Model
  Building and Phenomenology in Grand Unified Theories}}, Ph.D. thesis,
  University Coll. London (2015).
\newline\urlprefix\url{http://discovery.ucl.ac.uk/id/eprint/1471160}

\bibitem{Albright:2016lpi}
C.~H. Albright, R.~P. Feger, T.~W. Kephart, {Unification of Gauge, Family, and
  Flavor Symmetries Illustrated in Gauged SU(12) Models}, Phys. Rev. D93~(7)
  (2016) 075032.
\newblock \href {http://arxiv.org/abs/1601.07523} {\path{arXiv:1601.07523}},
  \href {http://dx.doi.org/10.1103/PhysRevD.93.075032}
  {\path{doi:10.1103/PhysRevD.93.075032}}.

\bibitem{Bajc:2016efj}
B.~Bajc, F.~Sannino, {Asymptotically Safe Grand Unification}, JHEP 12 (2016)
  141.
\newblock \href {http://arxiv.org/abs/1610.09681} {\path{arXiv:1610.09681}},
  \href {http://dx.doi.org/10.1007/JHEP12(2016)141}
  {\path{doi:10.1007/JHEP12(2016)141}}.

\bibitem{Schwichtenberg:2017xhv}
J.~Schwichtenberg, {Dark matter in E$_{6}$ Grand unification}, JHEP 02 (2018)
  016.
\newblock \href {http://arxiv.org/abs/1704.04219} {\path{arXiv:1704.04219}},
  \href {http://dx.doi.org/10.1007/JHEP02(2018)016}
  {\path{doi:10.1007/JHEP02(2018)016}}.

\bibitem{Yamatsu:2017mei}
N.~Yamatsu, {Special Grand Unification}, PTEP 2017~(6) (2017) 061B01.
\newblock \href {http://arxiv.org/abs/1704.08827} {\path{arXiv:1704.08827}},
  \href {http://dx.doi.org/10.1093/ptep/ptx088}
  {\path{doi:10.1093/ptep/ptx088}}.

\bibitem{Benli:2017eld}
S.~Benli, T.~Dereli, {Masses and Mixing of Neutral Leptons in a Grand Unified
  E$_{6}$ Model with Intermediate Pati-Salam Symmetry}, Int. J. Theor. Phys.
  57~(8) (2018) 2343--2358.
\newblock \href {http://arxiv.org/abs/1707.03144} {\path{arXiv:1707.03144}},
  \href {http://dx.doi.org/10.1007/s10773-018-3757-8}
  {\path{doi:10.1007/s10773-018-3757-8}}.

\bibitem{Ernst:2018bib}
A.~Ernst, A.~Ringwald, C.~Tamarit, {Axion Predictions in $SO(10)\times
  U(1)_{\rm PQ}$ Models}, JHEP 02 (2018) 103.
\newblock \href {http://arxiv.org/abs/1801.04906} {\path{arXiv:1801.04906}},
  \href {http://dx.doi.org/10.1007/JHEP02(2018)103}
  {\path{doi:10.1007/JHEP02(2018)103}}.

\bibitem{Chala:2018ari}
M.~Chala, C.~Krause, G.~Nardini, {Signals of the electroweak phase transition
  at colliders and gravitational wave observatories}, JHEP 07 (2018) 062.
\newblock \href {http://arxiv.org/abs/1802.02168} {\path{arXiv:1802.02168}},
  \href {http://dx.doi.org/10.1007/JHEP07(2018)062}
  {\path{doi:10.1007/JHEP07(2018)062}}.

\bibitem{Yamatsu:2018tnv}
N.~Yamatsu, {Family Unification in Special Grand Unification}, PTEP 2018~(9)
  (2018) 091B01.
\newblock \href {http://arxiv.org/abs/1807.10855} {\path{arXiv:1807.10855}},
  \href {http://dx.doi.org/10.1093/ptep/pty100}
  {\path{doi:10.1093/ptep/pty100}}.

\bibitem{Fonseca:2013qka}
R.~M. Sousa~da Fonseca, {Renormalization in supersymmetric models}, Ph.D.
  thesis, Lisbon, CENTRA (2013).
\newblock \href {http://arxiv.org/abs/1310.1296} {\path{arXiv:1310.1296}}.

\bibitem{Chen:2018wep}
Z.~Chen, W.~Gu, H.~Parsian, E.~Sharpe, {Two-dimensional supersymmetric gauge
  theories with exceptional gauge groups }\href
  {http://arxiv.org/abs/1808.04070} {\path{arXiv:1808.04070}}.

\bibitem{Agarwal:2018oxb}
P.~Agarwal, {On dimensional reduction of 4d N=1 Lagrangians for Argyres-Douglas
  theories}, JHEP 03 (2019) 011.
\newblock \href {http://arxiv.org/abs/1809.10534} {\path{arXiv:1809.10534}},
  \href {http://dx.doi.org/10.1007/JHEP03(2019)011}
  {\path{doi:10.1007/JHEP03(2019)011}}.

\bibitem{Abzalov:2015ega}
A.~Abzalov, I.~Bakhmatov, E.~T. Musaev, {Exceptional field theory: $SO(5,5)$},
  JHEP 06 (2015) 088.
\newblock \href {http://arxiv.org/abs/1504.01523} {\path{arXiv:1504.01523}},
  \href {http://dx.doi.org/10.1007/JHEP06(2015)088}
  {\path{doi:10.1007/JHEP06(2015)088}}.

\bibitem{Panico:2015jxa}
G.~Panico, A.~Wulzer, {The Composite Nambu-Goldstone Higgs}, Lect. Notes Phys.
  913 (2016) pp.1--316.
\newblock \href {http://arxiv.org/abs/1506.01961} {\path{arXiv:1506.01961}},
  \href {http://dx.doi.org/10.1007/978-3-319-22617-0}
  {\path{doi:10.1007/978-3-319-22617-0}}.

\bibitem{Kephart:2015oaa}
T.~W. Kephart, T.-C. Yuan, {Origins of Inert Higgs Doublets}, Nucl. Phys. B906
  (2016) 549--560.
\newblock \href {http://arxiv.org/abs/1508.00673} {\path{arXiv:1508.00673}},
  \href {http://dx.doi.org/10.1016/j.nuclphysb.2016.03.023}
  {\path{doi:10.1016/j.nuclphysb.2016.03.023}}.

\bibitem{Danielsson:2015rca}
U.~Danielsson, G.~Dibitetto, {Type IIB on $S^{3}\times S^{3}$ through $Q$ \&
  $P$ fluxes}, JHEP 01 (2016) 057.
\newblock \href {http://arxiv.org/abs/1507.04476} {\path{arXiv:1507.04476}},
  \href {http://dx.doi.org/10.1007/JHEP01(2016)057}
  {\path{doi:10.1007/JHEP01(2016)057}}.

\bibitem{Nibbelink:2016wms}
S.~Groot~Nibbelin, F.~Ruehle, {Line bundle embeddings for heterotic theories},
  JHEP 04 (2016) 186.
\newblock \href {http://arxiv.org/abs/1601.00676} {\path{arXiv:1601.00676}},
  \href {http://dx.doi.org/10.1007/JHEP04(2016)186}
  {\path{doi:10.1007/JHEP04(2016)186}}.

\bibitem{Ahn:2017noo}
C.~Ahn, {Wolf space coset spectrum in the large ${\cal N}=4$ holography}, J.
  Phys. A51~(43) (2018) 435402.
\newblock \href {http://arxiv.org/abs/1711.07599} {\path{arXiv:1711.07599}},
  \href {http://dx.doi.org/10.1088/1751-8121/aae15d}
  {\path{doi:10.1088/1751-8121/aae15d}}.

\bibitem{Apruzzi:2017nck}
F.~Apruzzi, M.~Fazzi, {AdS$_{7}$/CFT$_{6}$ with orientifolds}, JHEP 01 (2018)
  124.
\newblock \href {http://arxiv.org/abs/1712.03235} {\path{arXiv:1712.03235}},
  \href {http://dx.doi.org/10.1007/JHEP01(2018)124}
  {\path{doi:10.1007/JHEP01(2018)124}}.

\bibitem{Ahn:2018lqx}
C.~Ahn, {Higher spin currents with manifest $SO(4)$ symmetry in the large
  ${\cal N}=4$ holography}, Int. J. Mod. Phys. A33~(35) (2018) 1850208.
\newblock \href {http://arxiv.org/abs/1805.02298} {\path{arXiv:1805.02298}},
  \href {http://dx.doi.org/10.1142/S0217751X18502081}
  {\path{doi:10.1142/S0217751X18502081}}.

\bibitem{deMedeiros:2013mca}
P.~de~Medeiros, S.~Hollands, {Superconformal quantum field theory in curved
  spacetime}, Class. Quant. Grav. 30 (2013) 175015.
\newblock \href {http://arxiv.org/abs/1305.0499} {\path{arXiv:1305.0499}},
  \href {http://dx.doi.org/10.1088/0264-9381/30/17/175015}
  {\path{doi:10.1088/0264-9381/30/17/175015}}.

\bibitem{Agarwal:2013uga}
P.~Agarwal, J.~Song, {New N=1 Dualities from M5-branes and Outer-automorphism
  Twists}, JHEP 03 (2014) 133.
\newblock \href {http://arxiv.org/abs/1311.2945} {\path{arXiv:1311.2945}},
  \href {http://dx.doi.org/10.1007/JHEP03(2014)133}
  {\path{doi:10.1007/JHEP03(2014)133}}.

\bibitem{Chacaltana:2014jba}
O.~Chacaltana, J.~Distler, A.~Trimm, {Tinkertoys for the E$_{6}$ theory}, JHEP
  09 (2015) 007.
\newblock \href {http://arxiv.org/abs/1403.4604} {\path{arXiv:1403.4604}},
  \href {http://dx.doi.org/10.1007/JHEP09(2015)007}
  {\path{doi:10.1007/JHEP09(2015)007}}.

\bibitem{Hwang:2014uwa}
C.~Hwang, J.~Kim, S.~Kim, J.~Park, {General instanton counting and 5d SCFT},
  JHEP 07 (2015) 063, [Addendum: JHEP04,094(2016)].
\newblock \href {http://arxiv.org/abs/1406.6793} {\path{arXiv:1406.6793}},
  \href {http://dx.doi.org/10.1007/JHEP07(2015)063, 10.1007/JHEP04(2016)094}
  {\path{doi:10.1007/JHEP07(2015)063, 10.1007/JHEP04(2016)094}}.

\bibitem{Mitev:2014jza}
V.~Mitev, E.~Pomoni, M.~Taki, F.~Yagi, {Fiber-Base Duality and Global Symmetry
  Enhancement}, JHEP 04 (2015) 052.
\newblock \href {http://arxiv.org/abs/1411.2450} {\path{arXiv:1411.2450}},
  \href {http://dx.doi.org/10.1007/JHEP04(2015)052}
  {\path{doi:10.1007/JHEP04(2015)052}}.

\bibitem{Chacaltana:2016shw}
O.~Chacaltana, J.~Distler, A.~Trimm, {Tinkertoys for the Z3-twisted D4 Theory
  }\href {http://arxiv.org/abs/1601.02077} {\path{arXiv:1601.02077}}.

\bibitem{Cordova:2016emh}
C.~Cordova, T.~T. Dumitrescu, K.~Intriligator, {Multiplets of Superconformal
  Symmetry in Diverse Dimensions}, JHEP 03 (2019) 163.
\newblock \href {http://arxiv.org/abs/1612.00809} {\path{arXiv:1612.00809}},
  \href {http://dx.doi.org/10.1007/JHEP03(2019)163}
  {\path{doi:10.1007/JHEP03(2019)163}}.

\bibitem{Karateev:2017yoq}
D.~Karateev, \href{http://hdl.handle.net/20.500.11767/57140}{{Kinematics of 4D
  Conformal Field Theories}}, Ph.D. thesis, SISSA, Trieste (2017).
\newline\urlprefix\url{http://hdl.handle.net/20.500.11767/57140}

\bibitem{Jefferson:2017ahm}
P.~Jefferson, H.-C. Kim, C.~Vafa, G.~Zafrir, {Towards Classification of 5d
  SCFTs: Single Gauge Node }\href {http://arxiv.org/abs/1705.05836}
  {\path{arXiv:1705.05836}}.

\bibitem{Kim:2014nqa}
S.-S. Kim, F.~Yagi, {5d E$_{n}$ Seiberg-Witten curve via toric-like diagram},
  JHEP 06 (2015) 082.
\newblock \href {http://arxiv.org/abs/1411.7903} {\path{arXiv:1411.7903}},
  \href {http://dx.doi.org/10.1007/JHEP06(2015)082}
  {\path{doi:10.1007/JHEP06(2015)082}}.

\bibitem{Danielsson:2015tsa}
U.~Danielsson, G.~Dibitetto, {Geometric non-geometry}, JHEP 04 (2015) 084.
\newblock \href {http://arxiv.org/abs/1501.03944} {\path{arXiv:1501.03944}},
  \href {http://dx.doi.org/10.1007/JHEP04(2015)084}
  {\path{doi:10.1007/JHEP04(2015)084}}.

\bibitem{Martucci:2015oaa}
L.~Martucci, T.~Weigand, {Hidden Selection Rules, M5-instantons and Fluxes in
  F-theory}, JHEP 10 (2015) 131.
\newblock \href {http://arxiv.org/abs/1507.06999} {\path{arXiv:1507.06999}},
  \href {http://dx.doi.org/10.1007/JHEP10(2015)131}
  {\path{doi:10.1007/JHEP10(2015)131}}.

\bibitem{Cvetic:2018xaq}
M.~Cvetič, J.~J. Heckman, L.~Lin, {Towards Exotic Matter and Discrete
  Non-Abelian Symmetries in F-theory}, JHEP 11 (2018) 001.
\newblock \href {http://arxiv.org/abs/1806.10594} {\path{arXiv:1806.10594}},
  \href {http://dx.doi.org/10.1007/JHEP11(2018)001}
  {\path{doi:10.1007/JHEP11(2018)001}}.

\bibitem{Raghuram:2018hjn}
N.~Raghuram, W.~Taylor, {Large U(1) charges in F-theory}, JHEP 10 (2018) 182.
\newblock \href {http://arxiv.org/abs/1809.01666} {\path{arXiv:1809.01666}},
  \href {http://dx.doi.org/10.1007/JHEP10(2018)182}
  {\path{doi:10.1007/JHEP10(2018)182}}.

\bibitem{Ahmed:2017mqq}
I.~Ahmed, R.~I. Nepomechie, C.~Wang, {Quantum group symmetries and completeness
  for $\boldsymbol {A}_{\boldsymbol {2n}}^{\boldsymbol{(2)}}$ open spin
  chains}, J. Phys. A50~(28) (2017) 284002.
\newblock \href {http://arxiv.org/abs/1702.01482} {\path{arXiv:1702.01482}},
  \href {http://dx.doi.org/10.1088/1751-8121/aa7606}
  {\path{doi:10.1088/1751-8121/aa7606}}.

\bibitem{Nepomechie:2017hgw}
R.~I. Nepomechie, R.~A. Pimenta, A.~L. Retore, {The integrable quantum group
  invariant $A_{2n-1}^(2)$ and $D_{n+1}^(2)$ open spin chains}, Nucl. Phys.
  B924 (2017) 86--127.
\newblock \href {http://arxiv.org/abs/1707.09260} {\path{arXiv:1707.09260}},
  \href {http://dx.doi.org/10.1016/j.nuclphysb.2017.09.004}
  {\path{doi:10.1016/j.nuclphysb.2017.09.004}}.

\bibitem{Nepomechie:2018dsn}
R.~I. Nepomechie, A.~L. Retore, {Surveying the quantum group symmetries of
  integrable open spin chains}, Nucl. Phys. B930 (2018) 91--134.
\newblock \href {http://arxiv.org/abs/1802.04864} {\path{arXiv:1802.04864}},
  \href {http://dx.doi.org/10.1016/j.nuclphysb.2018.02.023}
  {\path{doi:10.1016/j.nuclphysb.2018.02.023}}.

\bibitem{Nepomechie:2018wzp}
R.~I. Nepomechie, R.~A. Pimenta, {New $D_{n+1}^{(2)}$ K-matrices with quantum
  group symmetry}, J. Phys. A51~(39) (2018) 39LT02.
\newblock \href {http://arxiv.org/abs/1805.10144} {\path{arXiv:1805.10144}},
  \href {http://dx.doi.org/10.1088/1751-8121/aad957}
  {\path{doi:10.1088/1751-8121/aad957}}.

\bibitem{Kim:2018gjo}
H.-C. Kim, J.~Kim, S.~Kim, K.-H. Lee, J.~Park, {6d strings and exceptional
  instantons }\href {http://arxiv.org/abs/1801.03579}
  {\path{arXiv:1801.03579}}.

\bibitem{Bazzanella:2014vla}
M.~Bazzanella, J.~Nilsson, {Non-Linear Methods in Strongly Correlated Electron
  Systems }\href {http://arxiv.org/abs/1405.5176} {\path{arXiv:1405.5176}}.

\bibitem{Kim:2012gu}
H.-C. Kim, S.-S. Kim, K.~Lee, {5-dim Superconformal Index with Enhanced En
  Global Symmetry}, JHEP 10 (2012) 142.
\newblock \href {http://arxiv.org/abs/1206.6781} {\path{arXiv:1206.6781}},
  \href {http://dx.doi.org/10.1007/JHEP10(2012)142}
  {\path{doi:10.1007/JHEP10(2012)142}}.

\bibitem{Lemos:2012ph}
M.~Lemos, W.~Peelaers, L.~Rastelli, {The superconformal index of class $S$
  theories of type $D$}, JHEP 05 (2014) 120.
\newblock \href {http://arxiv.org/abs/1212.1271} {\path{arXiv:1212.1271}},
  \href {http://dx.doi.org/10.1007/JHEP05(2014)120}
  {\path{doi:10.1007/JHEP05(2014)120}}.

\bibitem{Cremonesi:2013lqa}
S.~Cremonesi, A.~Hanany, A.~Zaffaroni, {Monopole operators and Hilbert series
  of Coulomb branches of $3d$ $\mathcal{N} = 4$ gauge theories}, JHEP 01 (2014)
  005.
\newblock \href {http://arxiv.org/abs/1309.2657} {\path{arXiv:1309.2657}},
  \href {http://dx.doi.org/10.1007/JHEP01(2014)005}
  {\path{doi:10.1007/JHEP01(2014)005}}.

\bibitem{Fortunato:2013cta}
L.~Fortunato, W.~A. de~Graaf, {$E_7 \subset Sp(56,R)$ irrep decompositions of
  interest for physical models }\href {http://arxiv.org/abs/1312.5151}
  {\path{arXiv:1312.5151}}.

\bibitem{Hanany:2014dia}
A.~Hanany, R.~Kalveks, {Highest Weight Generating Functions for Hilbert
  Series}, JHEP 10 (2014) 152.
\newblock \href {http://arxiv.org/abs/1408.4690} {\path{arXiv:1408.4690}},
  \href {http://dx.doi.org/10.1007/JHEP10(2014)152}
  {\path{doi:10.1007/JHEP10(2014)152}}.

\bibitem{Kim:2015jxc}
J.~Kim, \href{http://s-space.snu.ac.kr/handle/10371/121526}{{Exact Results on
  Higher Dimensional Quantum Field Theories}}, Ph.D. thesis, Seoul Natl. U.,
  Dept. Phys. Astron. (2015).
\newline\urlprefix\url{http://s-space.snu.ac.kr/handle/10371/121526}

\bibitem{Lehman:2015via}
L.~Lehman, A.~Martin, {Hilbert Series for Constructing Lagrangians: expanding
  the phenomenologist's toolbox}, Phys. Rev. D91 (2015) 105014.
\newblock \href {http://arxiv.org/abs/1503.07537} {\path{arXiv:1503.07537}},
  \href {http://dx.doi.org/10.1103/PhysRevD.91.105014}
  {\path{doi:10.1103/PhysRevD.91.105014}}.

\bibitem{Hanany:2015hxa}
A.~Hanany, R.~Kalveks, {Construction and Deconstruction of Single Instanton
  Hilbert Series}, JHEP 12 (2015) 118.
\newblock \href {http://arxiv.org/abs/1509.01294} {\path{arXiv:1509.01294}},
  \href {http://dx.doi.org/10.1007/JHEP12(2015)118}
  {\path{doi:10.1007/JHEP12(2015)118}}.

\bibitem{Hoehn:2015zom}
P.~A. Höhn, C.~S.~P. Wever, {Quantum theory from questions}, Phys. Rev.
  A95~(1) (2017) 012102.
\newblock \href {http://arxiv.org/abs/1511.01130} {\path{arXiv:1511.01130}},
  \href {http://dx.doi.org/10.1103/PhysRevA.95.012102}
  {\path{doi:10.1103/PhysRevA.95.012102}}.

\bibitem{Matassa:2015gla}
M.~Matassa, {An analogue of Weyl’s law for quantized irreducible generalized
  flag manifolds}, J. Math. Phys. 56~(9) (2015) 091704.
\newblock \href {http://arxiv.org/abs/1410.8029} {\path{arXiv:1410.8029}},
  \href {http://dx.doi.org/10.1063/1.4931606} {\path{doi:10.1063/1.4931606}}.

\bibitem{Linander:2016jyf}
H.~Linander,
  \href{http://publications.lib.chalmers.se/publication/233715-20-theory-and-higher-spin-twisting-turning-and-spinning-towards-higher-energies}{{(2,0)
  Theory and Higher Spin: Twisting, Turning and Spinning Towards Higher
  Energies}}, Ph.D. thesis, Chalmers U. Tech. (2016).
\newline\urlprefix\url{http://publications.lib.chalmers.se/publication/233715-20-theory-and-higher-spin-twisting-turning-and-spinning-towards-higher-energies}

\bibitem{Urichuk:2016xau}
A.~Urichuk, M.~A. Walton, {Adjoint affine fusion and tadpoles}, J. Math. Phys.
  57~(6) (2016) 061702.
\newblock \href {http://arxiv.org/abs/1606.03842} {\path{arXiv:1606.03842}},
  \href {http://dx.doi.org/10.1063/1.4954909} {\path{doi:10.1063/1.4954909}}.

\bibitem{Allys:2016kbq}
E.~Allys, P.~Peter, Y.~Rodriguez, {Generalized SU(2) Proca Theory}, Phys. Rev.
  D94~(8) (2016) 084041.
\newblock \href {http://arxiv.org/abs/1609.05870} {\path{arXiv:1609.05870}},
  \href {http://dx.doi.org/10.1103/PhysRevD.94.084041}
  {\path{doi:10.1103/PhysRevD.94.084041}}.

\bibitem{Allys:2016hfl}
E.~Allys, {New terms for scalar multi-Galileon models and application to SO(N)
  and SU(N) group representations}, Phys. Rev. D95~(6) (2017) 064051.
\newblock \href {http://arxiv.org/abs/1612.01972} {\path{arXiv:1612.01972}},
  \href {http://dx.doi.org/10.1103/PhysRevD.95.064051}
  {\path{doi:10.1103/PhysRevD.95.064051}}.

\bibitem{Watanabe:2017bmi}
N.~Watanabe, {Schur indices with class S line operators from networks and
  further skein relations }\href {http://arxiv.org/abs/1701.04090}
  {\path{arXiv:1701.04090}}.

\bibitem{Hayashi:2017jze}
H.~Hayashi, K.~Ohmori, {5d/6d DE instantons from trivalent gluing of web
  diagrams}, JHEP 06 (2017) 078.
\newblock \href {http://arxiv.org/abs/1702.07263} {\path{arXiv:1702.07263}},
  \href {http://dx.doi.org/10.1007/JHEP06(2017)078}
  {\path{doi:10.1007/JHEP06(2017)078}}.

\bibitem{Niehoff:2017mbk}
B.~E. Niehoff, {Faster Tensor Canonicalization}, Comput. Phys. Commun. 228
  (2018) 123--145.
\newblock \href {http://arxiv.org/abs/1702.08114} {\path{arXiv:1702.08114}},
  \href {http://dx.doi.org/10.1016/j.cpc.2018.02.014}
  {\path{doi:10.1016/j.cpc.2018.02.014}}.

\bibitem{Moriyama:2017nbw}
S.~Moriyama, T.~Nosaka, K.~Yano, {Superconformal Chern-Simons Theories from del
  Pezzo Geometries}, JHEP 11 (2017) 089.
\newblock \href {http://arxiv.org/abs/1707.02420} {\path{arXiv:1707.02420}},
  \href {http://dx.doi.org/10.1007/JHEP11(2017)089}
  {\path{doi:10.1007/JHEP11(2017)089}}.

\bibitem{Mukhi:2017ugw}
S.~Mukhi, G.~Muralidhara, {Universal RCFT Correlators from the Holomorphic
  Bootstrap}, JHEP 02 (2018) 028.
\newblock \href {http://arxiv.org/abs/1708.06772} {\path{arXiv:1708.06772}},
  \href {http://dx.doi.org/10.1007/JHEP02(2018)028}
  {\path{doi:10.1007/JHEP02(2018)028}}.

\bibitem{Shahlaei:2018lth}
A.~Shahlaei, S.~Rafibakhsh, {$F_4$ , $E_6$ and $G_2$ exceptional gauge groups
  in the vacuum domain structure model}, Phys. Rev. D97~(5) (2018) 056015.
\newblock \href {http://arxiv.org/abs/1802.02905} {\path{arXiv:1802.02905}},
  \href {http://dx.doi.org/10.1103/PhysRevD.97.056015}
  {\path{doi:10.1103/PhysRevD.97.056015}}.

\bibitem{Liendo:2018ukf}
P.~Liendo, C.~Meneghelli, V.~Mitev, {Bootstrapping the half-BPS line defect},
  JHEP 10 (2018) 077.
\newblock \href {http://arxiv.org/abs/1806.01862} {\path{arXiv:1806.01862}},
  \href {http://dx.doi.org/10.1007/JHEP10(2018)077}
  {\path{doi:10.1007/JHEP10(2018)077}}.

\bibitem{Cheng:2018wll}
S.~Cheng, S.-S. Kim, {Refined topological vertex for 5d $Sp(N)$ gauge theories
  with antisymmetric matter }\href {http://arxiv.org/abs/1809.00629}
  {\path{arXiv:1809.00629}}.

\bibitem{Kim:1982jg}
C.-H. Kim, I.-G. Koh, Y.-J. Park, K.~Y. Kim, Y.~Kim, {Generalized Projection
  Matrices for Nonsupersymmetric and Supersymmetric Grand Unified Theories},
  Phys. Rev. D27 (1983) 1932.
\newblock \href {http://dx.doi.org/10.1103/PhysRevD.27.1932}
  {\path{doi:10.1103/PhysRevD.27.1932}}.

\bibitem{Gilmore}
R.~Gilmore, {Lie groups, Lie algebras, and some of their applications}, Wiley,
  New York, 1974.

\bibitem{klimyk_orbit_2006}
A.~Klimyk, J.~Patera, {Orbit Functions}, SIGMA 2 (2006) 6--66.
\newblock \href {http://arxiv.org/abs/math-ph/0601037}
  {\path{arXiv:math-ph/0601037}}, \href
  {http://dx.doi.org/10.3842/SIGMA.2006.006}
  {\path{doi:10.3842/SIGMA.2006.006}}.

\bibitem{Moody:1982}
R.~V. Moody, J.~Patera, {Fast Recursion Formula for Weight Multiplicities},
  Bull.Amer.Math.Soc.(N.S.) 7~(1) (1982) 237--242.

\bibitem{Fonseca:2011sy}
R.~M. Fonseca, Calculating the renormalisation group equations of a susy model
  with susyno, Computer Physics Communications 183~(10) (2012) 2298 -- 2306.
\newblock \href {http://arxiv.org/abs/1106.5016} {\path{arXiv:1106.5016}},
  \href {http://dx.doi.org/10.1016/j.cpc.2012.05.017}
  {\path{doi:10.1016/j.cpc.2012.05.017}}.

\bibitem{klimyk_1967}
A.~Klimyk, {Multiplicities of Weights of Representations and Multiplicities of
  Representations of Semisimple Lie Algebras}, Doklady Akad. Nauk SSSR 177
  (1967) 1001--1004.

\end{thebibliography}

\newpage

\appendix
\colorlet{tableoverheadcolor}{gray!37.5}
% \colorlet{tableheadcolor}{gray!25}
% \colorlet{tablerowcolor}{gray!12.5}
\colorlet{tableheadcolor}{gray!40}
\colorlet{tablerowcolor}{gray!20}

\rowcolors{4}{}{tablerowcolor}

\renewcommand*{\appendixname}{}
% \renewcommand*{\thesection}{\Alph{section}}

% \makeatletter
% \gdef\thetable{\@Alph\c@section.\arabic{subsection}.\arabic{table}}%
% \makeatother

\section{Tables}
\setcounter{table}{0}
We present here tables of properties of irreps, such as Dynkin labels,
dimensional names, indices, congruency classes and the number of singlets in
various subalgebra branchings in Section \ref{ssec:IrrepProperties}, as well as
tables of tensor products in Section \ref{ssec:TensorProducts} and subalgebra
branching rules in Section \ref{ssec:BranchingRules} for many classical and all
exceptional Lie algebras. In presentation style, selection of irreps and
subalgebra branching we closely follow \cite{Slansky:1981yr}, which has been the
definitive reference for unified model building since its publication. The
tables were created by the supplemental package \texttt{Tables.m}, which uses LieART
for the computation. The tables can also be found as Mathematica notebooks in the
LieART documentation integrated into the Wolfram Documentation as
\texttt{Representation Properties}, \texttt{Tensor Products} and \texttt{Branching Rules} under
the section \texttt{Tables} on the LieART documentation home. Since LieART comes with
the functions that generate the tables, the user may extend them to the limit of
his or her computer power.

{
\renewcommand{\arraystretch}{1.3}
\begin{longtable}[!!h]{l|ll|ll|ll}
% \begin{center}
% \begin{tabular}{l|ll|ll|ll}
    \toprule\rowcolor{tableheadcolor}
    &\multicolumn{2}{>{\columncolor{tableheadcolor}}l|}{\textbf{Irrep Properties}} & \multicolumn{2}{>{\columncolor{tableheadcolor}}l|}{\textbf{Tensor Products}} & \multicolumn{2}{>{\columncolor{tableheadcolor}}l}{\textbf{Branching Rules}}\\
    \rowcolor{tableheadcolor}\textbf{Algebra} & \textbf{Number} & \textbf{Page} & \textbf{Number} & \textbf{Page} & \textbf{Number} & \textbf{Page}\\
    \midrule
    \SU2    & \ref{tab:SU2Irreps}  & \pageref{tab:SU2Irreps}  & \ref{tab:SU2TensorProducts}  & \pageref{tab:SU2TensorProducts}  & \ref{tab:SU2BranchingRules}  
    &\pageref{tab:SU2BranchingRules} 
    \\
    \SU3    & \ref{tab:SU3Irreps}  & \pageref{tab:SU3Irreps}  & \ref{tab:SU3TensorProducts}  & \pageref{tab:SU3TensorProducts}  & \ref{tab:SU3BranchingRules}  & \pageref{tab:SU3BranchingRules} \\
    \SU4    & \ref{tab:SU4Irreps}  & \pageref{tab:SU4Irreps}  & \ref{tab:SU4TensorProducts}  & \pageref{tab:SU4TensorProducts}  & \ref{tab:SU4BranchingRules}  & \pageref{tab:SU4BranchingRules} \\
    \SU5    & \ref{tab:SU5Irreps}  & \pageref{tab:SU5Irreps}  & \ref{tab:SU5TensorProducts}  & \pageref{tab:SU5TensorProducts}  & \ref{tab:SU5BranchingRules}  & \pageref{tab:SU5BranchingRules} \\
    \SU6    & \ref{tab:SU6Irreps}  & \pageref{tab:SU6Irreps}  & \ref{tab:SU6TensorProducts}  & \pageref{tab:SU6TensorProducts}  & \ref{tab:SU6BranchingRules}  & \pageref{tab:SU6BranchingRules} \\
    \SU7    & \ref{tab:SU7Irreps}  & \pageref{tab:SU7Irreps}  & \ref{tab:SU7TensorProducts}  & \pageref{tab:SU7TensorProducts}  & \ref{tab:SU7BranchingRules}
    & \pageref{tab:SU7BranchingRules} 
    \\
    \SU8    & \ref{tab:SU8Irreps}  & \pageref{tab:SU8Irreps}  & \ref{tab:SU8TensorProducts}  & \pageref{tab:SU8TensorProducts}  & \ref{tab:SU8BranchingRules}  & \pageref{tab:SU8BranchingRules} \\
    \SU9    & \ref{tab:SU9Irreps}  & \pageref{tab:SU9Irreps}  & \ref{tab:SU9TensorProducts}  & \pageref{tab:SU9TensorProducts}  & \ref{tab:SU9BranchingRules}  & \pageref{tab:SU9BranchingRules} \\
    \SU{10} & \ref{tab:SU10Irreps} & \pageref{tab:SU10Irreps} & \ref{tab:SU10TensorProducts} & \pageref{tab:SU10TensorProducts} & \ref{tab:SU10BranchingRules} & \pageref{tab:SU10BranchingRules}\\
    \SU{11} & \ref{tab:SU11Irreps} & \pageref{tab:SU11Irreps} & \ref{tab:SU11TensorProducts} & \pageref{tab:SU11TensorProducts} & \ref{tab:SU11BranchingRules} & \pageref{tab:SU11BranchingRules}\\
    \SU{12} & \ref{tab:SU12Irreps} & \pageref{tab:SU12Irreps} & \ref{tab:SU12TensorProducts} & \pageref{tab:SU12TensorProducts} & \ref{tab:SU12BranchingRules}
    & \pageref{tab:SU12BranchingRules}
    \\
    \SU{13} & \ref{tab:SU13Irreps} & \pageref{tab:SU13Irreps} & \ref{tab:SU13TensorProducts} & \pageref{tab:SU13TensorProducts} & \ref{tab:SU13BranchingRules}
    & \pageref{tab:SU13BranchingRules}
    \\
    \SU{14} & \ref{tab:SU14Irreps} & \pageref{tab:SU14Irreps} & \ref{tab:SU14TensorProducts} & \pageref{tab:SU14TensorProducts} & \ref{tab:SU14BranchingRules}
    & \pageref{tab:SU14BranchingRules}
    \\
    \SU{15} & \ref{tab:SU15Irreps} & \pageref{tab:SU15Irreps} & \ref{tab:SU15TensorProducts} & \pageref{tab:SU15TensorProducts} & \ref{tab:SU15BranchingRules} & \pageref{tab:SU15BranchingRules}\\
    \SU{16} & \ref{tab:SU16Irreps} & \pageref{tab:SU16Irreps} & \ref{tab:SU16TensorProducts} & \pageref{tab:SU16TensorProducts} & \ref{tab:SU16BranchingRules} & \pageref{tab:SU16BranchingRules}\\
    \midrule
    \SO7    & \ref{tab:SO7Irreps}  & \pageref{tab:SO7Irreps}  & \ref{tab:SO7TensorProducts}  & \pageref{tab:SO7TensorProducts}  & \ref{tab:SO7BranchingRules}  & \pageref{tab:SO7BranchingRules} \\
    \SO8    & \ref{tab:SO8Irreps}  & \pageref{tab:SO8Irreps}  & \ref{tab:SO8TensorProducts}  & \pageref{tab:SO8TensorProducts}  & \ref{tab:SO8BranchingRules}  & \pageref{tab:SO8BranchingRules} \\
    \SO9    & \ref{tab:SO9Irreps}  & \pageref{tab:SO9Irreps}  & \ref{tab:SO9TensorProducts}  & \pageref{tab:SO9TensorProducts}  & \ref{tab:SO9BranchingRules}  & \pageref{tab:SO9BranchingRules} \\
    \SO{10} & \ref{tab:SO10Irreps} & \pageref{tab:SO10Irreps} & \ref{tab:SO10TensorProducts} & \pageref{tab:SO10TensorProducts} & \ref{tab:SO10BranchingRules} & \pageref{tab:SO10BranchingRules}\\
    \SO{11} & \ref{tab:SO11Irreps} & \pageref{tab:SO11Irreps} & \ref{tab:SO11TensorProducts} & \pageref{tab:SO11TensorProducts} & \ref{tab:SO11BranchingRules} & \pageref{tab:SO11BranchingRules}  \\
    \SO{12} & \ref{tab:SO12Irreps} & \pageref{tab:SO12Irreps} & \ref{tab:SO12TensorProducts} & \pageref{tab:SO12TensorProducts} & \ref{tab:SO12BranchingRules} & \pageref{tab:SO12BranchingRules}                \\
    \SO{13} & \ref{tab:SO13Irreps} & \pageref{tab:SO13Irreps} & \ref{tab:SO13TensorProducts} & \pageref{tab:SO13TensorProducts} & \ref{tab:SO13BranchingRules} & \pageref{tab:SO13BranchingRules} \\
    \SO{14} & \ref{tab:SO14Irreps} & \pageref{tab:SO14Irreps} & \ref{tab:SO14TensorProducts} & \pageref{tab:SO14TensorProducts} & \ref{tab:SO14BranchingRules} & \pageref{tab:SO14BranchingRules}\\
    \SO{15} & \ref{tab:SO15Irreps} & \pageref{tab:SO15Irreps} & \ref{tab:SO15TensorProducts} & \pageref{tab:SO15TensorProducts} & \ref{tab:SO15BranchingRules} & \pageref{tab:SO15BranchingRules}\\
    \SO{16} & \ref{tab:SO16Irreps} & \pageref{tab:SO16Irreps} & \ref{tab:SO16TensorProducts} & \pageref{tab:SO16TensorProducts} & \ref{tab:SO16BranchingRules} & \pageref{tab:SO16BranchingRules}\\
    \SO{17} & \ref{tab:SO17Irreps} & \pageref{tab:SO17Irreps} & \ref{tab:SO17TensorProducts} & \pageref{tab:SO17TensorProducts} & \ref{tab:SO17BranchingRules} & \pageref{tab:SO17BranchingRules}\\
    \SO{18} & \ref{tab:SO18Irreps} & \pageref{tab:SO18Irreps} & \ref{tab:SO18TensorProducts} & \pageref{tab:SO18TensorProducts} & \ref{tab:SO18BranchingRules} 
    & \pageref{tab:SO18BranchingRules}
    \\
    \SO{19} & \ref{tab:SO19Irreps} & \pageref{tab:SO19Irreps} & \ref{tab:SO19TensorProducts} & \pageref{tab:SO19TensorProducts} & \ref{tab:SO19BranchingRules} & \pageref{tab:SO19BranchingRules}\\
    \SO{20} & \ref{tab:SO20Irreps} & \pageref{tab:SO20Irreps} & \ref{tab:SO20TensorProducts} & \pageref{tab:SO20TensorProducts} & \ref{tab:SO20BranchingRules} & \pageref{tab:SO20BranchingRules}\\
    \SO{21} & \ref{tab:SO21Irreps} & \pageref{tab:SO21Irreps} & \ref{tab:SO21TensorProducts} & \pageref{tab:SO21TensorProducts} & \ref{tab:SO21BranchingRules} & \pageref{tab:SO21BranchingRules}\\
    \SO{22} & \ref{tab:SO22Irreps} & \pageref{tab:SO22Irreps} & \ref{tab:SO22TensorProducts} & \pageref{tab:SO22TensorProducts} & \ref{tab:SO22BranchingRules} 
    & \pageref{tab:SO22BranchingRules}
    \\
    \SO{23} & \ref{tab:SO23Irreps} & \pageref{tab:SO23Irreps} & \ref{tab:SO23TensorProducts} & \pageref{tab:SO23TensorProducts} & \ref{tab:SO23BranchingRules} & \pageref{tab:SO23BranchingRules}\\
    \SO{24} & \ref{tab:SO24Irreps} & \pageref{tab:SO24Irreps} & \ref{tab:SO24TensorProducts} & \pageref{tab:SO24TensorProducts} & \ref{tab:SO24BranchingRules} & \pageref{tab:SO24BranchingRules}\\
    \SO{25} & \ref{tab:SO25Irreps} & \pageref{tab:SO25Irreps} & \ref{tab:SO25TensorProducts} & \pageref{tab:SO25TensorProducts} & \ref{tab:SO25BranchingRules} & \pageref{tab:SO25BranchingRules}\\
    \SO{26} & \ref{tab:SO26Irreps} & \pageref{tab:SO26Irreps} & \ref{tab:SO26TensorProducts} & \pageref{tab:SO26TensorProducts} & \ref{tab:SO26BranchingRules} 
    & \pageref{tab:SO26BranchingRules}
    \\
    \SO{27} & \ref{tab:SO27Irreps} & \pageref{tab:SO27Irreps} & \ref{tab:SO27TensorProducts} & \pageref{tab:SO27TensorProducts} & \ref{tab:SO27BranchingRules} & \pageref{tab:SO27BranchingRules}\\
    \SO{28} & \ref{tab:SO28Irreps} & \pageref{tab:SO28Irreps} & \ref{tab:SO28TensorProducts} & \pageref{tab:SO28TensorProducts} & \ref{tab:SO28BranchingRules} & \pageref{tab:SO28BranchingRules}\\
    \SO{29} & \ref{tab:SO29Irreps} & \pageref{tab:SO29Irreps} & \ref{tab:SO29TensorProducts} & \pageref{tab:SO29TensorProducts} & \ref{tab:SO29BranchingRules} & \pageref{tab:SO29BranchingRules}\\
    \SO{30} & \ref{tab:SO30Irreps} & \pageref{tab:SO30Irreps} & \ref{tab:SO30TensorProducts} & \pageref{tab:SO30TensorProducts} & \ref{tab:SO30BranchingRules} & \pageref{tab:SO30BranchingRules}\\
    \SO{31} & \ref{tab:SO31Irreps} & \pageref{tab:SO31Irreps} & \ref{tab:SO31TensorProducts} & \pageref{tab:SO31TensorProducts} & \ref{tab:SO31BranchingRules} & \pageref{tab:SO31BranchingRules}\\
    \SO{32} & \ref{tab:SO32Irreps} & \pageref{tab:SO32Irreps} & \ref{tab:SO32TensorProducts} & \pageref{tab:SO32TensorProducts} & \ref{tab:SO32BranchingRules} & \pageref{tab:SO32BranchingRules}\\
    \midrule
    \Sp4    & \ref{tab:Sp4Irreps}  & \pageref{tab:Sp4Irreps}  & \ref{tab:Sp4TensorProducts}  & \pageref{tab:Sp4TensorProducts}  & \ref{tab:Sp4BranchingRules}  & \pageref{tab:Sp4BranchingRules}       \\
    \Sp6    & \ref{tab:Sp6Irreps}  & \pageref{tab:Sp6Irreps}  & \ref{tab:Sp6TensorProducts}  & \pageref{tab:Sp6TensorProducts}  & \ref{tab:Sp6BranchingRules}  & \pageref{tab:Sp6BranchingRules}     \\
    \Sp8    & \ref{tab:Sp8Irreps}  & \pageref{tab:Sp8Irreps}  & \ref{tab:Sp8TensorProducts}  & \pageref{tab:Sp8TensorProducts}  & \ref{tab:Sp8BranchingRules}  & \pageref{tab:Sp8BranchingRules}       \\
    \Sp{10} & \ref{tab:Sp10Irreps} & \pageref{tab:Sp10Irreps} & \ref{tab:Sp10TensorProducts} & \pageref{tab:Sp10TensorProducts} & \ref{tab:Sp10BranchingRules}  & \pageref{tab:Sp10BranchingRules}      \\
    \Sp{12} & \ref{tab:Sp12Irreps} & \pageref{tab:Sp12Irreps} & \ref{tab:Sp12TensorProducts} & \pageref{tab:Sp12TensorProducts} & \ref{tab:Sp12BranchingRules}  & \pageref{tab:Sp12BranchingRules}        \\
    \Sp{14} & \ref{tab:Sp14Irreps} & \pageref{tab:Sp14Irreps} & \ref{tab:Sp14TensorProducts} & \pageref{tab:Sp14TensorProducts} & \ref{tab:Sp14BranchingRules}  & \pageref{tab:Sp14BranchingRules}        \\
    \Sp{16} & \ref{tab:Sp16Irreps} & \pageref{tab:Sp16Irreps} & \ref{tab:Sp16TensorProducts} & \pageref{tab:Sp16TensorProducts} & \ref{tab:Sp16BranchingRules}  & \pageref{tab:Sp16BranchingRules}        \\
    \Sp{18} & \ref{tab:Sp18Irreps} & \pageref{tab:Sp18Irreps} & \ref{tab:Sp18TensorProducts} & \pageref{tab:Sp18TensorProducts} & \ref{tab:Sp18BranchingRules}  & \pageref{tab:Sp18BranchingRules}        \\
    \Sp{20} & \ref{tab:Sp20Irreps} & \pageref{tab:Sp20Irreps} & \ref{tab:Sp20TensorProducts} & \pageref{tab:Sp20TensorProducts} & \ref{tab:Sp20BranchingRules}  & \pageref{tab:Sp20BranchingRules}        \\
    \Sp{22} & \ref{tab:Sp22Irreps} & \pageref{tab:Sp22Irreps} & \ref{tab:Sp22TensorProducts} & \pageref{tab:Sp22TensorProducts} & \ref{tab:Sp22BranchingRules}  & \pageref{tab:Sp22BranchingRules}        \\
    \Sp{24} & \ref{tab:Sp24Irreps} & \pageref{tab:Sp24Irreps} & \ref{tab:Sp24TensorProducts} & \pageref{tab:Sp24TensorProducts} & \ref{tab:Sp24BranchingRules}  & \pageref{tab:Sp24BranchingRules}        \\
    \Sp{26} & \ref{tab:Sp26Irreps} & \pageref{tab:Sp26Irreps} & \ref{tab:Sp26TensorProducts} & \pageref{tab:Sp26TensorProducts} & \ref{tab:Sp26BranchingRules}  & \pageref{tab:Sp26BranchingRules}        \\
    \Sp{28} & \ref{tab:Sp28Irreps} & \pageref{tab:Sp28Irreps} & \ref{tab:Sp28TensorProducts} & \pageref{tab:Sp28TensorProducts} & \ref{tab:Sp28BranchingRules}  & \pageref{tab:Sp28BranchingRules}        \\
    \Sp{30} & \ref{tab:Sp30Irreps} & \pageref{tab:Sp30Irreps} & \ref{tab:Sp30TensorProducts} & \pageref{tab:Sp30TensorProducts} & \ref{tab:Sp30BranchingRules}  & \pageref{tab:Sp30BranchingRules}        \\
    \midrule
    \E6     & \ref{tab:E6Irreps}   & \pageref{tab:E6Irreps}   & \ref{tab:E6TensorProducts}   & \pageref{tab:E6TensorProducts}   & \ref{tab:E6BranchingRules}   
    &\pageref{tab:E6BranchingRules} 
    \\
    \E7     & \ref{tab:E7Irreps}   & \pageref{tab:E7Irreps}   & \ref{tab:E7TensorProducts}   & \pageref{tab:E7TensorProducts}   & \ref{tab:E7BranchingRules}   & \pageref{tab:E7BranchingRules} \\
    \E8     & \ref{tab:E8Irreps}   & \pageref{tab:E8Irreps}   & \ref{tab:E8TensorProducts}   & \pageref{tab:E8TensorProducts}   & \ref{tab:E8BranchingRules}   & \pageref{tab:E8BranchingRules} \\
    \F4     & \ref{tab:F4Irreps}   & \pageref{tab:F4Irreps}   & \ref{tab:F4TensorProducts}   & \pageref{tab:F4TensorProducts}   & \ref{tab:F4BranchingRules}   & \pageref{tab:F4BranchingRules}   \\
    \G2     & \ref{tab:G2Irreps}   & \pageref{tab:G2Irreps}   & \ref{tab:G2TensorProducts}   & \pageref{tab:G2TensorProducts}   & \ref{tab:G2BranchingRules}   & \pageref{tab:G2BranchingRules}       \\
    \bottomrule
% \end{tabular}
% \end{center}
\rowcolor{white}\caption{Table of tables}
\end{longtable}
}
\newpage

\subsection{Properties of Irreducible Representations}
\label{ssec:IrrepProperties}

{
	\rowcolors{2}{tablerowcolor}{}
    \newcommand\starred[1]{#1\makebox[0pt][l]{${}^\ast$}}
    \renewcommand{\irrep}[2][0]{\ensuremath{\irrepbase{#2}\makebox[0pt][l]{${}^{\primes{#1}{\prime}}$}}}
    \renewcommand{\irrepbar}[2][0]{\ensuremath{\irrepbarbase{#2}}\makebox[0pt][l]{${}^{\primes{#1}{\prime}}$}}
    \renewcommand{\irrepsub}[3][0]{\ensuremath{\irrep[#1]{#2}\makebox[0pt][l]{${}_\text{#3}$}}}
    \renewcommand{\irrepbarsub}[3][0]{\ensuremath{\irrepbar[#1]{#2}}\makebox[0pt][l]{${}_\text{#3}$}}%
    \subsubsection{\SU{N}}
\enlargethispage{10pt}
\rowcolors{2}{tablerowcolor}{}
\begin{longtable}{lrrc}
\rowcolor{white}
\caption{SU(2) Irreps\label{tab:SU2Irreps}}\\
\toprule
\rowcolor{tableheadcolor}
Dynkin & Dimension & l & \\
\rowcolor{tableheadcolor}label & (name) & (index) & Duality\\
\midrule
\endfirsthead
\caption[]{SU(2) Irreps (continued)}\\
\toprule
\rowcolor{tableheadcolor}
Dynkin & Dimension & l & \\
\rowcolor{tableheadcolor}label & (name) & (index) & Duality\\
\midrule
\endhead
\endfoot
\bottomrule
\endlastfoot
\dynkin{1} & \irrep{2} & 1 & 1\\
\dynkin{2} & \irrep{3} & 4 & 0\\
\dynkin{3} & \irrep{4} & 10 & 1\\
\dynkin{4} & \irrep{5} & 20 & 0\\
\dynkin{5} & \irrep{6} & 35 & 1\\
\dynkin{6} & \irrep{7} & 56 & 0\\
\dynkin{7} & \irrep{8} & 84 & 1\\
\dynkin{8} & \irrep{9} & 120 & 0\\
\end{longtable}
\newpage

\rowcolors{2}{tablerowcolor}{}
\begin{longtable}{lrrcc}
\rowcolor{white}
\caption{\label{tab:SU3Irreps}SU(3) Irreps}\\
\toprule
\rowcolor{tableheadcolor}
Dynkin & Dimension & l &  & SU(2)\\
\rowcolor{tableheadcolor}label & (name) & (index) & Triality & singlets\\
\midrule
\endfirsthead
\rowcolor{white}
\caption[]{SU(3) Irreps (continued)}\\
\toprule
\rowcolor{tableheadcolor}
Dynkin & Dimension & l &  & SU(2)\\
\rowcolor{tableheadcolor}label & (name) & (index) & Triality & singlets\\
\midrule
\endhead
\rowcolor{white}\multicolumn{5}{l}{\footnotesize $^\ast$SU(2)${\times}$U(1) singlet.}
\endfoot
\bottomrule
\rowcolor{white}\multicolumn{5}{l}{\footnotesize $^\ast$SU(2)${\times}$U(1) singlet.}
\endlastfoot
\dynkin{1, 0} & \irrep{3} & 1 & 1 & 1\\
\dynkin{2, 0} & \irrep{6} & 5 & 2 & 1\\
\dynkin{1, 1} & \irrep{8} & 6 & 0 & \starred{1}\\
\dynkin{3, 0} & \irrep{10} & 15 & 0 & 1\\
\dynkin{2, 1} & \irrep{15} & 20 & 1 & 1\\
\dynkin{4, 0} & \irrep[1]{15} & 35 & 1 & 1\\
\dynkin{0, 5} & \irrep{21} & 70 & 1 & 1\\
\dynkin{1, 3} & \irrep{24} & 50 & 1 & 1\\
\end{longtable}
\newpage

\rowcolors{2}{tablerowcolor}{}
\begin{longtable}{lrrcc}
\caption{\label{tab:SU4Irreps}SU(4) Irreps}\\
\toprule
\rowcolor{tableheadcolor}
Dynkin & Dimension & l &  & SU(3)\\
\rowcolor{tableheadcolor}label & (name) & (index) & Quadrality & singlets\\
\midrule
\endfirsthead
\rowcolor{white}
\caption[]{SU(4) Irreps (continued)}\\
\toprule
\rowcolor{tableheadcolor}
Dynkin & Dimension & l &  & SU(3)\\
\rowcolor{tableheadcolor}label & (name) & (index) & Quadrality & singlets\\
\midrule
\endhead
\rowcolor{white}\multicolumn{5}{l}{\footnotesize $^\ast$SU(3)${\times}$U(1) singlet.}
\endfoot
\bottomrule
\rowcolor{white}\multicolumn{5}{l}{\footnotesize $^\ast$SU(3)${\times}$U(1) singlet.}
\endlastfoot
\dynkin{1, 0, 0} & \irrep{4} & 1 & 1 & 1\\
\dynkin{0, 1, 0} & \irrep{6} & 2 & 2 & 0\\
\dynkin{2, 0, 0} & \irrep{10} & 6 & 2 & 1\\
\dynkin{1, 0, 1} & \irrep{15} & 8 & 0 & \starred{1}\\
\dynkin{0, 1, 1} & \irrep{20} & 13 & 1 & 0\\
\dynkin{0, 2, 0} & \irrep[1]{20} & 16 & 0 & 0\\
\dynkin{0, 0, 3} & \irrep[2]{20} & 21 & 1 & 1\\
\dynkin{4, 0, 0} & \irrep{35} & 56 & 0 & 1\\
\end{longtable}
\newpage

\rowcolors{2}{tablerowcolor}{}
\begin{longtable}{lrrccc}
\caption{\label{tab:SU5Irreps}SU(5) Irreps}\\
\toprule
\rowcolor{tableheadcolor}
Dynkin & Dimension & l &  & SU(4) & SU(3)${\times}$SU(2)\\
\rowcolor{tableheadcolor}label & (name) & (index) & Quintality & singlets & singlets\\
\midrule
\endfirsthead
\rowcolor{white}
\caption[]{SU(5) Irreps (continued)}\\
\toprule
\rowcolor{tableheadcolor}
Dynkin & Dimension & l &  & SU(4) & SU(3)${\times}$SU(2)\\
\rowcolor{tableheadcolor}label & (name) & (index) & Quintality & singlets & singlets\\
\midrule
\endhead
\rowcolor{white}\multicolumn{6}{l}{\footnotesize $^\ast$SU(4)${\times}$U(1) and SU(3)${\times}$SU(2)${\times}$U(1) singlets resp.}
\endfoot
\bottomrule
\rowcolor{white}\multicolumn{6}{l}{\footnotesize $^\ast$SU(4)${\times}$U(1) and SU(3)${\times}$SU(2)${\times}$U(1) singlets resp.}
\endlastfoot
\dynkin{1, 0, 0, 0} & \irrep{5} & 1 & 1 & 1 & 0\\
\dynkin{0, 1, 0, 0} & \irrep{10} & 3 & 2 & 0 & 1\\
\dynkin{2, 0, 0, 0} & \irrep{15} & 7 & 2 & 1 & 0\\
\dynkin{1, 0, 0, 1} & \irrep{24} & 10 & 0 & \starred{1} & \starred{1}\\
\dynkin{0, 0, 0, 3} & \irrep{35} & 28 & 2 & 1 & 0\\
\dynkin{0, 0, 1, 1} & \irrep{40} & 22 & 2 & 0 & 0\\
\dynkin{0, 1, 0, 1} & \irrep{45} & 24 & 1 & 0 & 0\\
\dynkin{0, 0, 2, 0} & \irrep{50} & 35 & 1 & 0 & 1\\
\end{longtable}
\newpage

\rowcolors{2}{tablerowcolor}{}
\begin{longtable}{lrrcccc}
\caption{\label{tab:SU6Irreps}SU(6) Irreps}\\
\toprule
\rowcolor{tableheadcolor}
Dynkin & Dimension & l &  & SU(5) & SU(4)${\times}$SU(2) & SU(3)${\times}$SU(3)\\
\rowcolor{tableheadcolor}label & (name) & (index) & Sextality & singlets & singlets & singlets\\
\midrule
\endfirsthead
\rowcolor{white}
\caption[]{SU(6) Irreps (continued)}\\
\toprule
\rowcolor{tableheadcolor}
Dynkin & Dimension & l &  & SU(5) & SU(4)${\times}$SU(2) & SU(3)${\times}$SU(3)\\
\rowcolor{tableheadcolor}label & (name) & (index) & Sextality & singlets & singlets & singlets\\
\midrule
\endhead
\rowcolor{white}\multicolumn{7}{l}{\footnotesize $^\ast$SU(5)${\times}$U(1) and SU(4)${\times}$SU(2)${\times}$U(1) and SU(3)${\times}$SU(3)${\times}$U(1) singlets resp.}
\endfoot
\bottomrule
\rowcolor{white}\multicolumn{7}{l}{\footnotesize $^\ast$SU(5)${\times}$U(1) and SU(4)${\times}$SU(2)${\times}$U(1) and SU(3)${\times}$SU(3)${\times}$U(1) singlets resp.}
\endlastfoot
\dynkin{1, 0, 0, 0, 0} & \irrep{6} & 1 & 1 & 1 & 0 & 0\\
\dynkin{0, 1, 0, 0, 0} & \irrep{15} & 4 & 2 & 0 & 1 & 0\\
\dynkin{0, 0, 1, 0, 0} & \irrep{20} & 6 & 3 & 0 & 0 & 2\\
\dynkin{2, 0, 0, 0, 0} & \irrep{21} & 8 & 2 & 1 & 0 & 0\\
\dynkin{1, 0, 0, 0, 1} & \irrep{35} & 12 & 0 & \starred{1} & \starred{1} & \starred{1}\\
\dynkin{3, 0, 0, 0, 0} & \irrep{56} & 36 & 3 & 1 & 0 & 0\\
\dynkin{1, 1, 0, 0, 0} & \irrep{70} & 33 & 3 & 0 & 0 & 0\\
\dynkin{0, 1, 0, 0, 1} & \irrep{84} & 38 & 1 & 0 & 0 & 0\\
\end{longtable}
\newpage

\rowcolors{2}{tablerowcolor}{}
\begin{longtable}{lrrcccc}
\caption{\label{tab:SU7Irreps}SU(7) Irreps}\\
\toprule
\rowcolor{tableheadcolor}
Dynkin & Dimension & l &  & SU(6) & SU(5)${\times}$SU(2) & SU(4)${\times}$SU(3)\\
\rowcolor{tableheadcolor}label & (name) & (index) & Septality & singlets & singlets & singlets\\
\midrule
\endfirsthead
\rowcolor{white}
\caption[]{SU(7) Irreps (continued)}\\
\toprule
\rowcolor{tableheadcolor}
Dynkin & Dimension & l &  & SU(6) & SU(5)${\times}$SU(2) & SU(4)${\times}$SU(3)\\
\rowcolor{tableheadcolor}label & (name) & (index) & Septality & singlets & singlets & singlets\\
\midrule
\endhead
\rowcolor{white}\multicolumn{7}{l}{\footnotesize $^\ast$SU(6)${\times}$U(1) and SU(5)${\times}$SU(2)${\times}$U(1) and SU(4)${\times}$SU(3)${\times}$U(1) singlets resp.}
\endfoot
\bottomrule
\rowcolor{white}\multicolumn{7}{l}{\footnotesize $^\ast$SU(6)${\times}$U(1) and SU(5)${\times}$SU(2)${\times}$U(1) and SU(4)${\times}$SU(3)${\times}$U(1) singlets resp.}
\endlastfoot
\dynkin{1, 0, 0, 0, 0, 0} & \irrep{7} & 1 & 1 & 1 & 0 & 0\\
\dynkin{0, 1, 0, 0, 0, 0} & \irrep{21} & 5 & 2 & 0 & 1 & 0\\
\dynkin{2, 0, 0, 0, 0, 0} & \irrep{28} & 9 & 2 & 1 & 0 & 0\\
\dynkin{0, 0, 1, 0, 0, 0} & \irrep{35} & 10 & 3 & 0 & 0 & 1\\
\dynkin{1, 0, 0, 0, 0, 1} & \irrep{48} & 14 & 0 & \starred{1} & \starred{1} & \starred{1}\\
\dynkin{3, 0, 0, 0, 0, 0} & \irrep{84} & 45 & 3 & 1 & 0 & 0\\
\dynkin{1, 1, 0, 0, 0, 0} & \irrep{112} & 46 & 3 & 0 & 0 & 0\\
\dynkin{0, 1, 0, 0, 0, 1} & \irrep{140} & 55 & 1 & 0 & 0 & 0\\
\end{longtable}
\newpage

\setlength{\tabcolsep}{5.5pt}
\rowcolors{2}{tablerowcolor}{}
\begin{longtable}{lrrccccc}
\caption{\label{tab:SU8Irreps}SU(8) Irreps}\\
\toprule
\rowcolor{tableheadcolor}
Dynkin & Dimension & l &  & SU(7) & SU(6)${\times}$SU(2) & SU(5)${\times}$SU(3) & SU(4)${\times}$SU(4)\\
\rowcolor{tableheadcolor}label & (name) & (index) & Octality & singlets & singlets & singlets & singlets\\
\midrule
\endfirsthead
\rowcolor{white}
\caption[]{SU(8) Irreps (continued)}\\
\toprule
\rowcolor{tableheadcolor}
Dynkin & Dimension & l &  & SU(7) & SU(6)${\times}$SU(2) & SU(5)${\times}$SU(3) & SU(4)${\times}$SU(4)\\
\rowcolor{tableheadcolor}label & (name) & (index) & Octality & singlets & singlets & singlets & singlets\\
\midrule
\endhead
\rowcolor{white}\multicolumn{8}{l}{\footnotesize $^\ast$SU(7)${\times}$U(1) and SU(6)${\times}$SU(2)${\times}$U(1) and SU(5)${\times}$SU(3)${\times}$U(1) and SU(4)${\times}$SU(4)${\times}$U(1) singlets resp.}
\endfoot
\bottomrule
\rowcolor{white}\multicolumn{8}{l}{\footnotesize $^\ast$SU(7)${\times}$U(1) and SU(6)${\times}$SU(2)${\times}$U(1) and SU(5)${\times}$SU(3)${\times}$U(1) and SU(4)${\times}$SU(4)${\times}$U(1) singlets resp.}
\endlastfoot
\dynkin{1, 0, 0, 0, 0, 0, 0} & \irrep{8} & 1 & 1 & 1 & 0 & 0 & 0\\
\dynkin{0, 1, 0, 0, 0, 0, 0} & \irrep{28} & 6 & 2 & 0 & 1 & 0 & 0\\
\dynkin{2, 0, 0, 0, 0, 0, 0} & \irrep{36} & 10 & 2 & 1 & 0 & 0 & 0\\
\dynkin{0, 0, 1, 0, 0, 0, 0} & \irrep{56} & 15 & 3 & 0 & 0 & 1 & 0\\
\dynkin{1, 0, 0, 0, 0, 0, 1} & \irrep{63} & 16 & 0 & \starred{1} & \starred{1} & \starred{1} & \starred{1}\\
\dynkin{0, 0, 0, 1, 0, 0, 0} & \irrep{70} & 20 & 4 & 0 & 0 & 0 & 2\\
\dynkin{3, 0, 0, 0, 0, 0, 0} & \irrep{120} & 55 & 3 & 1 & 0 & 0 & 0\\
\dynkin{1, 1, 0, 0, 0, 0, 0} & \irrep{168} & 61 & 3 & 0 & 0 & 0 & 0\\
\end{longtable}
\newpage

\setlength{\tabcolsep}{5pt}
\enlargethispage{20pt}
{
\rowcolors{2}{tablerowcolor}{}
\begin{longtable}{lrrccccc}
\caption{\label{tab:SU9Irreps}SU(9) Irreps}\\
\toprule
\rowcolor{tableheadcolor}
Dynkin & Dimension & l &  & SU(8) & SU(7)${\times}$SU(2) & SU(6)${\times}$SU(3) & SU(5)${\times}$SU(4)\\
\rowcolor{tableheadcolor}label & (name) & (index) & Nonality & singlets & singlets & singlets & singlets\\
\midrule
\endfirsthead
\rowcolor{white}
\caption[]{SU(9) Irreps (continued)}\\
\toprule
\rowcolor{tableheadcolor}
Dynkin & Dimension & l &  & SU(8) & SU(7)${\times}$SU(2) & SU(6)${\times}$SU(3) & SU(5)${\times}$SU(4)\\
\rowcolor{tableheadcolor}label & (name) & (index) & Nonality & singlets & singlets & singlets & singlets\\
\midrule
\endhead
\rowcolor{white}\multicolumn{8}{l}{\footnotesize $^\ast$SU(8)${\times}$U(1) and SU(7)${\times}$SU(2)${\times}$U(1) and SU(6)${\times}$SU(3)${\times}$U(1) and SU(5)${\times}$SU(4)${\times}$U(1) singlets resp.}
\endfoot
\bottomrule
\rowcolor{white}\multicolumn{8}{l}{\footnotesize $^\ast$SU(8)${\times}$U(1) and SU(7)${\times}$SU(2)${\times}$U(1) and SU(6)${\times}$SU(3)${\times}$U(1) and SU(5)${\times}$SU(4)${\times}$U(1) singlets resp.}
\endlastfoot
\dynkin{1, 0, 0, 0, 0, 0, 0, 0} & \irrep{9} & 1 & 1 & 1 & 0 & 0 & 0\\
\dynkin{0, 1, 0, 0, 0, 0, 0, 0} & \irrep{36} & 7 & 2 & 0 & 1 & 0 & 0\\
\dynkin{2, 0, 0, 0, 0, 0, 0, 0} & \irrep{45} & 11 & 2 & 1 & 0 & 0 & 0\\
\dynkin{1, 0, 0, 0, 0, 0, 0, 1} & \irrep{80} & 18 & 0 & \starred{1} & \starred{1} & \starred{1} & \starred{1}\\
\dynkin{0, 0, 1, 0, 0, 0, 0, 0} & \irrep{84} & 21 & 3 & 0 & 0 & 1 & 0\\
\dynkin{0, 0, 0, 1, 0, 0, 0, 0} & \irrep{126} & 35 & 4 & 0 & 0 & 0 & 1\\
\dynkin{3, 0, 0, 0, 0, 0, 0, 0} & \irrep{165} & 66 & 3 & 1 & 0 & 0 & 0\\
\dynkin{1, 1, 0, 0, 0, 0, 0, 0} & \irrep{240} & 78 & 3 & 0 & 0 & 0 & 0\\
\end{longtable}
\vspace{-20pt}

\setlength{\tabcolsep}{1pt}
\rowcolors{2}{tablerowcolor}{}
\begin{longtable}{lrrcccccc}
\caption{\label{tab:SU10Irreps}SU(10) Irreps}\\
\toprule
\rowcolor{tableheadcolor}
Dynkin & Dimension & l &  & SU(9) & SU(8)${\times}$SU(2) & SU(7)${\times}$SU(3) & SU(6)${\times}$SU(4) & SU(5)${\times}$SU(5)\\
\rowcolor{tableheadcolor}label & (name) & (index) & Decality & singlets & singlets & singlets & singlets & singlets\\
\midrule
\endfirsthead
\rowcolor{white}
\caption[]{SU(10) Irreps (continued)}\\
\toprule
\rowcolor{tableheadcolor}
Dynkin & Dimension & l &  & SU(9) & SU(8)${\times}$SU(2) & SU(7)${\times}$SU(3) & SU(6)${\times}$SU(4) & SU(5)${\times}$SU(5)\\
\rowcolor{tableheadcolor}label & (name) & (index) & Decality & singlets & singlets & singlets & singlets & singlets\\
\midrule
\endhead
\rowcolor{white}\multicolumn{9}{>{\raggedright\arraybackslash}p{0.8\textwidth}}{\footnotesize $^\ast$SU(9)${\times}$U(1) and SU(8)${\times}$SU(2)${\times}$U(1) and SU(7)${\times}$SU(3)${\times}$U(1) and SU(6)${\times}$SU(4)${\times}$U(1) and SU(5)${\times}$SU(5)${\times}$U(1) singlets resp.}
\endfoot
\bottomrule
\rowcolor{white}\multicolumn{9}{>{\raggedright\arraybackslash}p{0.8\textwidth}}{\footnotesize $^\ast$SU(9)${\times}$U(1) and SU(8)${\times}$SU(2)${\times}$U(1) and SU(7)${\times}$SU(3)${\times}$U(1) and SU(6)${\times}$SU(4)${\times}$U(1) and SU(5)${\times}$SU(5)${\times}$U(1) singlets resp.}
\endlastfoot
\dynkin{1, 0, 0, 0, 0, 0, 0, 0, 0} & \irrep{10} & 1 & 1 & 1 & 0 & 0 & 0 & 0\\
\dynkin{0, 1, 0, 0, 0, 0, 0, 0, 0} & \irrep{45} & 8 & 2 & 0 & 1 & 0 & 0 & 0\\
\dynkin{2, 0, 0, 0, 0, 0, 0, 0, 0} & \irrep{55} & 12 & 2 & 1 & 0 & 0 & 0 & 0\\
\dynkin{1, 0, 0, 0, 0, 0, 0, 0, 1} & \irrep{99} & 20 & 0 & \starred{1} & \starred{1} & \starred{1} & \starred{1} & \starred{1}\\
\dynkin{0, 0, 1, 0, 0, 0, 0, 0, 0} & \irrep{120} & 28 & 3 & 0 & 0 & 1 & 0 & 0\\
\dynkin{0, 0, 0, 1, 0, 0, 0, 0, 0} & \irrep{210} & 56 & 4 & 0 & 0 & 0 & 1 & 0\\
\dynkin{3, 0, 0, 0, 0, 0, 0, 0, 0} & \irrep{220} & 78 & 3 & 1 & 0 & 0 & 0 & 0\\
\dynkin{0, 0, 0, 0, 1, 0, 0, 0, 0} & \irrep{252} & 70 & 5 & 0 & 0 & 0 & 0 & 2\\
\end{longtable}
}
\newpage

\setlength{\tabcolsep}{0.5pt}
\rowcolors{2}{tablerowcolor}{}
\begin{longtable}{lrrcccccc}
\caption{\label{tab:SU11Irreps}SU(11) Irreps}\\
\toprule
\rowcolor{tableheadcolor}
Dynkin & Dimension & l &  & SU(10) & SU(9)${\times}$SU(2) & SU(8)${\times}$SU(3) & SU(7)${\times}$SU(4) & SU(6)${\times}$SU(5)\\
\rowcolor{tableheadcolor}label & (name) & (index) & Decality & singlets & singlets & singlets & singlets & singlets\\
\midrule
\endfirsthead
\rowcolor{white}
\caption[]{SU(11) Irreps (continued)}\\
\toprule
\rowcolor{tableheadcolor}
Dynkin & Dimension & l &  & SU(10) & SU(9)${\times}$SU(2) & SU(8)${\times}$SU(3) & SU(7)${\times}$SU(4) & SU(6)${\times}$SU(5)\\
\rowcolor{tableheadcolor}label & (name) & (index) & Decality & singlets & singlets & singlets & singlets & singlets\\
\midrule
\endhead
\rowcolor{white}\multicolumn{9}{>{\raggedright\arraybackslash}p{0.8\textwidth}}{\footnotesize $^\ast$SU(10)${\times}$U(1) and SU(9)${\times}$SU(2)${\times}$U(1) and SU(8)${\times}$SU(3)${\times}$U(1) and SU(7)${\times}$SU(4)${\times}$U(1) and SU(6)${\times}$SU(5)${\times}$U(1) singlets resp.}
\endfoot
\bottomrule
\rowcolor{white}\multicolumn{9}{>{\raggedright\arraybackslash}p{0.8\textwidth}}{\footnotesize $^\ast$SU(10)${\times}$U(1) and SU(9)${\times}$SU(2)${\times}$U(1) and SU(8)${\times}$SU(3)${\times}$U(1) and SU(7)${\times}$SU(4)${\times}$U(1) and SU(6)${\times}$SU(5)${\times}$U(1) singlets resp.}
\endlastfoot
\dynkin{1, 0, 0, 0, 0, 0, 0, 0, 0, 0} & \irrep{11} & 1 & 1 & 1 & 0 & 0 & 0 & 0\\
\dynkin{0, 1, 0, 0, 0, 0, 0, 0, 0, 0} & \irrep{55} & 9 & 2 & 0 & 1 & 0 & 0 & 0\\
\dynkin{2, 0, 0, 0, 0, 0, 0, 0, 0, 0} & \irrep{66} & 13 & 2 & 1 & 0 & 0 & 0 & 0\\
\dynkin{1, 0, 0, 0, 0, 0, 0, 0, 0, 1} & \irrep{120} & 22 & 0 & \starred{1} & \starred{1} & \starred{1} & \starred{1} & \starred{1}\\
\dynkin{0, 0, 1, 0, 0, 0, 0, 0, 0, 0} & \irrep{165} & 36 & 3 & 0 & 0 & 1 & 0 & 0\\
\dynkin{3, 0, 0, 0, 0, 0, 0, 0, 0, 0} & \irrep{286} & 91 & 3 & 1 & 0 & 0 & 0 & 0\\
\dynkin{0, 0, 0, 1, 0, 0, 0, 0, 0, 0} & \irrep{330} & 84 & 4 & 0 & 0 & 0 & 1 & 0\\
\dynkin{1, 1, 0, 0, 0, 0, 0, 0, 0, 0} & \irrep{440} & 118 & 3 & 0 & 0 & 0 & 0 & 0\\
\end{longtable}

\rowcolors{2}{tablerowcolor}{}
\begin{longtable}{lrrcccccc}
\caption{\label{tab:SU12Irreps}SU(12) Irreps}\\
\toprule
\rowcolor{tableheadcolor}
Dynkin & Dimension & l &  & SU(11) & SU(10)${\times}$SU(2) & SU(9)${\times}$SU(3) & SU(7)${\times}$SU(5) & SU(6)${\times}$SU(6)\\
\rowcolor{tableheadcolor}label & (name) & (index) & Decality & singlets & singlets & singlets & singlets & singlets\\
\midrule
\endfirsthead
\rowcolor{white}
\caption[]{SU(12) Irreps (continued)}\\
\toprule
\rowcolor{tableheadcolor}
Dynkin & Dimension & l &  & SU(11) & SU(10)${\times}$SU(2) & SU(9)${\times}$SU(3) & SU(7)${\times}$SU(5) & SU(6)${\times}$SU(6)\\
\rowcolor{tableheadcolor}label & (name) & (index) & Decality & singlets & singlets & singlets & singlets & singlets\\
\midrule
\endhead
\rowcolor{white}\multicolumn{9}{>{\raggedright\arraybackslash}p{0.8\textwidth}}{\footnotesize $^\ast$SU(11)${\times}$U(1) and SU(10)${\times}$SU(2)${\times}$U(1) and SU(9)${\times}$SU(3)${\times}$U(1) and SU(7)${\times}$SU(5)${\times}$U(1) and SU(6)${\times}$SU(6)${\times}$U(1) singlets resp.}
\endfoot
\bottomrule
\rowcolor{white}\multicolumn{9}{>{\raggedright\arraybackslash}p{0.8\textwidth}}{\footnotesize $^\ast$SU(11)${\times}$U(1) and SU(10)${\times}$SU(2)${\times}$U(1) and SU(9)${\times}$SU(3)${\times}$U(1) and SU(7)${\times}$SU(5)${\times}$U(1) and SU(6)${\times}$SU(6)${\times}$U(1) singlets resp.}
\endlastfoot
\dynkin{1, 0, 0, 0, 0, 0, 0, 0, 0, 0, 0} & \irrep{12} & 1 & 1 & 1 & 0 & 0 & 0 & 0\\
\dynkin{0, 1, 0, 0, 0, 0, 0, 0, 0, 0, 0} & \irrep{66} & 10 & 2 & 0 & 1 & 0 & 0 & 0\\
\dynkin{2, 0, 0, 0, 0, 0, 0, 0, 0, 0, 0} & \irrep{78} & 14 & 2 & 1 & 0 & 0 & 0 & 0\\
\dynkin{1, 0, 0, 0, 0, 0, 0, 0, 0, 0, 1} & \irrep{143} & 24 & 0 & \starred{1} & \starred{1} & \starred{1} & \starred{1} & \starred{1}\\
\dynkin{0, 0, 1, 0, 0, 0, 0, 0, 0, 0, 0} & \irrep{220} & 45 & 3 & 0 & 0 & 1 & 0 & 0\\
\dynkin{3, 0, 0, 0, 0, 0, 0, 0, 0, 0, 0} & \irrep{364} & 105 & 3 & 1 & 0 & 0 & 0 & 0\\
\dynkin{0, 0, 0, 1, 0, 0, 0, 0, 0, 0, 0} & \irrep{495} & 120 & 4 & 0 & 0 & 0 & 0 & 0\\
\dynkin{1, 1, 0, 0, 0, 0, 0, 0, 0, 0, 0} & \irrep{572} & 141 & 3 & 0 & 0 & 0 & 0 & 0\\
\end{longtable}

\rowcolors{2}{tablerowcolor}{}
\begin{longtable}{lrrcccccc}
\caption{\label{tab:SU13Irreps}SU(13) Irreps}\\
\toprule
\rowcolor{tableheadcolor}
Dynkin & Dimension & l &  & SU(12) & SU(11)${\times}$SU(2) & SU(10)${\times}$SU(3) & SU(9)${\times}$SU(4) & SU(8)${\times}$SU(5)\\
\rowcolor{tableheadcolor}label & (name) & (index) & Decality &
singlets & singlets & singlets & singlets & singlets\\
\midrule 
\endfirsthead
\caption[]{SU(13) Irreps (continued)}\\
\toprule
\rowcolor{tableheadcolor}
Dynkin & Dimension & l &  & SU(12) & SU(11)${\times}$SU(2) & SU(10)${\times}$SU(3) & SU(9)${\times}$SU(4) & SU(8)${\times}$SU(5) \\
\rowcolor{tableheadcolor}label & (name) & (index) & Decality &
singlets & singlets & singlets & singlets & singlets \\
\midrule 
\endhead
\multicolumn{9}{l}{\footnotesize $^\ast$SU(12)${\times}$U(1) and
SU(11)${\times}$SU(2)${\times}$U(1) and SU(10)${\times}$SU(3)${\times}$U(1) and SU(9)${\times}$SU(4)${\times}$U(1) and SU(8)${\times}$SU(5)${\times}$U(1)
singlets resp.}
\endfoot
\bottomrule
\rowcolor{white}\multicolumn{9}{l}{\footnotesize $^\ast$SU(12)${\times}$U(1) and
SU(11)${\times}$SU(2)${\times}$U(1) and SU(10)${\times}$SU(3)${\times}$U(1) and SU(9)${\times}$SU(4)${\times}$U(1) and SU(8)${\times}$SU(5)${\times}$U(1) singlets resp.}
\endlastfoot
\dynkin{1, 0, 0, 0, 0, 0, 0, 0, 0, 0, 0, 0} & \irrep{13} & 1 & 1 & 1 
& 0 & 0 & 0 & 0\\
\dynkin{0, 1, 0, 0, 0, 0, 0, 0, 0, 0, 0, 0} & \irrep{78} & 11 & 2 & 0 
& 1 & 0 & 0 & 0\\
\dynkin{2, 0, 0, 0, 0, 0, 0, 0, 0, 0, 0, 0} & \irrep{91} & 15 & 2 & 1 
& 0 & 0 & 0 & 0\\
\dynkin{1, 0, 0, 0, 0, 0, 0, 0, 0, 0, 0, 1} & \irrep{168} & 26 & 0 & 
\starred{1} & \starred{1} & \starred{1} & \starred{1} & \starred{1} \\
\dynkin{0, 0, 1, 0, 0, 0, 0, 0, 0, 0, 0, 0} & \irrep{286} & 55 & 3 & 
0 & 0 & 1 & 0 & 0\\
\dynkin{3, 0, 0, 0, 0, 0, 0, 0, 0, 0, 0, 0} & \irrep{455} & 120 & 3 & 
1 & 0 & 0 & 0 & 0\\
\dynkin{0, 0, 0, 1, 0, 0, 0, 0, 0, 0, 0, 0} & \irrep{715} & 165 & 4 & 
0 & 0 & 0 & 1 & 0\\
\dynkin{1, 1, 0, 0, 0, 0, 0, 0, 0, 0, 0, 0} & \irrep{728} & 166 & 3 & 
0 & 0 & 0 & 0 & 0\\
\end{longtable}

\rowcolors{2}{tablerowcolor}{}
\begin{longtable}{lrrcccccc}
\caption{\label{tab:SU14Irreps}SU(14) Irreps}\\
\toprule
\rowcolor{tableheadcolor}
Dynkin & Dimension & l &  & SU(13) & SU(12)${\times}$SU(2) & SU(11)${
\times}$SU(3) & SU(10)${\times}$SU(4) & SU(9)${\times}$SU(5)\\
\rowcolor{tableheadcolor}label & (name) & (index) & Decality & 
singlets & singlets & singlets & singlets & 
singlets\\
\midrule 
\endfirsthead
\caption[]{SU(14) Irreps (continued)}\\
\toprule
\rowcolor{tableheadcolor}
Dynkin & Dimension & l &  & SU(13) & SU(12)${\times}$SU(2) & SU(11)${
\times}$SU(3) & SU(10)${\times}$SU(4) & SU(9)${\times}$SU(5)\\
\rowcolor{tableheadcolor}label & (name) & (index) & Decality & 
singlets & singlets & singlets & singlets & singlets\\
\midrule 
\endhead
\multicolumn{9}{l}{\footnotesize $^\ast$SU(13)${\times}$U(1) and 
SU(12)${\times}$SU(2)${\times}$U(1) and SU(11)${\times}$SU(3)${\times}$U(1) and SU(10)${\times}$SU(4)${\times}$U(1) and SU(9)${\times}$SU(5)${\times}$U(1) singlets resp.}
\endfoot
\bottomrule
\rowcolor{white}\multicolumn{9}{l}{\footnotesize $^\ast$SU(13)${\times}$U(1) and 
SU(12)${\times}$SU(2)${\times}$U(1) and SU(11)${\times}$SU(3)${\times}$U(1) and SU(10)${\times}$SU(4)${\times}$U(1) and SU(9)${\times}$SU(5)${\times}$U(1) singlets resp.}
\endlastfoot
\dynkin{1, 0, 0, 0, 0, 0, 0, 0, 0, 0, 0, 0, 0} & \irrep{14} & 1 & 1 & 
1 & 0 & 0 & 0 & 0\\
\dynkin{0, 1, 0, 0, 0, 0, 0, 0, 0, 0, 0, 0, 0} & \irrep{91} & 12 & 2 
& 0 & 1 & 0 & 0 & 0\\
\dynkin{2, 0, 0, 0, 0, 0, 0, 0, 0, 0, 0, 0, 0} & \irrep{105} & 16 & 2 
& 1 & 0 & 0 & 0 & 0\\
\dynkin{1, 0, 0, 0, 0, 0, 0, 0, 0, 0, 0, 0, 1} & \irrep{195} & 28 & 0 
& \starred{1} & \starred{1} & \starred{1} & \starred{1} & \starred{1} 
\\
\dynkin{0, 0, 1, 0, 0, 0, 0, 0, 0, 0, 0, 0, 0} & \irrep{364} & 66 & 3 
& 0 & 0 & 1 & 0 & 0\\
\dynkin{1, 1, 0, 0, 0, 0, 0, 0, 0, 0, 0, 0, 0} & \irrep{910} & 193 & 3 & 0 & 0 & 0 & 0 & 0\\
\end{longtable}

\rowcolors{2}{tablerowcolor}{}
\begin{longtable}{lrrcccccc}
\caption{\label{tab:SU15Irreps}SU(15) Irreps}\\
\toprule
\rowcolor{tableheadcolor}
Dynkin & Dimension & l &  & SU(14) & SU(13)${\times}$SU(2) & SU(12)${
\times}$SU(3) & SU(11)${\times}$SU(4) & SU(10)${\times}$SU(5)\\
\rowcolor{tableheadcolor}label & (name) & (index) & Decality & 
singlets & singlets & singlets & singlets & singlets\\
\midrule 
\endfirsthead
\caption[]{SU(15) Irreps (continued)}\\
\toprule
\rowcolor{tableheadcolor}
Dynkin & Dimension & l &  & SU(14) & SU(13)${\times}$SU(2) & SU(12)${
\times}$SU(3) & SU(11)${\times}$SU(4) & SU(10)${\times}$SU(5)\\
\rowcolor{tableheadcolor}label & (name) & (index) & Decality & 
singlets & singlets & singlets & singlets & singlets\\
\midrule 
\endhead
\multicolumn{9}{l}{\footnotesize $^\ast$SU(14)${\times}$U(1) and 
SU(13)${\times}$SU(2)${\times}$U(1) and SU(12)${\times}$SU(3)${\times}$U(1) and SU(11)${\times}$SU(4)${\times}$U(1) and SU(10)${\times}$SU(5)${\times}$U(1) singlets resp.}
\endfoot
\bottomrule
\rowcolor{white}\multicolumn{9}{l}{\footnotesize $^\ast$SU(14)${\times}$U(1) and 
SU(13)${\times}$SU(2)${\times}$U(1) and SU(12)${\times}$SU(3)${\times}$U(1) and SU(11)${\times}$SU(4)${\times}$U(1) and SU(10)${\times}$SU(5)${\times}$U(1) singlets resp.}
\endlastfoot
\dynkin{1, 0, 0, 0, 0, 0, 0, 0, 0, 0, 0, 0, 0, 0} & \irrep{15} & 1 &
1 & 1 & 0 & 0 & 0 & 0\\
\dynkin{0, 1, 0, 0, 0, 0, 0, 0, 0, 0, 0, 0, 0, 0} & \irrep{105} & 13
& 2 & 0 & 1 & 0 & 0 & 0\\
\dynkin{2, 0, 0, 0, 0, 0, 0, 0, 0, 0, 0, 0, 0, 0} & \irrep{120} & 17
& 2 & 1 & 0 & 0 & 0 & 0\\
\dynkin{1, 0, 0, 0, 0, 0, 0, 0, 0, 0, 0, 0, 0, 1} & \irrep{224} & 30
& 0 & \starred{1} & \starred{1} & \starred{1} & \starred{1} & \
\starred{1}\\
\dynkin{0, 0, 1, 0, 0, 0, 0, 0, 0, 0, 0, 0, 0, 0} & \irrep{455} & 78
& 3 & 0 & 0 & 1 & 0 & 0\\
\dynkin{1, 1, 0, 0, 0, 0, 0, 0, 0, 0, 0, 0, 0, 0} & \irrep{1120} & 222 & 3 & 0 & 0 & 0 & 0 & 0\\
\dynkin{0, 0, 0, 1, 0, 0, 0, 0, 0, 0, 0, 0, 0, 0} & \irrep{1365} & 286 & 4 & 0 & 0 & 0 & 1 & 0\\
\dynkin{0, 1, 0, 0, 0, 0, 0, 0, 0, 0, 0, 0, 0, 1} & \irrep{1560} & 299 & 1 & 0 & 0 & 0 & 0 & 0\\
\end{longtable}

\rowcolors{2}{tablerowcolor}{}
\begin{longtable}{lrrcccccc}
\caption{\label{tab:SU16Irreps}SU(16) Irreps}\\
\toprule
\rowcolor{tableheadcolor}
Dynkin & Dimension & l &  & SU(15) & SU(14)${\times}$SU(2) & SU(13)${\times}$SU(3) & SU(12)${\times}$SU(4) & SU(11)${\times}$SU(5)\\
\rowcolor{tableheadcolor}label & (name) & (index) & Decality & singlets & singlets & singlets & singlets & singlets\\
\midrule 
\endfirsthead
\caption[]{SU(16) Irreps (continued)}\\
\toprule
\rowcolor{tableheadcolor}
Dynkin & Dimension & l &  & SU(15) & SU(14)${\times}$SU(2) & SU(13)${\times}$SU(3) & SU(12)${\times}$SU(4) & SU(11)${\times}$SU(5)\\
\rowcolor{tableheadcolor}label & (name) & (index) & Decality & singlets & singlets & singlets & singlets & singlets\\
\midrule 
\endhead
\multicolumn{9}{l}{\footnotesize $^\ast$SU(15)${\times}$U(1) and SU(14)${\times}$SU(2)${\times}$U(1) and SU(13)${\times}$SU(3)${\times}$U(1) and SU(12)${\times}$SU(4)${\times}$U(1) and SU(11)${\times}$SU(5)${\times}$U(1) singlets resp.}
\endfoot
\bottomrule
\rowcolor{white}\multicolumn{9}{l}{\footnotesize $^\ast$SU(15)${\times}$U(1) and SU(14)${\times}$SU(2)${\times}$U(1) and SU(13)${\times}$SU(3)${\times}$U(1) and SU(12)${\times}$SU(4)${\times}$U(1) and SU(11)${\times}$SU(5)${\times}$U(1) singlets resp.}
\endlastfoot
\dynkin{1, 0, 0, 0, 0, 0, 0, 0, 0, 0, 0, 0, 0, 0, 0} & \irrep{16} & 1 & 1 & 1 & 0 & 0 & 0 & 0\\
\dynkin{0, 1, 0, 0, 0, 0, 0, 0, 0, 0, 0, 0, 0, 0, 0} & \irrep{120} & 14 & 2 & 0 & 1 & 0 & 0 & 0\\
\dynkin{2, 0, 0, 0, 0, 0, 0, 0, 0, 0, 0, 0, 0, 0, 0} & \irrep{136} & 18 & 2 & 1 & 0 & 0 & 0 & 0\\
\dynkin{1, 0, 0, 0, 0, 0, 0, 0, 0, 0, 0, 0, 0, 0, 1} & \irrep{255} & 32 & 0 & \starred{1} & \starred{1} & \starred{1} & \starred{1} & \starred{1}\\
\dynkin{0, 0, 1, 0, 0, 0, 0, 0, 0, 0, 0, 0, 0, 0, 0} & \irrep{560} & 91 & 3 & 0 & 0 & 1 & 0 & 0\\
\end{longtable}
\newpage

\subsubsection{\SO{N}}

\setlength{\tabcolsep}{7pt}

\rowcolors{2}{tablerowcolor}{}
%
}
\newpage

\subsubsection{\Sp{N}}

\rowcolors{2}{tablerowcolor}{}
\begin{longtable}{lrrc}
\caption{\label{tab:Sp4Irreps}Sp(4) Irreps}\\
\toprule
\rowcolor{tableheadcolor}
Dynkin & Dimension & l & Congruency\\
\rowcolor{tableheadcolor}label & (name) & (index) & class\\
\midrule
\endfirsthead
\rowcolor{white}
\caption[]{Sp(4) Irreps (continued)}\\
\toprule
\rowcolor{tableheadcolor}
Dynkin & Dimension & l & Congruency\\
\rowcolor{tableheadcolor}label & (name) & (index) & class\\
\midrule
\endhead
\endfoot
\bottomrule
\endlastfoot
\dynkin{1, 0} & \irrep{4} & 1 & 1\\
\dynkin{0, 1} & \irrep{5} & 2 & 0\\
\dynkin{2, 0} & \irrep{10} & 6 & 0\\
\dynkin{0, 2} & \irrep{14} & 14 & 0\\
\dynkin{1, 1} & \irrep{16} & 12 & 1\\
\dynkin{3, 0} & \irrep{20} & 21 & 1\\
\dynkin{0, 3} & \irrep{30} & 54 & 0\\
\dynkin{2, 1} & \irrep{35} & 42 & 0\\
\end{longtable}
\newpage

\rowcolors{2}{tablerowcolor}{}
\begin{longtable}{lrrc}
\caption{\label{tab:Sp6Irreps}Sp(6) Irreps}\\
\toprule
\rowcolor{tableheadcolor}
Dynkin & Dimension & l & Congruency\\
\rowcolor{tableheadcolor}label & (name) & (index) & class\\
\midrule
\endfirsthead
\rowcolor{white}
\caption[]{Sp(6) Irreps (continued)}\\
\toprule
\rowcolor{tableheadcolor}
Dynkin & Dimension & l & Congruency\\
\rowcolor{tableheadcolor}label & (name) & (index) & class\\
\midrule
\endhead
\endfoot
\bottomrule
\endlastfoot
\dynkin{1, 0, 0} & \irrep{6} & 1 & 1\\
\dynkin{0, 1, 0} & \irrep{14} & 4 & 0\\
\dynkin{0, 0, 1} & \irrep[1]{14} & 5 & 1\\
\dynkin{2, 0, 0} & \irrep{21} & 8 & 0\\
\dynkin{3, 0, 0} & \irrep{56} & 36 & 1\\
\dynkin{1, 1, 0} & \irrep{64} & 32 & 1\\
\dynkin{1, 0, 1} & \irrep{70} & 40 & 0\\
\dynkin{0, 0, 2} & \irrep{84} & 72 & 0\\
\end{longtable}
\newpage

\rowcolors{2}{tablerowcolor}{}
\begin{longtable}{lrrc}
\caption{\label{tab:Sp8Irreps}Sp(8) Irreps}\\
\toprule
\rowcolor{tableheadcolor}
Dynkin & Dimension & l & Congruency\\
\rowcolor{tableheadcolor}label & (name) & (index) & class\\
\midrule
\endfirsthead
\rowcolor{white}
\caption[]{Sp(8) Irreps (continued)}\\
\toprule
\rowcolor{tableheadcolor}
Dynkin & Dimension & l & Congruency\\
\rowcolor{tableheadcolor}label & (name) & (index) & class\\
\midrule
\endhead
\endfoot
\bottomrule
\endlastfoot
\dynkin{1, 0, 0, 0} & \irrep{8} & 1 & 1\\
\dynkin{0, 1, 0, 0} & \irrep{27} & 6 & 0\\
\dynkin{2, 0, 0, 0} & \irrep{36} & 10 & 0\\
\dynkin{0, 0, 0, 1} & \irrep{42} & 14 & 0\\
\dynkin{0, 0, 1, 0} & \irrep{48} & 14 & 1\\
\dynkin{3, 0, 0, 0} & \irrep{120} & 55 & 1\\
\dynkin{1, 1, 0, 0} & \irrep{160} & 60 & 1\\
\dynkin{1, 0, 0, 1} & \irrep{288} & 140 & 1\\
\end{longtable}
\newpage

\rowcolors{2}{tablerowcolor}{}
\begin{longtable}{lrrc}
\caption{\label{tab:Sp10Irreps}Sp(10) Irreps}\\
\toprule
\rowcolor{tableheadcolor}
Dynkin & Dimension & l & Congruency\\
\rowcolor{tableheadcolor}label & (name) & (index) & class\\
\midrule
\endfirsthead
\rowcolor{white}
\caption[]{Sp(10) Irreps (continued)}\\
\toprule
\rowcolor{tableheadcolor}
Dynkin & Dimension & l & Congruency\\
\rowcolor{tableheadcolor}label & (name) & (index) & class\\
\midrule
\endhead
\endfoot
\bottomrule
\endlastfoot
\dynkin{1, 0, 0, 0, 0} & \irrep{10} & 1 & 1\\
\dynkin{0, 1, 0, 0, 0} & \irrep{44} & 8 & 0\\
\dynkin{2, 0, 0, 0, 0} & \irrep{55} & 12 & 0\\
\dynkin{0, 0, 1, 0, 0} & \irrep{110} & 27 & 1\\
\dynkin{0, 0, 0, 0, 1} & \irrep{132} & 42 & 1\\
\dynkin{0, 0, 0, 1, 0} & \irrep{165} & 48 & 0\\
\dynkin{3, 0, 0, 0, 0} & \irrep{220} & 78 & 1\\
\dynkin{1, 1, 0, 0, 0} & \irrep{320} & 96 & 1\\
\end{longtable}
\newpage

\rowcolors{2}{tablerowcolor}{}
\begin{longtable}{lrrc}
\caption{\label{tab:Sp12Irreps}Sp(12) Irreps}\\
\toprule
\rowcolor{tableheadcolor}
Dynkin & Dimension & l & Congruency\\
\rowcolor{tableheadcolor}label & (name) & (index) & class\\
\midrule
\endfirsthead
\rowcolor{white}
\caption[]{Sp(12) Irreps (continued)}\\
\toprule
\rowcolor{tableheadcolor}
Dynkin & Dimension & l & Congruency\\
\rowcolor{tableheadcolor}label & (name) & (index) & class\\
\midrule
\endhead
\endfoot
\bottomrule
\endlastfoot
\dynkin{1, 0, 0, 0, 0, 0} & \irrep{12} & 1 & 1\\
\dynkin{0, 1, 0, 0, 0, 0} & \irrep{65} & 10 & 0\\
\dynkin{2, 0, 0, 0, 0, 0} & \irrep{78} & 14 & 0\\
\dynkin{0, 0, 1, 0, 0, 0} & \irrep{208} & 44 & 1\\
\dynkin{3, 0, 0, 0, 0, 0} & \irrep{364} & 105 & 1\\
\dynkin{0, 0, 0, 1, 0, 0} & \irrep{429} & 110 & 0\\
\dynkin{0, 0, 0, 0, 0, 1} & \irrep[1]{429} & 132 & 0\\
\dynkin{1, 1, 0, 0, 0, 0} & \irrep{560} & 140 & 1\\
\end{longtable}
\newpage

\rowcolors{2}{tablerowcolor}{}
\begin{longtable}{lrrc}
\caption{\label{tab:Sp14Irreps}Sp(14) Irreps}\\
\toprule
\rowcolor{tableheadcolor}
Dynkin & Dimension & l & Congruency\\
\rowcolor{tableheadcolor}label & (name) & (index) & class\\
\midrule
\endfirsthead
\rowcolor{white}
\caption[]{Sp(14) Irreps (continued)}\\
\toprule
\rowcolor{tableheadcolor}
Dynkin & Dimension & l & Congruency\\
\rowcolor{tableheadcolor}label & (name) & (index) & class\\
\midrule
\endhead
\endfoot
\bottomrule
\endlastfoot
\dynkin{1, 0, 0, 0, 0, 0, 0} & \irrep{14} & 1 & 1\\
\dynkin{0, 1, 0, 0, 0, 0, 0} & \irrep{90} & 12 & 0\\
\dynkin{2, 0, 0, 0, 0, 0, 0} & \irrep{105} & 16 & 0\\
\dynkin{0, 0, 1, 0, 0, 0, 0} & \irrep{350} & 65 & 1\\
\dynkin{3, 0, 0, 0, 0, 0, 0} & \irrep{560} & 136 & 1\\
\dynkin{1, 1, 0, 0, 0, 0, 0} & \irrep{896} & 192 & 1\\
\dynkin{0, 0, 0, 1, 0, 0, 0} & \irrep{910} & 208 & 0\\
\dynkin{0, 0, 0, 0, 0, 0, 1} & \irrep{1430} & 429 & 1\\
\dynkin{0, 0, 0, 0, 1, 0, 0} & \irrep{1638} & 429 & 1\\
\dynkin{0, 0, 0, 0, 0, 1, 0} & \irrep{2002} & 572 & 0\\
\dynkin{4, 0, 0, 0, 0, 0, 0} & \irrep{2380} & 816 & 0\\
\dynkin{0, 2, 0, 0, 0, 0, 0} & \irrep{3094} & 884 & 0\\
\dynkin{1, 0, 1, 0, 0, 0, 0} & \irrep{3900} & 1040 & 0\\
\dynkin{2, 1, 0, 0, 0, 0, 0} & \irrep{5355} & 1632 & 0\\
\dynkin{5, 0, 0, 0, 0, 0, 0} & \irrep{8568} & 3876 & 1\\
\dynkin{1, 0, 0, 1, 0, 0, 0} & \irrep{10752} & 3328 & 1\\
\dynkin{0, 1, 1, 0, 0, 0, 0} & \irrep{17850} & 6035 & 1\\
\dynkin{1, 0, 0, 0, 0, 0, 1} & \irrep{18018} & 6864 & 0\\
\dynkin{1, 0, 0, 0, 1, 0, 0} & \irrep{20020} & 6864 & 0\\
\dynkin{3, 1, 0, 0, 0, 0, 0} & \irrep{24192} & 9792 & 1\\
\dynkin{1, 2, 0, 0, 0, 0, 0} & \irrep{24570} & 9243 & 1\\
\dynkin{2, 0, 1, 0, 0, 0, 0} & \irrep{24752} & 8840 & 1\\
\dynkin{1, 0, 0, 0, 0, 1, 0} & \irrep{24960} & 9152 & 1\\
\end{longtable}

\rowcolors{2}{tablerowcolor}{}
\begin{longtable}{lrrc}
\caption{\label{tab:Sp16Irreps}Sp(16) Irreps}\\
\toprule
\rowcolor{tableheadcolor}
Dynkin & Dimension & l & Congruency\\
\rowcolor{tableheadcolor}label & (name) & (index) & class\\
\midrule 
\endfirsthead
\caption[]{Sp(16) Irreps (continued)}\\
\toprule
\rowcolor{tableheadcolor}
Dynkin & Dimension & l & Congruency\\
\rowcolor{tableheadcolor}label & (name) & (index) & class\\
\midrule 
\endhead
\endfoot
\bottomrule
\endlastfoot
\dynkin{1, 0, 0, 0, 0, 0, 0, 0} & \irrep{16} & 1 & 1\\
\dynkin{0, 1, 0, 0, 0, 0, 0, 0} & \irrep{119} & 14 & 0\\
\dynkin{2, 0, 0, 0, 0, 0, 0, 0} & \irrep{136} & 18 & 0\\
\dynkin{0, 0, 1, 0, 0, 0, 0, 0} & \irrep{544} & 90 & 1\\
\dynkin{3, 0, 0, 0, 0, 0, 0, 0} & \irrep{816} & 171 & 1\\
\dynkin{1, 1, 0, 0, 0, 0, 0, 0} & \irrep{1344} & 252 & 1\\
\dynkin{0, 0, 0, 1, 0, 0, 0, 0} & \irrep{1700} & 350 & 0\\
\dynkin{0, 0, 0, 0, 1, 0, 0, 0} & \irrep{3808} & 910 & 1\\
\dynkin{4, 0, 0, 0, 0, 0, 0, 0} & \irrep{3876} & 1140 & 0\\
\dynkin{0, 0, 0, 0, 0, 0, 0, 1} & \irrep{4862} & 1430 & 0\\
\dynkin{0, 2, 0, 0, 0, 0, 0, 0} & \irrep{5320} & 1330 & 0\\
\dynkin{0, 0, 0, 0, 0, 1, 0, 0} & \irrep{6188} & 1638 & 0\\
\dynkin{1, 0, 1, 0, 0, 0, 0, 0} & \irrep{6885} & 1620 & 0\\
\dynkin{0, 0, 0, 0, 0, 0, 1, 0} & \irrep{7072} & 2002 & 1\\
\dynkin{2, 1, 0, 0, 0, 0, 0, 0} & \irrep{9044} & 2394 & 0\\
\dynkin{1, 0, 0, 1, 0, 0, 0, 0} & \irrep{22848} & 6300 & 1\\
\end{longtable}
\newpage

\rowcolors{2}{tablerowcolor}{}
\begin{longtable}{lrrc}
\caption{\label{tab:Sp18Irreps}Sp(18) Irreps}\\
\toprule
\rowcolor{tableheadcolor}
Dynkin & Dimension & l & Congruency\\
\rowcolor{tableheadcolor}label & (name) & (index) & class\\
\midrule 
\endfirsthead
\caption[]{Sp(18) Irreps (continued)}\\
\toprule
\rowcolor{tableheadcolor}
Dynkin & Dimension & l & Congruency\\
\rowcolor{tableheadcolor}label & (name) & (index) & class\\
\midrule 
\endhead
\endfoot
\bottomrule
\endlastfoot
\dynkin{1, 0, 0, 0, 0, 0, 0, 0, 0} & \irrep{18} & 1 & 1\\
\dynkin{0, 1, 0, 0, 0, 0, 0, 0, 0} & \irrep{152} & 16 & 0\\
\dynkin{2, 0, 0, 0, 0, 0, 0, 0, 0} & \irrep{171} & 20 & 0\\
\dynkin{0, 0, 1, 0, 0, 0, 0, 0, 0} & \irrep{798} & 119 & 1\\
\dynkin{3, 0, 0, 0, 0, 0, 0, 0, 0} & \irrep{1140} & 210 & 1\\
\dynkin{1, 1, 0, 0, 0, 0, 0, 0, 0} & \irrep{1920} & 320 & 1\\
\dynkin{0, 0, 0, 1, 0, 0, 0, 0, 0} & \irrep{2907} & 544 & 0\\
\dynkin{4, 0, 0, 0, 0, 0, 0, 0, 0} & \irrep{5985} & 1540 & 0\\
\dynkin{0, 0, 0, 0, 1, 0, 0, 0, 0} & \irrep{7752} & 1700 & 1\\
\dynkin{0, 2, 0, 0, 0, 0, 0, 0, 0} & \irrep{8568} & 1904 & 0\\
\dynkin{1, 0, 1, 0, 0, 0, 0, 0, 0} & \irrep{11305} & 2380 & 0\\
\dynkin{2, 1, 0, 0, 0, 0, 0, 0, 0} & \irrep{14364} & 3360 & 0\\
\dynkin{0, 0, 0, 0, 0, 1, 0, 0, 0} & \irrep{15504} & 3808 & 0\\
\dynkin{0, 0, 0, 0, 0, 0, 0, 0, 1} & \irrep{16796} & 4862 & 1\\
\end{longtable}

\rowcolors{2}{tablerowcolor}{}
\begin{longtable}{lrrc}
\caption{\label{tab:Sp20Irreps}Sp(20) Irreps}\\
\toprule
\rowcolor{tableheadcolor}
Dynkin & Dimension & l & Congruency\\
\rowcolor{tableheadcolor}label & (name) & (index) & class\\
\midrule 
\endfirsthead
\caption[]{Sp(20) Irreps (continued)}\\
\toprule
\rowcolor{tableheadcolor}
Dynkin & Dimension & l & Congruency\\
\rowcolor{tableheadcolor}label & (name) & (index) & class\\
\midrule 
\endhead
\endfoot
\bottomrule
\endlastfoot
\dynkin{1, 0, 0, 0, 0, 0, 0, 0, 0, 0} & \irrep{20} & 1 & 1\\
\dynkin{0, 1, 0, 0, 0, 0, 0, 0, 0, 0} & \irrep{189} & 18 & 0\\
\dynkin{2, 0, 0, 0, 0, 0, 0, 0, 0, 0} & \irrep{210} & 22 & 0\\
\dynkin{0, 0, 1, 0, 0, 0, 0, 0, 0, 0} & \irrep{1120} & 152 & 1\\
\dynkin{3, 0, 0, 0, 0, 0, 0, 0, 0, 0} & \irrep{1540} & 253 & 1\\
\dynkin{1, 1, 0, 0, 0, 0, 0, 0, 0, 0} & \irrep{2640} & 396 & 1\\
\dynkin{0, 0, 0, 1, 0, 0, 0, 0, 0, 0} & \irrep{4655} & 798 & 0\\
\dynkin{4, 0, 0, 0, 0, 0, 0, 0, 0, 0} & \irrep{8855} & 2024 & 0\\
\end{longtable}

\rowcolors{2}{tablerowcolor}{}
\begin{longtable}{lrrc}
\caption{\label{tab:Sp22Irreps}Sp(22) Irreps}\\
\toprule
\rowcolor{tableheadcolor}
Dynkin & Dimension & l & Congruency\\
\rowcolor{tableheadcolor}label & (name) & (index) & class\\
\midrule 
\endfirsthead
\caption[]{Sp(22) Irreps (continued)}\\
\toprule
\rowcolor{tableheadcolor}
Dynkin & Dimension & l & Congruency\\
\rowcolor{tableheadcolor}label & (name) & (index) & class\\
\midrule 
\endhead
\endfoot
\bottomrule
\endlastfoot
\dynkin{1, 0, 0, 0, 0, 0, 0, 0, 0, 0, 0} & \irrep{22} & 1 & 1\\
\dynkin{0, 1, 0, 0, 0, 0, 0, 0, 0, 0, 0} & \irrep{230} & 20 & 0\\
\dynkin{2, 0, 0, 0, 0, 0, 0, 0, 0, 0, 0} & \irrep{253} & 24 & 0\\
\dynkin{0, 0, 1, 0, 0, 0, 0, 0, 0, 0, 0} & \irrep{1518} & 189 & 1\\
\dynkin{3, 0, 0, 0, 0, 0, 0, 0, 0, 0, 0} & \irrep{2024} & 300 & 1\\
\dynkin{1, 1, 0, 0, 0, 0, 0, 0, 0, 0, 0} & \irrep{3520} & 480 & 1\\
\dynkin{0, 0, 0, 1, 0, 0, 0, 0, 0, 0, 0} & \irrep{7084} & 1120 & 0\\
\dynkin{0, 2, 0, 0, 0, 0, 0, 0, 0, 0, 0} & \irrep{19250} & 3500 & 0\\
\dynkin{0, 0, 0, 0, 1, 0, 0, 0, 0, 0, 0} & \irrep{24794} & 4655 & 1\\
\end{longtable}
\newpage

\rowcolors{2}{tablerowcolor}{}
\begin{longtable}{lrrc}
\caption{\label{tab:Sp24Irreps}Sp(24) Irreps}\\
\toprule
\rowcolor{tableheadcolor}
Dynkin & Dimension & l & Congruency\\
\rowcolor{tableheadcolor}label & (name) & (index) & class\\
\midrule 
\endfirsthead
\caption[]{Sp(24) Irreps (continued)}\\
\toprule
\rowcolor{tableheadcolor}
Dynkin & Dimension & l & Congruency\\
\rowcolor{tableheadcolor}label & (name) & (index) & class\\
\midrule 
\endhead
\endfoot
\bottomrule
\endlastfoot
\dynkin{1, 0, 0, 0, 0, 0, 0, 0, 0, 0, 0, 0} & \irrep{24} & 1 & 1\\
\dynkin{0, 1, 0, 0, 0, 0, 0, 0, 0, 0, 0, 0} & \irrep{275} & 22 & 0\\
\dynkin{2, 0, 0, 0, 0, 0, 0, 0, 0, 0, 0, 0} & \irrep{300} & 26 & 0\\
\dynkin{0, 0, 1, 0, 0, 0, 0, 0, 0, 0, 0, 0} & \irrep{2000} & 230 & 1\\
\dynkin{1, 1, 0, 0, 0, 0, 0, 0, 0, 0, 0, 0} & \irrep{4576} & 572 & 1\\
\dynkin{0, 0, 0, 1, 0, 0, 0, 0, 0, 0, 0, 0} & \irrep{10350} & 1518 & 0\\
\dynkin{0, 2, 0, 0, 0, 0, 0, 0, 0, 0, 0, 0} & \irrep{27324} & 4554 & 0\\
\end{longtable}

\rowcolors{2}{tablerowcolor}{}
\begin{longtable}{lrrc}
\caption{\label{tab:Sp26Irreps}Sp(26) Irreps}\\
\toprule
\rowcolor{tableheadcolor}
Dynkin & Dimension & l & Congruency\\
\rowcolor{tableheadcolor}label & (name) & (index) & class\\
\midrule 
\endfirsthead
\caption[]{Sp(26) Irreps (continued)}\\
\toprule
\rowcolor{tableheadcolor}
Dynkin & Dimension & l & Congruency\\
\rowcolor{tableheadcolor}label & (name) & (index) & class\\
\midrule 
\endhead
\endfoot
\bottomrule
\endlastfoot
\dynkin{1, 0, 0, 0, 0, 0, 0, 0, 0, 0, 0, 0, 0} & \irrep{26} & 1 & 1\\
\dynkin{0, 1, 0, 0, 0, 0, 0, 0, 0, 0, 0, 0, 0} & \irrep{324} & 24 & 0\\
\dynkin{2, 0, 0, 0, 0, 0, 0, 0, 0, 0, 0, 0, 0} & \irrep{351} & 28 & 0\\
\dynkin{0, 0, 1, 0, 0, 0, 0, 0, 0, 0, 0, 0, 0} & \irrep{2574} & 275 & 1\\
\dynkin{1, 1, 0, 0, 0, 0, 0, 0, 0, 0, 0, 0, 0} & \irrep{5824} & 672 & 1\\
\dynkin{0, 0, 0, 1, 0, 0, 0, 0, 0, 0, 0, 0, 0} & \irrep{14625} & 2000 & 0\\
\dynkin{0, 2, 0, 0, 0, 0, 0, 0, 0, 0, 0, 0, 0} & \irrep{37700} & 5800 & 0\\
\end{longtable}

\rowcolors{2}{tablerowcolor}{}
\begin{longtable}{lrrc}
\caption{\label{tab:Sp28Irreps}Sp(28) Irreps}\\
\toprule
\rowcolor{tableheadcolor}
Dynkin & Dimension & l & Congruency\\
\rowcolor{tableheadcolor}label & (name) & (index) & class\\
\midrule 
\endfirsthead
\caption[]{Sp(28) Irreps (continued)}\\
\toprule
\rowcolor{tableheadcolor}
Dynkin & Dimension & l & Congruency\\
\rowcolor{tableheadcolor}label & (name) & (index) & class\\
\midrule 
\endhead
\endfoot
\bottomrule
\endlastfoot
\dynkin{1, 0, 0, 0, 0, 0, 0, 0, 0, 0, 0, 0, 0, 0} & \irrep{28} & 1 & 1\\
\dynkin{0, 1, 0, 0, 0, 0, 0, 0, 0, 0, 0, 0, 0, 0} & \irrep{377} & 26 & 0\\
\dynkin{0, 0, 1, 0, 0, 0, 0, 0, 0, 0, 0, 0, 0, 0} & \irrep{3248} & 324 & 1\\
\dynkin{1, 1, 0, 0, 0, 0, 0, 0, 0, 0, 0, 0, 0, 0} & \irrep{7280} & 780 & 1\\
\dynkin{0, 0, 0, 1, 0, 0, 0, 0, 0, 0, 0, 0, 0, 0} & \irrep{20097} & 2574 & 0\\
\dynkin{1, 0, 1, 0, 0, 0, 0, 0, 0, 0, 0, 0, 0, 0} & \irrep{70470} & 9720 & 0\\
\dynkin{0, 0, 0, 0, 1, 0, 0, 0, 0, 0, 0, 0, 0, 0} & \irrep{95004} & 14625 & 1\\
\dynkin{0, 0, 0, 0, 0, 1, 0, 0, 0, 0, 0, 0, 0, 0} & \irrep{356265} & 63180 & 0\\
\end{longtable}

\rowcolors{2}{tablerowcolor}{}
\begin{longtable}{lrrc}
\caption{\label{tab:Sp30Irreps}Sp(30) Irreps}\\
\toprule
\rowcolor{tableheadcolor}
Dynkin & Dimension & l & Congruency\\
\rowcolor{tableheadcolor}label & (name) & (index) & class\\
\midrule 
\endfirsthead
\caption[]{Sp(30) Irreps (continued)}\\
\toprule
\rowcolor{tableheadcolor}
Dynkin & Dimension & l & Congruency\\
\rowcolor{tableheadcolor}label & (name) & (index) & class\\
\midrule 
\endhead
\endfoot
\bottomrule
\endlastfoot
\dynkin{1, 0, 0, 0, 0, 0, 0, 0, 0, 0, 0, 0, 0, 0, 0} & \irrep{30} & 1 & 1\\
\dynkin{0, 1, 0, 0, 0, 0, 0, 0, 0, 0, 0, 0, 0, 0, 0} & \irrep{434} & 28 & 0\\
\dynkin{0, 0, 1, 0, 0, 0, 0, 0, 0, 0, 0, 0, 0, 0, 0} & \irrep{4030} & 377 & 1\\
\dynkin{1, 1, 0, 0, 0, 0, 0, 0, 0, 0, 0, 0, 0, 0, 0} & \irrep{8960} & 896 & 1\\
\dynkin{0, 0, 0, 1, 0, 0, 0, 0, 0, 0, 0, 0, 0, 0, 0} & \irrep{26970} & 3248 & 0\\
\dynkin{1, 0, 1, 0, 0, 0, 0, 0, 0, 0, 0, 0, 0, 0, 0} & \irrep{93496} & 12064 & 0\\
\dynkin{0, 0, 0, 0, 1, 0, 0, 0, 0, 0, 0, 0, 0, 0, 0} & \irrep{138446} & 20097 & 1\\
\end{longtable}
\newpage

\subsubsection{Exceptional Algebras}

\enlargethispage{10pt}
\rowcolors{2}{tablerowcolor}{}
\begin{longtable}{lrrcccc}
\caption{\label{tab:E6Irreps}\E6 Irreps}\\
\toprule
\rowcolor{tableheadcolor}
Dynkin & Dimension & l/6 &  & SO(10) & SU(6)${\times}$SU(2) & SU(3)${\times}$SU(3)${\times}$SU(3)\\
\rowcolor{tableheadcolor}label & (name) & (index) & Triality & singlets & singlets & singlets\\
\midrule
\endfirsthead
\rowcolor{white}
\caption[]{$E_6$ Irreps (continued)}\\
\toprule
\rowcolor{tableheadcolor}
Dynkin & Dimension & l/6 &  & SO(10) & SU(6)${\times}$SU(2) & SU(3)${\times}$SU(3)${\times}$SU(3)\\
\rowcolor{tableheadcolor}label & (name) & (index) & Triality & singlets & singlets & singlets\\
\midrule
\endhead
\rowcolor{white}\multicolumn{7}{l}{\footnotesize $^\ast$SO(10)${\times}$U(1) singlets resp.}
\endfoot
\bottomrule
\rowcolor{white}\multicolumn{7}{l}{\footnotesize $^\ast$SO(10)${\times}$U(1) singlets resp.}
\endlastfoot
\dynkin{1, 0, 0, 0, 0, 0} & \irrep{27} & 1 & 1 & 1 & 0 & 0\\
\dynkin{0, 0, 0, 0, 0, 1} & \irrep{78} & 4 & 0 & \starred{1} & 0 & 0\\
\dynkin{0, 0, 0, 1, 0, 0} & \irrep{351} & 25 & 1 & 0 & 0 & 0\\
\dynkin{0, 0, 0, 0, 2, 0} & \irrep[1]{351} & 28 & 1 & 1 & 0 & 0\\
\dynkin{1, 0, 0, 0, 1, 0} & \irrep{650} & 50 & 0 & \starred{1} & 1 & 2\\
\dynkin{1, 0, 0, 0, 0, 1} & \irrep{1728} & 160 & 1 & 1 & 0 & 0\\
\dynkin{0, 0, 0, 0, 0, 2} & \irrep{2430} & 270 & 0 & \starred{1} & 1 & 1\\
\dynkin{0, 0, 1, 0, 0, 0} & \irrep{2925} & 300 & 0 & 0 & 0 & 1\\
\end{longtable}
\newpage

\rowcolors{2}{tablerowcolor}{}
\begin{longtable}{lrrc}
\caption{\label{tab:E7Irreps}\E7 Irreps}\\
\toprule
\rowcolor{tableheadcolor}
Dynkin & Dimension & l/12 & Congruency\\
\rowcolor{tableheadcolor}label & (name) & (index) & class\\
\midrule
\endfirsthead
\rowcolor{white}
\caption[]{$E_7$ Irreps (continued)}\\
\toprule
\rowcolor{tableheadcolor}
Dynkin & Dimension & l/12 & Congruency\\
\rowcolor{tableheadcolor}label & (name) & (index) & class\\
\midrule
\endhead
\endfoot
\bottomrule
\endlastfoot
\dynkin{0, 0, 0, 0, 0, 1, 0} & \irrep{56} & 1 & 1\\
\dynkin{1, 0, 0, 0, 0, 0, 0} & \irrep{133} & 3 & 0\\
\dynkin{0, 0, 0, 0, 0, 0, 1} & \irrep{912} & 30 & 1\\
\dynkin{0, 0, 0, 0, 0, 2, 0} & \irrep{1463} & 55 & 0\\
\dynkin{0, 0, 0, 0, 1, 0, 0} & \irrep{1539} & 54 & 0\\
\dynkin{1, 0, 0, 0, 0, 1, 0} & \irrep{6480} & 270 & 1\\
\dynkin{2, 0, 0, 0, 0, 0, 0} & \irrep{7371} & 351 & 0\\
\dynkin{0, 1, 0, 0, 0, 0, 0} & \irrep{8645} & 390 & 0\\
\end{longtable}
\vspace{-10pt}

\rowcolors{2}{tablerowcolor}{}
\begin{longtable}{lrrc}
\caption{\label{tab:E8Irreps}\E8 Irreps}\\
\toprule
\rowcolor{tableheadcolor}
Dynkin & Dimension & l/60 & Congruency\\
\rowcolor{tableheadcolor}label & (name) & (index) & class\\
\midrule
\endfirsthead
\rowcolor{white}
\caption[]{$E_8$ Irreps (continued)}\\
\toprule
\rowcolor{tableheadcolor}
Dynkin & Dimension & l/60 & Congruency\\
\rowcolor{tableheadcolor}label & (name) & (index) & class\\
\midrule
\endhead
\endfoot
\bottomrule
\endlastfoot
\dynkin{0, 0, 0, 0, 0, 0, 1, 0} & \irrep{248} & 1 & 0\\
\dynkin{1, 0, 0, 0, 0, 0, 0, 0} & \irrep{3875} & 25 & 0\\
\dynkin{0, 0, 0, 0, 0, 0, 2, 0} & \irrep{27000} & 225 & 0\\
\dynkin{0, 0, 0, 0, 0, 1, 0, 0} & \irrep{30380} & 245 & 0\\
\dynkin{0, 0, 0, 0, 0, 0, 0, 1} & \irrep{147250} & 1425 & 0\\
\dynkin{1, 0, 0, 0, 0, 0, 1, 0} & \irrep{779247} & 8379 & 0\\
\dynkin{0, 0, 0, 0, 0, 0, 3, 0} & \irrep{1763125} & 22750 & 0\\
\dynkin{0, 0, 0, 0, 1, 0, 0, 0} & \irrep{2450240} & 29640 & 0\\
\dynkin{0, 0, 0, 0, 0, 1, 1, 0} & \irrep{4096000} & 51200 & 0\\
\dynkin{2, 0, 0, 0, 0, 0, 0, 0} & \irrep{4881384} & 65610 & 0\\
\dynkin{0, 1, 0, 0, 0, 0, 0, 0} & \irrep{6696000} & 88200 & 0\\
\dynkin{0, 0, 0, 0, 0, 0, 1, 1} & \irrep{26411008} & 372736 & 0\\
\dynkin{1, 0, 0, 0, 0, 0, 2, 0} & \irrep{70680000} & 1083000 & 0\\
\dynkin{1, 0, 0, 0, 0, 1, 0, 0} & \irrep{76271625} & 1148175 & 0\\
\end{longtable}
\newpage

\rowcolors{2}{tablerowcolor}{}
\begin{longtable}{lrrc}
\caption{\label{tab:F4Irreps}\F4 Irreps}\\
\toprule
\rowcolor{tableheadcolor}
Dynkin & Dimension & l/6 & Congruency\\
\rowcolor{tableheadcolor}label & (name) & (index) & class\\
\midrule
\endfirsthead
\rowcolor{white}
\caption[]{$F_4$ Irreps (continued)}\\
\toprule
\rowcolor{tableheadcolor}
Dynkin & Dimension & l/6 & Congruency\\
\rowcolor{tableheadcolor}label & (name) & (index) & class\\
\midrule
\endhead
\endfoot
\bottomrule
\endlastfoot
\dynkin{0, 0, 0, 1} & \irrep{26} & 1 & 0\\
\dynkin{1, 0, 0, 0} & \irrep{52} & 3 & 0\\
\dynkin{0, 0, 1, 0} & \irrep{273} & 21 & 0\\
\dynkin{0, 0, 0, 2} & \irrep{324} & 27 & 0\\
\dynkin{1, 0, 0, 1} & \irrep{1053} & 108 & 0\\
\dynkin{2, 0, 0, 0} & \irrep[1]{1053} & 135 & 0\\
\dynkin{0, 1, 0, 0} & \irrep{1274} & 147 & 0\\
\dynkin{0, 0, 0, 3} & \irrep{2652} & 357 & 0\\
\end{longtable}
\newpage

\rowcolors{2}{tablerowcolor}{}
\begin{longtable}{lrrc}
\caption{\label{tab:G2Irreps}\G2 Irreps}\\
\toprule
\rowcolor{tableheadcolor}
Dynkin & Dimension & l/2 & Congruency\\
\rowcolor{tableheadcolor}label & (name) & (index) & class\\
\midrule
\endfirsthead
\rowcolor{white}
\caption[]{$G_2$ Irreps (continued)}\\
\toprule
\rowcolor{tableheadcolor}
Dynkin & Dimension & l/2 & Congruency\\
\rowcolor{tableheadcolor}label & (name) & (index) & class\\
\midrule
\endhead
\endfoot
\bottomrule
\endlastfoot
\dynkin{1, 0} & \irrep{7} & 1 & 0\\
\dynkin{0, 1} & \irrep{14} & 4 & 0\\
\dynkin{2, 0} & \irrep{27} & 9 & 0\\
\dynkin{1, 1} & \irrep{64} & 32 & 0\\
\dynkin{3, 0} & \irrep{77} & 44 & 0\\
\dynkin{0, 2} & \irrep[1]{77} & 55 & 0\\
\dynkin{4, 0} & \irrep{182} & 156 & 0\\
\dynkin{2, 1} & \irrep{189} & 144 & 0\\
\end{longtable}
\newpage

}
\newpage

\subsection{Tensor Products}
\label{ssec:TensorProducts}

\setlength\extrarowheight{1.3pt}
\renewcommand{\tabcolsep}{2pt}
\subsubsection{\SU{N}}

\vspace{-10pt}
\enlargethispage{10pt}
\rowcolors{2}{}{tablerowcolor}\begin{longtable}{rcl}
\rowcolor{white}\caption{\label{tab:SU2TensorProducts}SU(2) Tensor Products}\\
\toprule
\endfirsthead
\rowcolor{white}\caption[]{SU(2) Tensor Products (continued)}\\
\endhead
\bottomrule
\endlastfoot
$\irrep{2}\times\irrep{2}$ & = & $\irrep{1}+\irrep{3}$\\
$\irrep{3}\times\irrep{2}$ & = & $\irrep{2}+\irrep{4}$\\
$\irrep{3}\times\irrep{3}$ & = & $\irrep{1}+\irrep{3}+\irrep{5}$\\
$\irrep{4}\times\irrep{2}$ & = & $\irrep{3}+\irrep{5}$\\
$\irrep{4}\times\irrep{3}$ & = & $\irrep{2}+\irrep{4}+\irrep{6}$\\
$\irrep{4}\times\irrep{4}$ & = & $\irrep{1}+\irrep{3}+\irrep{5}+\irrep{7}$\\
$\irrep{5}\times\irrep{2}$ & = & $\irrep{4}+\irrep{6}$\\
$\irrep{5}\times\irrep{3}$ & = & $\irrep{3}+\irrep{5}+\irrep{7}$\\
\end{longtable}
\newpage

\rowcolors{2}{}{tablerowcolor}\begin{longtable}{rcl}
\rowcolor{white}\caption{\label{tab:SU3TensorProducts}SU(3) Tensor Products}\\
\toprule
\endfirsthead
\rowcolor{white}\caption[]{SU(3) Tensor Products (continued)}\\
\endhead
\bottomrule
\endlastfoot
$\irrepbar{3}\times\irrep{3}$ & = & $\irrep{1}+\irrep{8}$\\
$\irrep{3}\times\irrep{3}$ & = & $\irrepbar{3}+\irrep{6}$\\
$\irrepbar{6}\times\irrep{3}$ & = & $\irrepbar{3}+\irrepbar{15}$\\
$\irrep{6}\times\irrep{3}$ & = & $\irrep{8}+\irrep{10}$\\
$\irrepbar{6}\times\irrepbar{6}$ & = & $\irrep{6}+\irrepbar{15}+\irrepbar[1]{15}$\\
$\irrep{6}\times\irrepbar{6}$ & = & $\irrep{1}+\irrep{8}+\irrep{27}$\\
$\irrep{8}\times\irrep{3}$ & = & $\irrep{3}+\irrepbar{6}+\irrep{15}$\\
$\irrep{8}\times\irrepbar{6}$ & = & $\irrep{3}+\irrepbar{6}+\irrep{15}+\irrep{24}$\\
\end{longtable}
\newpage

\enlargethispage{10pt}
\rowcolors{2}{}{tablerowcolor}\begin{longtable}{rcl}
\rowcolor{white}\caption{\label{tab:SU4TensorProducts}SU(4) Tensor Products}\\
\toprule
\endfirsthead
\rowcolor{white}\caption[]{SU(4) Tensor Products (continued)}\\
\endhead
\bottomrule
\endlastfoot
$\irrepbar{4}\times\irrep{4}$ & = & $\irrep{1}+\irrep{15}$\\
$\irrep{4}\times\irrep{4}$ & = & $\irrep{6}+\irrep{10}$\\
$\irrep{6}\times\irrep{4}$ & = & $\irrepbar{4}+\irrepbar{20}$\\
$\irrep{6}\times\irrep{6}$ & = & $\irrep{1}+\irrep{15}+\irrep[1]{20}$\\
$\irrepbar{10}\times\irrep{4}$ & = & $\irrepbar{4}+\irrepbar{36}$\\
$\irrep{10}\times\irrep{4}$ & = & $\irrepbar{20}+\irrepbar[2]{20}$\\
$\irrep{10}\times\irrep{6}$ & = & $\irrep{15}+\irrep{45}$\\
$\irrepbar{10}\times\irrep{10}$ & = & $\irrep{1}+\irrep{15}+\irrep{84}$\\
\end{longtable}
\newpage

\rowcolors{2}{}{tablerowcolor}\begin{longtable}{rcl}
\rowcolor{white}\caption{\label{tab:SU5TensorProducts}SU(5) Tensor Products}\\
\toprule
\endfirsthead
\rowcolor{white}\caption[]{SU(5) Tensor Products (continued)}\\
\endhead
\bottomrule
\endlastfoot
$\irrepbar{5}\times\irrep{5}$ & = & $\irrep{1}+\irrep{24}$\\
$\irrep{5}\times\irrep{5}$ & = & $\irrep{10}+\irrep{15}$\\
$\irrepbar{10}\times\irrep{5}$ & = & $\irrepbar{5}+\irrepbar{45}$\\
$\irrep{10}\times\irrep{5}$ & = & $\irrepbar{10}+\irrepbar{40}$\\
$\irrepbar{10}\times\irrep{10}$ & = & $\irrep{1}+\irrep{24}+\irrep{75}$\\
$\irrep{10}\times\irrep{10}$ & = & $\irrepbar{5}+\irrepbar{45}+\irrepbar{50}$\\
$\irrepbar{15}\times\irrep{5}$ & = & $\irrepbar{5}+\irrepbar{70}$\\
$\irrep{15}\times\irrep{5}$ & = & $\irrepbar{35}+\irrepbar{40}$\\
\end{longtable}%
\newpage

\subsubsection{\SO{N}}

\rowcolors{2}{}{tablerowcolor}%

\newpage

\subsubsection{\Sp{N}}
\enlargethispage{10pt}
\rowcolors{2}{}{tablerowcolor}%
\newpage

\subsubsection{Exceptional Algebras}

\rowcolors{2}{}{tablerowcolor}%
}

\newpage

\subsection{Branching Rules}
\label{ssec:BranchingRules}

\input{tables/BranchingRulesTables}

\end{document}